\documentclass[prx,epsf,showpacs,twocolumn,superscriptaddress,footinbib]{revtex4-2}

\usepackage[pdftex]{graphicx}
\usepackage{dcolumn}
\usepackage{bm}
\usepackage{epsfig}
\usepackage{latexsym}
\usepackage{amsmath}
\usepackage{amsfonts}
\usepackage[position=top,caption=false,singlelinecheck=off, justification=raggedright]{subfig}
\usepackage{color}
\usepackage{array}
\usepackage{braket}
\usepackage{bbold}
\usepackage{dsfont}
\usepackage{mathrsfs}
\usepackage{tikz}
\usetikzlibrary{quantikz}
\usepackage{hyperref}
\hypersetup{
    colorlinks,
    citecolor=blue,
    filecolor=red,
    linkcolor=magenta,
    urlcolor=blue
}

\renewcommand{\[}{\left[}

\newcommand{\eps}{\epsilon}

\newcommand\sbullet[1][.5]{\mathbin{\vcenter{\hbox{\scalebox{#1}{$\bullet$}}}}}

\makeatletter
\DeclareFontFamily{OMX}{MnSymbolE}{}
\DeclareSymbolFont{MnLargeSymbols}{OMX}{MnSymbolE}{m}{n}
\SetSymbolFont{MnLargeSymbols}{bold}{OMX}{MnSymbolE}{b}{n}
\DeclareFontShape{OMX}{MnSymbolE}{m}{n}{
    <-6>  MnSymbolE5
   <6-7>  MnSymbolE6
   <7-8>  MnSymbolE7
   <8-9>  MnSymbolE8
   <9-10> MnSymbolE9
  <10-12> MnSymbolE10
  <12->   MnSymbolE12
}{}
\DeclareFontShape{OMX}{MnSymbolE}{b}{n}{
    <-6>  MnSymbolE-Bold5
   <6-7>  MnSymbolE-Bold6
   <7-8>  MnSymbolE-Bold7
   <8-9>  MnSymbolE-Bold8
   <9-10> MnSymbolE-Bold9
  <10-12> MnSymbolE-Bold10
  <12->   MnSymbolE-Bold12
}{}

\let\llangle\@undefined
\let\rrangle\@undefined
\DeclareMathDelimiter{\llangle}{\mathopen}%
                     {MnLargeSymbols}{'164}{MnLargeSymbols}{'164}
\DeclareMathDelimiter{\rrangle}{\mathclose}%
                     {MnLargeSymbols}{'171}{MnLargeSymbols}{'171}

\newcommand{\beginappendix}{%
        \setcounter{table}{0}
        \renewcommand{\thetable}{A\arabic{table}}%
        \renewcommand{\theHtable}{Appendix\thetable}
        \setcounter{equation}{0}
        \renewcommand{\theequation}{A\arabic{equation}}%
        \renewcommand{\theHequation}{Appendix\theequation}
        \setcounter{figure}{0}
        \renewcommand{\thefigure}{A\arabic{figure}}%
        \renewcommand{\theHfigure}{A\thefigure}
        \setcounter{section}{0}
        \renewcommand{\thesection}{A\arabic{section}}%
     }
\newcommand{\fref}[1]{Fig.~\ref{#1}}
\newcommand{\eref}[1]{Eq.~\eqref{#1}}
\newcommand{\sref}[1]{Sec.~\ref{#1}}

\begin{document}

\title{The computational power of random quantum circuits in arbitrary geometries}

\author{M. DeCross}
\email{matthew.decross@quantinuum.com}
\affiliation{Quantinuum, Broomfield, CO 80021, USA}
\author{R. Haghshenas}
\affiliation{Quantinuum, Broomfield, CO 80021, USA}
\author{M. Liu}
\affiliation{Global Technology Applied Research, JPMorgan Chase, New York, NY 10017, USA}
\affiliation{Computational Science Division, Argonne National Laboratory, Lemont, IL 60439, USA}
\author{E. Rinaldi}
\affiliation{Quantinuum K.K., Otemachi Financial City Grand Cube 3F, 1-9-2 Otemachi, Chiyoda-ku, Tokyo, Japan}
\author{J.~Gray}
\affiliation{Division of Chemistry and Chemical Engineering, California Institute of Technology, Pasadena, USA 91125}

\author{Y. Alexeev}
\affiliation{Computational Science Division, Argonne National Laboratory, Lemont, IL 60439, USA}
\author{C. H. Baldwin}
\affiliation{Quantinuum, Broomfield, CO 80021, USA}
\author{J. P. Bartolotta}
\affiliation{Quantinuum, Broomfield, CO 80021, USA}
\author{M. Bohn}
\affiliation{Quantinuum, Broomfield, CO 80021, USA}
\author{E. Chertkov}
\affiliation{Quantinuum, Broomfield, CO 80021, USA}
\author{J. Cline}
\affiliation{Quantinuum, Broomfield, CO 80021, USA}
\author{J. Colina}
\affiliation{Quantinuum, Broomfield, CO 80021, USA}
\author{D. DelVento}
\affiliation{Quantinuum, Broomfield, CO 80021, USA}
\author{J.~M.~Dreiling}
\affiliation{Quantinuum, Broomfield, CO 80021, USA}
\author{C.~Foltz}
\affiliation{Quantinuum, Broomfield, CO 80021, USA}
\author{J.~P.~Gaebler}
\affiliation{Quantinuum, Broomfield, CO 80021, USA}
\author{T. M. Gatterman}
\affiliation{Quantinuum, Broomfield, CO 80021, USA}
\author{C. N. Gilbreth}
\affiliation{Quantinuum, Broomfield, CO 80021, USA}
\author{J. Giles}
\affiliation{Quantinuum, Broomfield, CO 80021, USA}
\author{D. Gresh}
\affiliation{Quantinuum, Broomfield, CO 80021, USA}
\author{A. Hall}
\affiliation{Quantinuum, Broomfield, CO 80021, USA}
\author{A. Hankin}
\affiliation{Quantinuum, Broomfield, CO 80021, USA}
\author{A. Hansen}
\affiliation{Quantinuum, Broomfield, CO 80021, USA}
\author{N. Hewitt}
\affiliation{Quantinuum, Broomfield, CO 80021, USA}
\author{I. Hoffman}
\affiliation{Quantinuum, Broomfield, CO 80021, USA}
\author{C. Holliman}
\affiliation{Quantinuum, Broomfield, CO 80021, USA}
\author{R. B. Hutson}
\affiliation{Quantinuum, Broomfield, CO 80021, USA}
\author{T. Jacobs}
\affiliation{Quantinuum, Broomfield, CO 80021, USA}
\author{J. Johansen}
\affiliation{Quantinuum, Broomfield, CO 80021, USA}
\author{P. J. Lee}
\affiliation{Quantinuum, Broomfield, CO 80021, USA}
\author{E. Lehman}
\affiliation{Quantinuum, Broomfield, CO 80021, USA}
\author{D. Lucchetti}
\affiliation{Quantinuum, Broomfield, CO 80021, USA}
\author{D. Lykov}
\affiliation{Computational Science Division, Argonne National Laboratory, Lemont, IL 60439, USA}
\author{I. S. Madjarov}
\affiliation{Quantinuum, Broomfield, CO 80021, USA}
\author{B. Mathewson}
\affiliation{Quantinuum, Broomfield, CO 80021, USA}
\author{K.~Mayer}
\affiliation{Quantinuum, Broomfield, CO 80021, USA}
\author{M.~Mills}
\affiliation{Quantinuum, Broomfield, CO 80021, USA}
\author{P. Niroula}
\affiliation{Global Technology Applied Research, JPMorgan Chase, New York, NY 10017, USA}
\author{J. M. Pino}
\affiliation{Quantinuum, Broomfield, CO 80021, USA}
\author{C. Roman}
\affiliation{Quantinuum, Broomfield, CO 80021, USA}
\author{M. Schecter}
\affiliation{Quantinuum, Broomfield, CO 80021, USA}
\author{P. E. Siegfried}
\affiliation{Quantinuum, Broomfield, CO 80021, USA}
\author{B. G. Tiemann}
\affiliation{Quantinuum, Broomfield, CO 80021, USA}
\author{C. Volin}
\affiliation{Quantinuum, Broomfield, CO 80021, USA}
\author{J. Walker}
\affiliation{Quantinuum, Broomfield, CO 80021, USA}

\author{R.~Shaydulin}
\affiliation{Global Technology Applied Research, JPMorgan Chase, New York, NY 10017, USA}
\author{M. Pistoia}
\affiliation{Global Technology Applied Research, JPMorgan Chase, New York, NY 10017, USA}
\author{S. A. Moses}
\affiliation{Quantinuum, Broomfield, CO 80021, USA}
\author{D. Hayes}
\affiliation{Quantinuum, Broomfield, CO 80021, USA}
\author{B. Neyenhuis}
\affiliation{Quantinuum, Broomfield, CO 80021, USA}
\author{R. P. Stutz}
\affiliation{Quantinuum, Broomfield, CO 80021, USA}
\author{M. Foss-Feig}
\email{michael.feig@quantinuum.com}
\affiliation{Quantinuum, Broomfield, CO 80021, USA}

\begin{abstract}
Empirical evidence for a gap between the computational powers of classical and quantum computers has been provided by experiments that sample the output distributions of two-dimensional quantum circuits.  Many attempts to close this gap have utilized classical simulations based on tensor network techniques, and their limitations shed light on the improvements to quantum hardware required to frustrate classical simulability.  In particular, quantum computers having in excess of $\sim 50$ qubits are primarily vulnerable to classical simulation due to restrictions on their gate fidelity and their connectivity, the latter determining how many gates are required (and therefore how much infidelity is suffered) in generating highly-entangled states.  Here, we describe recent hardware upgrades to Quantinuum's H2 quantum computer enabling it to operate on up to $56$ qubits with arbitrary connectivity and $99.843(5)\%$ two-qubit gate fidelity.  Utilizing the flexible connectivity of H2, we present data from random circuit sampling in highly connected geometries, doing so at unprecedented fidelities and a scale that appears to be beyond the capabilities of state-of-the-art classical algorithms. The considerable difficulty of classically simulating H2 is likely limited only by qubit number, demonstrating the promise and scalability of the QCCD architecture as continued progress is made towards building larger machines.
\end{abstract}
\maketitle

\section{Introduction}

It is widely expected that the computational power of quantum computers will only be fully realized by using quantum error correction (QEC). However, a significant gap exists between the capabilities of \emph{all} existing quantum computing architectures (e.g., trapped-ions, superconducting qubits, neutral atoms, etc.) and the requirements to perform large scale error-corrected algorithms.  To close this gap, desirable physical properties for a quantum computing architecture can readily be identified: all else being equal, it is generally better to have more qubits, lower physical error rates, higher connectivity, and faster clock speed. An honest assessment of the state of the field~---~supported empirically by the sheer variety of technologies under development~---~is that no one architecture is uniformly superior to the others with respect to all of these properties, each having strengths that can plausibly compensate for its weaknesses. For example, recent research into low-density parity check codes~\cite{Breuckmann2021} suggests that high connectivity may greatly reduce the qubit resource requirements for QEC, potentially alleviating challenges in manufacturability or scalability. Similarly, a slow clock-speed may be compensated by advantages that translate into higher fidelity, since a quantum computer can then produce the correct answer to a given problem with a higher probability, requiring fewer trials to succeed.

As quantum computing technologies mature, many researchers have begun to explore whether quantum computational advantage might be obtainable \emph{without} quantum error correction. A minimal demonstration of the viability of this program was provided in the pioneering work of Ref.~\cite{arute2019} on random circuit sampling (RCS), in which a quantum computer performed a (contrived) task that appeared to be infeasible to simulate on even the most powerful classical computers.  It is interesting to note that the same desirable qualities for achieving large-scale error-corrected quantum computing~---~speed, connectivity, fidelity, and size~---~all play a fundamental role in determining the relative separation between the computational power of quantum computers and classical computers for RCS.  For example, tensor-network methods based on compressed contraction allow classical computers to simulate quantum circuits with imperfect fidelity \cite{PhysRevX.10.041038,PRXQuantum.4.020304}, with the fidelities achievable in simulation determined primarily by the connectivity of the circuit (higher connectivity generally leading to lower effective fidelity given fixed classical resources). Therefore, increasing a quantum computer's connectivity, fidelity, or both can strongly frustrate such simulation methods.

\begin{figure*}[!t]
\centering
\includegraphics[width=2.05\columnwidth]{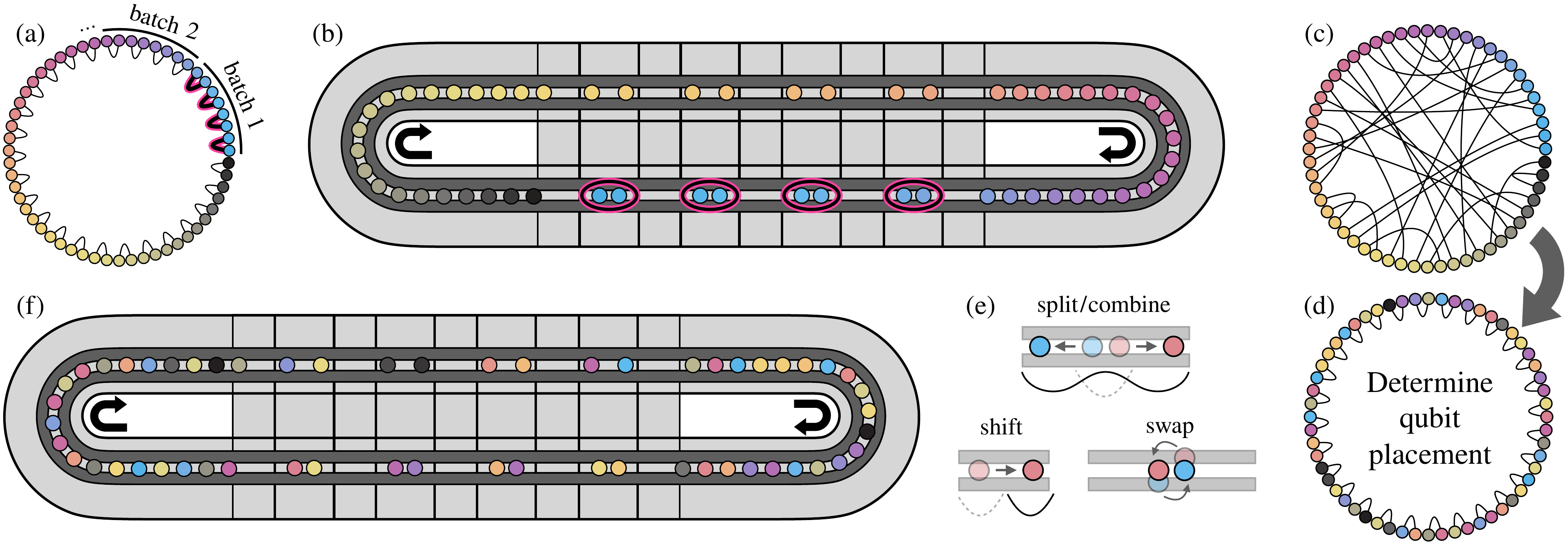}
\caption{Demonstration of how a densely-gated circuit ($N/2$ 2Q gates per layer) with arbitrary connectivity is executed on the H2 quantum computer. (a) The first layer is executed by assigning each qubit to a unique $^{171}{\rm Yb}^{+}$ ion (colored disk) such that gated qubits are co-located; each ion is labeled by a unique color, and the black lines connecting neighbors indicate that the associated qubits will be gated. Since H2 is currently configured with 4 active gate zones, the two-qubit (2Q) gates are executed in 7 batches of 4 parallel gates. (b) The first batch of four gates [highlighted in magenta in (a)] are executed in parallel in the bottom row of gate zones, and qubits are then shuffled around the trap in a ``rolodex'' fashion until all gates (including 1Q gates) in the first layer have been applied. (c) The next layer of gates can act on arbitrary pairs. (d) An automated compilation step decides where to locate qubits in the trap (placement) and how to get them there (routing), resulting in a new assignment of qubit positions that ensures all pairs of qubits to be gated in this layer are once again co-located. Given the placement determined in (d), sequences of voltages are applied to the trap electrodes that, via a combination of the split/combine, shift, and swap operations shown in (e) achieve the desired ion placement for this layer of gates shown in (f). The gates can once again be executed in a rolodex fashion, and this entire process is repeated until all layers of gates have been applied. Note that additional coolant ions are omitted in these illustrations; as described in Ref.~\cite{Pino2020,moses2023race} each $^{171}{\rm Yb}^{+}$ qubit ion is paired with a $^{138}{\rm Ba}^{+}$ coolant ion.
\label{fig:H2-1}}
\end{figure*}

In this paper, we describe upgrades to Quantinuum's H2 trapped-ion QCCD quantum computer that enable it to operate with up to $56$ qubits while maintaining arbitrary connectivity and improving upon the high two-qubit gate fidelities reported in Ref.~\cite{moses2023race} (Sec.~\ref{sec:h2}). As a test of the H2 quantum computer's capabilities, we implement RCS in randomly assigned geometries. In Sec.~\ref{sec:difficulty}, we show that the arbitrary connectivity of H2 enables the execution of flexibly programmed circuits that appear to be extremely challenging to simulate classically even at relatively low circuit depths. In Sec.~\ref{sec:exp_results}, we show that the combination of low depth requirements (afforded by arbitrary connectivity) together with the high gate fidelities achieved on the H2 quantum computer enable sampling from classically-challenging circuits in an unprecedented range of fidelity: Circuits deep enough to saturate the cost of classical simulation by exact tensor-network contraction (assuming no memory constraints) can be executed without a single error about $35\%$ of the time. Historically, the most significant loophole in the claim that RCS is classically hard in practice stems from the low circuit fidelities that have been achievable for circuits deep enough to become hard to simulate classically. The high fidelities achieved in this work appear to firmly close this loophole.  Unlike previous RCS demonstrations, in which circuits have been carefully defined in reference to the most performant achievable gates, our circuits are comprised of natural perfect entanglers (equivalent to control-$Z$ gates up to single-qubit rotations) and Haar-random single-qubit gates.  All circuits are run ``full stack'' with the default settings that would be applied to jobs submitted by any user of H2, without any special purpose compilation or calibration. Our primary conclusion is that even with $56$ qubits, the computational power of H2 for RCS is strongly limited by qubit number and not fidelity or clock speed, with the implication that the separation of computational power between QCCD-based trapped-ion quantum computers and classical computers will continue to grow very rapidly as the qubit number continues to be scaled up.

\section{The H2 quantum computer} \label{sec:h2}

The H2-series quantum computers are built around a race track-shaped surface-electrode trap, illustrated in Fig.~\ref{fig:H2-1}.  Similar to previous work \cite{moses2023race}, each qubit is encoded in the $^{2}S_{1/2}$ $\ket{F=0(1),m_{F}=0}\equiv \ket{0(1)}$ hyperfine states of a $^{171}{\rm Yb}^{+}$ ion, and we use $^{138}{\rm Ba}^{+}$ ions as sympathetic coolants.  Each of the colored circles in \fref{fig:H2-1} represents a Yb ion (qubit), which is stored in a two-ion crystal with a single Ba coolant ion (the crystal order alternating between Yb-Ba and Ba-Yb around the track).  Prior to application of a two-qubit gate or the physical swapping of two qubits, the two crystals associated with those two qubits are merged into a single Yb-Ba-Ba-Yb or Ba-Yb-Yb-Ba crystal (with gating only performed in the former configuration). An initial 32-qubit configuration of H2 was described in detail in Ref.~\cite{moses2023race}, though the trap described in that work was designed to eventually support larger qubit numbers.  Here we describe the system performance following a recent operational upgrade to $56$ qubits. The primary technical challenge in scaling from $N=32$ to $N=56$ was the development of transport waveforms for a larger number of potential wells. The new waveforms use the same basic primitives to rearrange qubits~---~linear transport, split/combine, and physical swap, see \fref{fig:H2-1}(e)~---~but the exact trajectories and spacing between potential wells have been reconfigured. 

All quantum operations including state preparation and measurement (SPAM), single-qubit (1Q) gates, and two-qubit (2Q) gates are performed as described in Ref.~\cite{moses2023race} using only the bottom 4 (out of 8 possible) gate zones.  A summary of benchmarking results for these primitive operations is given in Table~\ref{tab:RB_avgs}. One-qubit and two-qubit randomized benchmarking (1Q and 2Q RB) were performed with standard Clifford randomizations. Transport 1Q RB was performed with all 56 available qubits in a similar method described in Ref.~\cite{moses2023race}, and is intended to capture the combined effects of memory errors and 1Q gate errors (i.e.\ all errors not originating from the 2Q gates) on the circuits considered in this paper. 

We emphasize that the fidelity of primitive operations (gates, state preparation, and measurement) has not degraded relative to the initial $N=32$ configuration of H2. In fact, various technical improvements including reductions in laser phase noise have decreased the average error of our native perfect entangler from $\epsilon_{\rm 2Q} =1.83(5)\times10^{-3}$ measured in Ref.~\cite{moses2023race} to $\epsilon_{\rm 2Q}=1.57(5)\times 10^{-3}$ reported in Tab.\,\ref{tab:RB_avgs} and used to estimate circuit fidelities in \sref{sec:exp_results}.  The primary detrimental impact of loading more qubits into the trap is that the average time to execute a layer of $N/2$ randomly selected gate pairs (depth-1 time) increased from $53\,$ms at $N=32$ to $80\,{\rm ms}$ at $N=56$, with an associated increase of the memory error per qubit per layer from $2.2(3) \times 10^{-4}$ at $N=32$ to $4.0(4) \times 10^{-4}$ at $N=56$. After taking the data for \fref{fig:data}, additional improvements including the implementation of more efficient laser cooling methods have reduced the depth-1 time to around 70 ms, while also lowering the 2Q gate error rate further to the $\epsilon_{\rm 2Q} = 1.28(8)\times 10^{-3}$ reported in Ref.~\cite{github_spec}.  We expect that increasing the detuning of the 2Q gate lasers (to operate at the same detuning used currently by the H1 quantum computer) will lead to about another $4\times 10^{-4}$ improvement, bringing the 2Q gate fidelities on H2 up to the $99.9\%$ level achieved on H1~\cite{github_spec}.

\begin{table}[]
\begin{ruledtabular}
\begin{tabular}{llc}
Label & Test                  & Value ($\times 10^{-4}$) \\[.3em]  \hline \\[-.6em]
$\epsilon_{1{\rm Q}}$ & 1Q RB                 & 0.29(4)             \\[.1em]
$\epsilon_{2{\rm Q}}$ & 2Q RB                 & 15.7(5)             \\[.1em]
$\epsilon_{\rm mem}$ & Transport 1Q RB       & 4.0(4)             \\[.1em]
$p_{\rm SPAM}$ & SPAM                  & 14.7(9)             \\[.1em]
\end{tabular}
\end{ruledtabular}
\caption{Average component benchmarking results for the tests outlined in Sec.~III of Ref.~\cite{moses2023race}.  All quantities labeled by subscripted $\epsilon$ are average infidelities. The value reported for $p_{\rm SPAM}$ is the mean of the probabilities $p(1|0)$ and $p(0|1)$, where $p(b|a)$ is the probability to measure $b$ after preparing a qubit in state $a$.}
\label{tab:RB_avgs}
\end{table}

\section{The difficulty of random circuit sampling} \label{sec:difficulty}

Random circuit sampling (RCS) \cite{RevModPhys.95.035001}, as first envisioned and carried out on a superconducting quantum computer in Ref.~\cite{arute2019}, is a computational task designed to highlight the difficulty of using a classical computer to simulate a quantum computer. There are four primary factors determining the separation between classical and quantum run-times for RCS: the quantum computer's speed, size (number of qubits), gate fidelity, and connectivity.  We note that until very recently \cite{Bluvstein2024}, existing work on random circuit sampling has been heavily focused on circuits with 2D connectivity \cite{arute2019,PhysRevLett.127.180501,ZHU2022240,morvan2023phase}. The primary reason is that connectivity beyond 2D is technologically difficult to achieve in quantum computing platforms with fixed qubit locations, and to date 2D arrays of superconducting qubits are the only quantum computing architecture to have reached the system sizes and gate fidelities required to make RCS classically hard in practice (though significant progress towards executing computationally complex circuits with neutral atom arrays has recently been made \cite{Bluvstein2024}). Below, we will show how improved connectivity specifically leads to dramatically more favorable scaling between quantum and classical run times for RCS.

The task of RCS begins with a description of a random quantum circuit $C$, which we take to act on $N$ qubits with $d$ dense layers of 2Q gates (``dense'' meaning that there are $N/2$ 2Q gates per layer). The quantum state $\ket{\Psi}$ obtained by applying $C$ to the input state $\ket{0}^{\otimes N}$ induces a probability distribution over output bitstrings $x_j$ given by $P(x_j)=|\!\bra{x_j}\Psi\rangle|^2$. Informally, the goal of RCS is simply to sample bitstrings from a distribution that is suitably close to the ideal distribution $P$. Since RCS refers only to producing samples, and not the actual state $\ket{\Psi}$, it can equally well be performed on a classical or quantum computer, making it a natural computational task for which their respective powers can be compared.

We first consider performing RCS with a quantum computer, which we assume can implement 2Q gates with an error probability (process infidelity) $\varepsilon$. For simplicity, we assume in this preliminary discussion that all operations other than 2Q gates are perfect, though later calculations will use a more sophisticated model that incorporates various measured sources of error into an \emph{effective} per-2Q gate process infidelity, as well as SPAM errors. The quantum computer executes the circuit $M$ times and provides a list of $M$ samples (length-$N$ bit strings) $x_1,x_2,\ldots, x_M$, each drawn from the probability distribution $\mathscr{P}$ induced by the \emph{noisy} circuit. One potential measure for the closeness of $\mathscr{P}$ to $P$~---~originally proposed in Ref.~\cite{arute2019} but widely adopted in experimental RCS efforts \cite{PhysRevLett.127.180501,ZHU2022240,morvan2023phase,Bluvstein2024} since then~---~is a suitably scaled and shifted version of the linear cross-entropy
\begin{align}
F_{\rm XEB}&= 2^N\bigg(\sum_{x}\mathscr{P}(x)P(x)\bigg)-1
\label{eq:fxeb_formal}\\
&\approx 2^N\bigg(\frac{1}{M}\sum_{j}P(x_j)\bigg)-1,
\label{eq:fxeb}
\end{align}
where the second line is an estimator of the first using the bitstrings sampled from $\mathscr{P}$. For sufficiently deep random quantum circuits, the probabilities $P$ are themselves expected to be distributed according to the Porter-Thomas distribution. As a consequence, noiseless sampling [replacing $\mathscr{P}\rightarrow P$ in \eref{eq:fxeb_formal}] can readily be shown to result in $F_{\rm XEB}=1$. On the other hand, if the circuit is so noisy that it outputs the completely mixed state, then the bitstrings are sampled uniformly at random ($\mathscr{P}(x)=1/2^{N}$), and \eref{eq:fxeb_formal} yields $F_{\rm XEB}=0$. 

While producing samples with a large $F_{\rm XEB}$ does not \emph{necessarily} imply faithful sampling, the above two limits suggest that $F_{\rm XEB}$ might be a reasonable estimator of the fidelity $F$ of the state output by the circuit; if so, then sampling from a quantum computer with a high $F_{\rm XEB}$ can serve as evidence of faithful RCS. A more careful justification for $F_{\rm XEB}\approx F$ can be made by treating the noise on the 2Q gates as stochastic Pauli error channels, such that with probability $\varepsilon$ each 2Q gate is accompanied by a non-identity Pauli error. For circuits that generate highly entangled states one expects the presence of even a single Pauli error to cause both $F$ and $F_{\rm XEB}$ to nearly vanish, in which case both must be approximately equal to the probability that no error has occurred in the entire circuit (up to small expected boundary effects \cite{gao2021limitations}, see Appendix~\ref{sec:XEB_v_fidelity}),
\begin{align}
F\approx F_{{\rm XEB}}\approx (1-\varepsilon)^{N d/2}.
\label{eq:3_fids}
\end{align}

Early work on RCS generally assumed that any classical algorithm capable of producing samples with a non-zero $F_{\rm XEB}$ must essentially achieve a simulation of the quantum circuit in question with fidelity comparable to $F_{\rm XEB}$. Since simulating complex circuits with high fidelity is believed to be classically hard, this assumption makes it natural to \emph{define} the task of RCS as simply the generation of samples with a particular non-zero value of $F_{\rm XEB}$. However, this definition has several potential drawbacks. First and foremost, it is susceptible to the possibility that while faithfully sampling a circuit's output distribution is classically hard \cite{bouland2018,bouland2022,Movassagh2023,aaronson2011computational}, producing large $F_{\rm XEB}$ values \emph{without} faithful sampling is not. Moreover, even if large $F_{\rm XEB}$ does indicate faithful sampling, computing $F_{\rm XEB}$ from quantum data in order to validate that RCS was performed is necessarily hard in the desired regime of quantum advantage because computing the probabilities $P(x_j)$ of observed bitstrings must be performed classically, and is at least as hard as classically sampling from the probability distribution $P$.

Several recent works have scrutinized the relationship between $F_{\rm XEB}$ and the output state fidelity $F$ in considerable detail  \cite{gao2021limitations,morvan2023phase,ware2023sharp}.  Part of the conclusion of these works is that $F_{\rm XEB}$ is a good estimator of fidelity in the so-called ``weak noise'' regime, in which the probability of error per layer, $\varepsilon N/2$, is sufficiently small, and RCS experiments in the weak noise regime were carried out in Ref.~\cite{morvan2023phase}. When the noise rate exceeds a threshold value that depends on the choice of quantum gates and circuit geometry, there is a sharp phase transition into a ``strong-noise'' regime in which noise prevents quantum correlations from spreading across the system, and $F_{\rm XEB}$ can dramatically overestimate fidelity. The analysis of Ref.~\cite{ware2023sharp} for circuits with random geometries and the same gate set considered here, along with noisy simulations using a detailed error model for our machine, suggests that our experiments are very deeply in the ``weak-noise'' limit. Therefore, as long as circuits are deep enough for the output probabilities to be Porter-Thomas distributed, \eref{eq:3_fids} is expected to hold (see Appendix~\ref{sec:XEB_v_fidelity} for further justification of this claim).

Reference \cite{gao2021limitations} also points out that in a more adversarial setting, in which noise is tailored to make circuits easier to simulate by essentially removing gates in order to cut qubits into disconnected subsystems, particularly large discrepancies between $F_{\rm XEB}$ and $F$ can be achieved in the limit that both are small.  These results suggest a route to classically sampling with $F_{\rm XEB}$ values that are anomalously large compared to the achievable simulation fidelity, calling into question the use of achieving high linear cross-entropy scores \emph{per se} as the definition of RCS. The circuits executed on H2 and reported in this paper should be robust against such classical methods due to a combination of their high fidelity (which allows very few gates to be removed) and their high connectivity (which ensures that simulation remains difficult unless a large number of gates are removed).  In particular, our circuits have their gates drawn from the edges of graphs with good edge-expansion properties (see Appendix~\ref{sup:bounds}), which ultimately ensures that it is not possible to disconnect the system into two equally sized subregions without cutting an extensive number of gates, even at constant depth.

In light of the discussion above, in this work we primarily focus on providing direct evidence for the ability of the H2 quantum computer to sample with high fidelity from circuits that are difficult to classically simulate. Nevertheless, we do compute $F_{\rm XEB}$ for system sizes that we are able to ($N\leq 40$), and confirm that it agrees well with other fidelity estimates. We also report samples from circuits for which we are unable (due to the severe classical difficulty of simulating the circuits) to compute $F_{\rm XEB}$.  We are generally unaware of \emph{any} classical methods to produce $F_{\rm XEB}$ numbers comparable to what we expect from these circuits without performing essentially exact numerical simulations of the circuit (up to the modest loss in fidelity observed experimentally, see \sref{sec:exp_results}).  While the above statement certainly warrants further scrutiny, we take this assumption as motivation to focus primarily on the difficulty of exact tensor network contraction in the remainder of this work. However, in \sref{sec:DMRG} we also analyze the performance of tensor-network-based approximate simulation techniques, and show that the random geometries explored here significantly increase (relative to local geometries) the classical resource requirements to achieve a given quality of approximation.

\subsection{Circuit geometry and the cost of exact tensor-network contraction} \label{sec:TNcost}

\begin{figure*}[!t]
\centering
\includegraphics[width=1.0\textwidth]{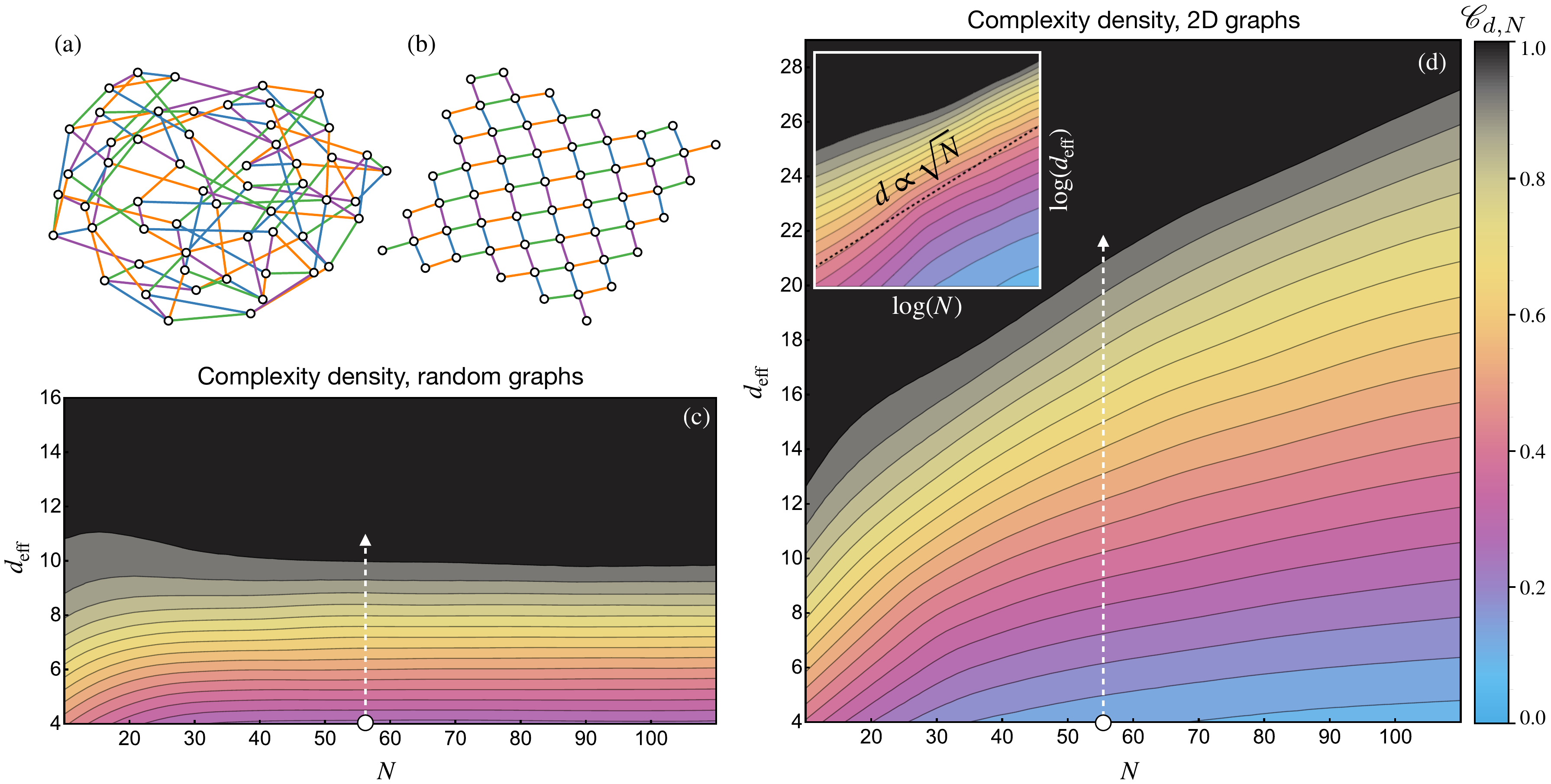}
\caption{Comparison of complexity density $\mathscr{C}_{d,N}$ for circuits with random geometries (RG) and 2D geometries. Figures (a) and (b) show the gating pattern for a depth-4 graph on $56$ qubits given random and 2D geometries, respectively. At each value $(N,d)$, an $N$-node graph is assigned a coloring using either $d$ colors (RG) or $4$ colors (2D). Each circuit is comprised of layers of $U_{\rm ZZ}(\pi/2)$ gates separated by layers of random $SU(2)$ gates on each qubit, with the $U_{\rm ZZ}$ gates applied to each pair of qubits whose associated vertices are joined by an edge of one color.  In (a) each layer corresponds to a unique color, while in (b) the $4$ colors are repeated cyclically until the desired depth is reached. Figures (c) and (d) show estimates of $\mathscr{C}_{d,N}$ for such RG and 2D circuits, respectively. While the complexity density for RG circuits is constant for $N\rightarrow\infty$ at fixed depth [evidenced by the horizontal contours in (c)], the log-log plot in the inset of (d) shows that $d\sim \sqrt{N}$ is required to maintain fixed complexity density in 2D. The white arrows in (c,d) highlight how far the depth needs to be increased at $N=56$ before the complexity density saturates (i.e. the TN contraction cost becomes roughly equivalent to the cost of doing brute-force $N=56$ state vector simulation). \label{fig:d_v_n} }
\end{figure*}

The most efficient known general-purpose classical method to simulate quantum circuits is to represent the probability of an output bit string as a tensor network (TN) and perform exact contraction of that network. Together with rejection sampling or other related methods, such calculations enable classical RCS at a cost lower-bounded by the cost of computing one such probability. Actually contracting the TN associated to a single probability incurs a cost that can be \textit{highly} dependent on the order in which the contraction of tensors is performed. For example, while there is always an ordering that leads to a time-like progression of the full statevector, with a computational cost that scales as $2^{N}$, circuits of low depth and/or low-connectivity will generally admit contraction orders with much lower cost.  Therefore, to assign a cost to the contraction of a given TN it is critical to minimize the contraction cost over all possible contraction orders.  While it may be computationally hard to determine the optimal contraction order in general \cite{markov2008simulating}, many good heuristic methods have been developed in recent years.  The contraction costs reported in this manuscript are obtained using cotengra \cite{Gray2021hyperoptimized}, a Python library that supports a variety of performant heuristic contraction order optimizers and hyperparameter optimization strategies to tune those optimizers.  For the results discussed in this section contraction order is optimized by targeting the minimization of floating-point operations (FLOPs) assuming no memory constraints. Since we have not performed exhaustive searches for FLOPs-optimized contraction orderings, all reported costs are, strictly speaking, only upper bounds on the true contraction cost. However, it is also important to note that the assumption of unconstrained memory can cause one to seriously \textit{underestimate} the cost of contraction in a more realistic memory-constrained setting. The deepest $56$-qubit circuits run in this paper are likely \emph{much} harder to contract in practice than the FLOPs requirements computed in this section would suggest. In \sref{sec:slicing} we investigate the complexity of memory-constrained TN contraction for our circuits in more detail, and show that it continues to grow extremely quickly with depth after the measure of complexity [see \eref{eq:C_finite_D_scaling}] plotted in \fref{fig:d_v_n}(c) saturates.

\begin{figure}[!t]
\centering
\includegraphics[width=0.47\textwidth]{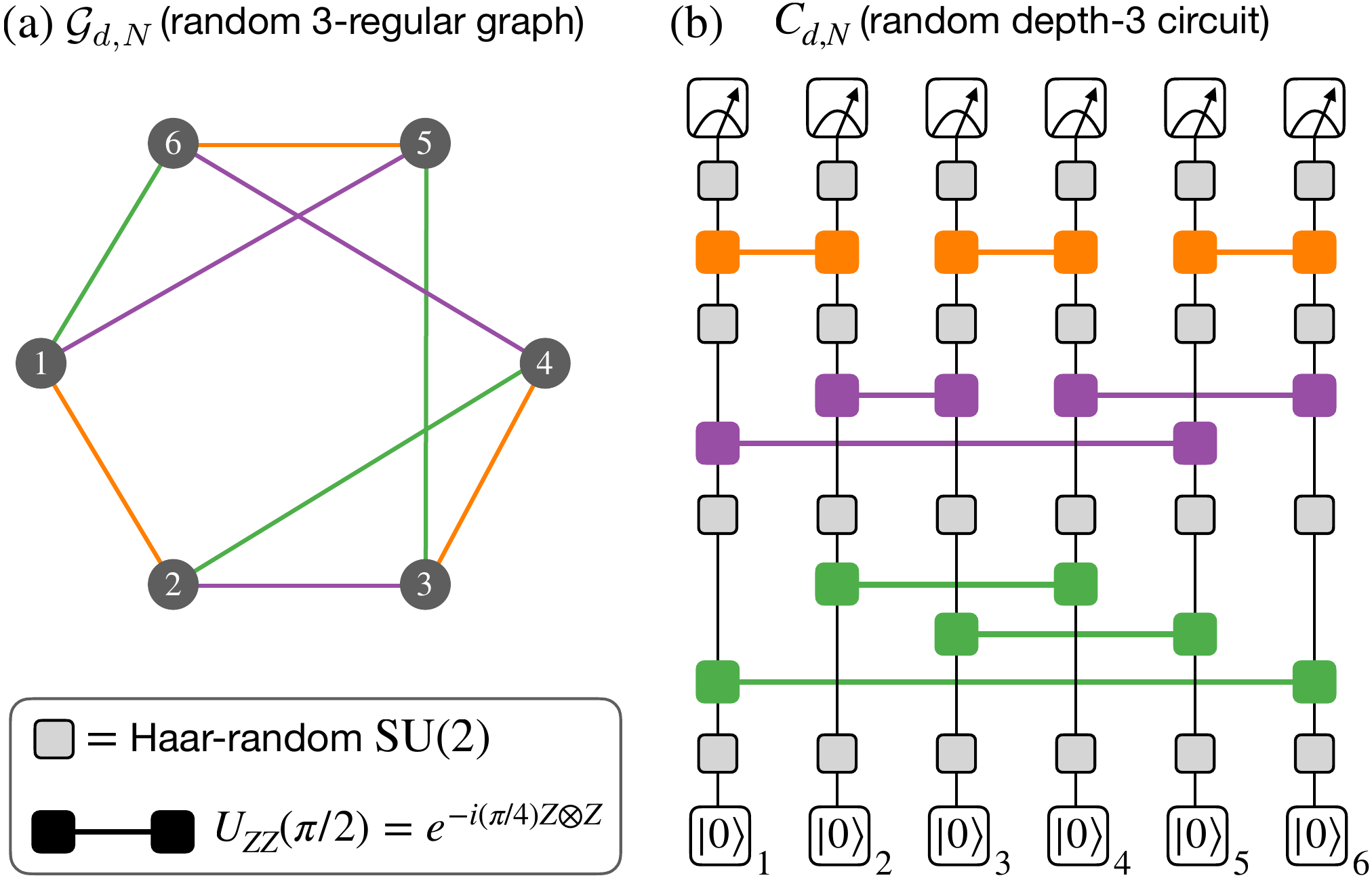}
\caption{The circuits considered in this paper have geometries induced by random regular graphs.  For a depth-$d$ circuit on $N$ qubits, denoted $C_{d,N}$, we sample a random $d$-regular graph on $N$ nodes, denoted $\mathcal{G}_{d,N}$ (the example shown above is for a depth-3 circuit on 6 qubits).  To arrive at a circuit we associate each vertex of $\mathcal{G}$ with a qubit in $C$, and then assign $\mathcal{G}$ an edge-coloring.  Each color is associated with a layer of 2Q gates in $C$, with each edge of that color corresponding to a 2Q gate in that layer.  Adjacent 2Q layers are separated by a layer of Haar-random 1Q gates on each qubit (with a single layer of 1Q gates immediately after state preparation and another immediately before measurement).  \label{fig:circuit} }
\end{figure}

A convenient way to parameterize the contraction cost of the TN corresponding to a quantum circuit is to take the base-2 logarithm of the required FLOPs per gate for performing the contraction \footnote{The denominator inside the $\log$ typically generates a weak (additive logarithmic) correction in $N$ and $d$, and is not particularly important for any of the conclusions that follow.},
\begin{align}
\mathscr{N}_{d,N}=\log_2\bigg(\frac{\rm cost~in~FLOPs}{Nd/2}\bigg).
\end{align}
Because the cost per gate (in FLOPs) of brute-force statevector simulation of $\mathscr{N}$ qubits scales as $2^{\mathscr{N}}$, $\mathscr{N}$ can be interpreted as an \emph{effective qubit number} in the following sense: A depth $d$ circuit $C$ with cost $\mathscr{N}_{d,N}$ may act on $N>\mathscr{N}_{d,N}$ qubits, but it is only (roughly) as difficult to simulate as the worst-case hardness of simulating $\mathscr{N}_{d,N}$ qubits. At any fixed depth $d$, a circuit that is spatially local in $D$-dimensions should have an effective qubit number scaling (asymptotically as $N\rightarrow\infty$) as the area of a minimal bisecting surface,
\begin{align}
\mathscr{N}_{d,N}\sim d\times N^{(D-1)/D}.
\label{eq:N_finite_D_scaling}
\end{align}
This result is perhaps most recognizable in dimensions $D=1,2$, with the implication that the simulation cost of finite depth circuits is asymptotically independent of $N$ for 1D circuits, and exponential in $\sqrt{N}$ for 2D circuits. 
It is also convenient to define a normalized effective qubit number
\begin{align}
\label{eq:C_finite_D_scaling}
\mathscr{C}_{d,N}\equiv\mathscr{N}_{d,N}/N\sim \frac{d}{N^{1/D}},
\end{align}
which we refer to as the \emph{complexity density} of the circuit; it quantifies what fraction of the qubits contribute to the worst-case complexity of simulation. Equation \eqref{eq:C_finite_D_scaling} implies that the asymptotic complexity density,
\begin{align}
\mathscr{C}_d\equiv\lim_{N\rightarrow\infty}\mathscr{C}_{d,N},
\end{align}
vanishes at any fixed depth for spatially local circuits, and therefore such circuits must be made deeper as they are made wider ($d\sim N$ in 1D or $d\sim \sqrt{N}$ in 2D) in order for the contribution of each qubit to the complexity of simulation to not vanish. To understand the behavior of very highly connected circuits, it is suggestive to consider the limit $D\rightarrow \infty$ in Eq.\,\eqref{eq:C_finite_D_scaling}; evidently the asymptotic complexity density will either be constant or decreasing more slowly with $N$ than any power-law. In this work, we will consider a family of highly connected geometries for which $\mathscr{C}_d$ is provably bounded below by a constant (see Appendix~\ref{sup:bounds}).

Given the flexibility of H2, we do not focus on a specific high-dimensional geometry but instead consider highly connected circuits with random geometries, a small example of which is shown in \fref{fig:circuit} (a detailed discussion of random-geometry circuit generation can be found in Appendix~\ref{sec:rand_circs}).  A random-geometry (RG) circuit $C_{d,N}$ of depth $d$ on $N$ qubits is generated by first selecting a random $d$-regular graph $\mathcal{G}_{d,N}$ with $N$ nodes ($d$-regular meaning that exactly $d$ edges impinge on each node). Each node of $\mathcal{G}_{d,N}$ is assigned to a qubit in $C_{d,N}$, and each edge to a 2Q gate.  The native 2Q gate of all Quantinuum H-series hardware is the parameterized entangler $U_{ZZ}(\theta)=\exp(-i(\theta/2)Z\otimes Z)$, and each of the $(N\times d/2)$ 2Q gates is chosen to be the perfect entangler $U_{ZZ}(\pi/2)$. The 2Q gates are then sorted into $d$ layers with $N/2$ gates per layer by finding a proper edge coloring of $\mathcal{G}_{d,N}$ and then assigning one of the $d$ colors to each of the $d$ layers of the circuit. Edges of a given color then have their associated 2Q gates executed in the layer to which that color is assigned.  A layer of Haar-random 1Q gates on every qubit is inserted immediately after qubit initialization, immediately before measurement, and between every layer of 2Q gates (for a total of $d+1$ 1Q layers).

To highlight the impact of connectivity on contraction complexity, in \fref{fig:d_v_n} we compare the complexity density of RG circuits to that of 2D circuits. The 2D circuits are similar in structure to the RG circuits described above, except that the gated pairs are drawn as edges of a 2D square lattice, as in \fref{fig:d_v_n}(b).  All gates corresponding to edges of a given color can be applied in a single layer, and for circuits with $d>4$ layers we simply proceed cyclically through the color list until the desired depth is reached.  To minimize contraction cost fluctuations due to boundary effects, we sample 10 2D grids at each $N$ and $d$, where each grid is obtained by starting with a nominal square lattice with a vertex centered at the origin, applying a random offset and rotation to the lattice, and then selecting the $N$ vertices with smallest Euclidean distance from the origin. The 2D circuits are slightly less densely gated than the RG circuits (which always have exactly $N/2$ gates per layer) because qubits at the edges of the 2D lattice are not gated in each layer. To make a fair comparison between RG and 2D circuits, we associate an effective circuit depth $d_{\rm eff}=n_{\rm 2Q}/(N/2)$ to each circuit, with $n_{\rm 2Q}$ the number of 2Q gates in the circuit (note that $d_{\rm eff}=d$ for the densely-gated RG circuits). Comparisons at fixed $d_{\rm eff}$ and $N$ then always amount to comparisons at a fixed total number of 2Q gates. It is also important to note that for this comparison, we have chosen the Quantinuum native $U_{ZZ}(\pi/2)$ gate as the 2Q gate. It is well known that $i$SWAP-like gates (gates with maximum entropy of their singular value decomposition) lead to relatively harder to simulate 2D circuits than $U_{\rm ZZ}(\pi/2)$ when compared at fixed depth; the contraction cost estimates reported here for 2D circuits should not be conflated with direct cost estimates of circuits run in specific experiments on superconducting qubits, which will generally be higher for a given $(N,d)$ given the use of rank-4 entangling gates.  Nevertheless, the scaling arguments made here apply equally well regardless of the 2Q gate choice.

As discussed above, constant complexity density is maintained by scaling $d\sim \sqrt{N}$ in 2D [see the inset of \fref{fig:d_v_n}(d)]. In contrast, RG circuits empirically achieve \emph{fixed} complexity density at constant depth as $N\rightarrow\infty$, evidenced by the flat contours in \fref{fig:d_v_n}(c). The constant asymptotic complexity density observed in \fref{fig:d_v_n}(c) is proven in the appendices, and is a consequence of the gate pairs being chosen in correspondence with the edges of graphs with good asymptotic expansion properties.  The white-dashed arrows in \fref{fig:d_v_n} correspond to the circuits achievable in the current 56-qubit configuration of H2, and at this system size the depth at which $\mathscr{C}_{d,56}$ saturates to near unity (i.e. where the contraction cost saturates to the statevector simulation cost of $56$ qubits) is roughly half as large ($d\approx 12$) for RG circuits as for 2D circuits ($d\approx 22$).

\subsection{Exact contraction cost with memory constraints \label{sec:slicing}}

Optimization of TN contraction solely targeting FLOP minimization can result in contraction orders that produce large intermediate tensors; the FLOP cost of a TN contraction is immaterial if those intermediate tensors cannot fit within available memory. While utilizing distributed storage for contractions is a theoretical possibility up to a point, in practice, all large scale brute-force TN simulations of random quantum circuits~\cite{pan2021simulating,Kalachev2021multi,liu2024verifying} have favored a technique called \emph{slicing}~\cite{chen2018classical,villalonga2019a} that instead breaks the computation into many independent tasks, each able to fit onto a single GPU. It is therefore natural to optimize the FLOP cost of sliced contraction subject to a constraint on the memory footprint of each individual slice. The potentially enormous reduction in memory footprint is not always free however, and at some point redundantly repeated operations introduce significant overhead. Below, we will compare the memory-unconstrained and sliced cost for exact computation of a single amplitude (`strong simulation') of the random quantum circuits considered here, which serves both as a baseline for various tasks such as sampling and XEB verification, and a useful comparison point to other circuits. We employ hyper-graph partitioning~\cite{kourtis2019fast,Gray2021hyperoptimized} with simulated annealing refinement~\cite{Kalachev2021multi} to find highly optimized TN contraction paths, and in the sliced case enforce a maximum tensor size, or `width' $W=2^{30}$, appropriate for the current generation of GPUs with 40-80GB of memory.
\begin{figure}[!t]
\centering
\includegraphics[width=0.98\columnwidth]{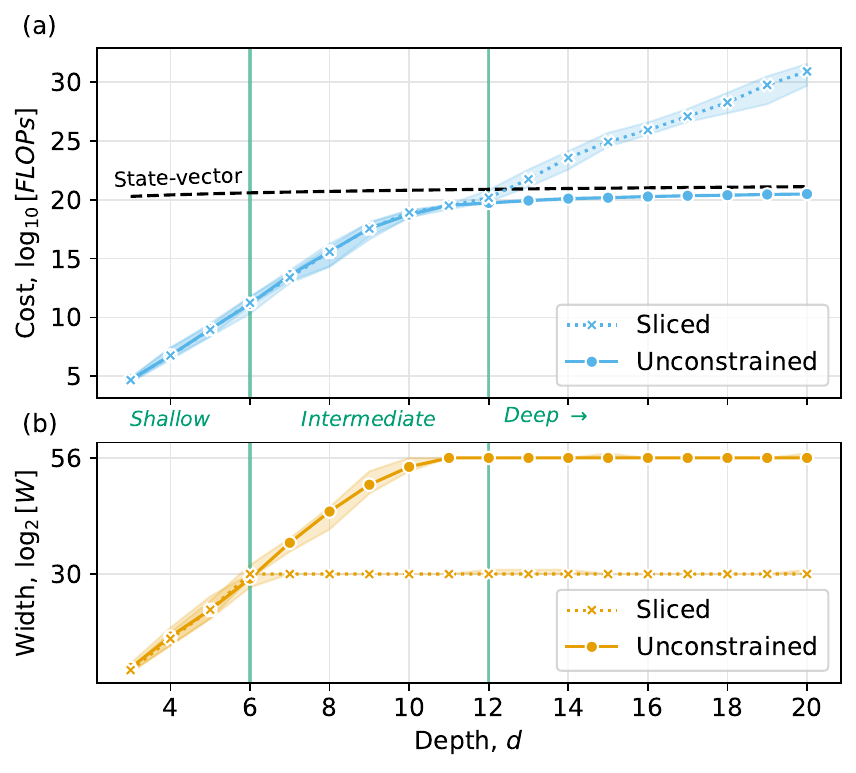}
\caption{
The impact of constraining memory on cost for TN contraction of a single amplitude for random quantum circuits with $N=56$.
(a) FLOP cost of unconstrained optimized contraction paths compared with those sliced to $W{=}2^{30}$ as a function of circuit depth, $d$. The cost of statevector simulation is marked for reference.
(b) Size of the largest intermediate tensor, or `contraction width', $W$, as a function of circuit depth.
The lines represent the median behavior across 20 circuit instances, with the bands showing the min/max range.
\label{fig:sliced_cost}
}
\end{figure}

In Fig.~\ref{fig:sliced_cost}(a,b) we plot the contraction cost in FLOPs (assuming complex tensors) and the contraction width as a function of circuit depth, $d$, averaged over 20 circuit instances at $N=56$. We also show the reference FLOPs of statevector contraction and identify three distinct regions. First a `shallow' region where both the cost and width increase exponentially but are still far below the statevector cost and no slicing is needed. Next, an `intermediate' regime where even though the slicing reduces $W$ by a factor of up to $2^{26} \sim$ 64~million, it introduces no significant overhead. Finally, a `deep' regime where the unconstrained FLOPs and width converge with statevector, but the cost of the sliced contraction continues to grow exponentially.

At the intermediate to deep transition point around $d=12$, where slicing overhead starts to emerge, the FLOP cost of ${\sim} 10^{20}$ is already very significant, although certainly within reach of the largest super-computers which can achieve ${\sim} 10^{18}$ FLOPs per second~\cite{atchley2023frontier}. However the median sliced cost continues to exponentially grow to ${\sim}10^{30}$ as the depth increases to $d=20$, putting the computation well out of reach. We note that while improved techniques (that make use of extensive caching for example) might bring this cost down somewhat, no such method is expected to fully remove the exponential cost of constraining $W \ll 2^N$ for deep circuits.

We also emphasize that these costs are meant primarily to situate the relative classical hardness of different depths with respect to other circuits. As discussed in \sref{sec:tts}, quantum RCS efforts generally produce $S$ samples from $C$ different circuits (for a total of $M=S\times C$ samples) such that the \emph{expected} value of $F_{\rm XEB}$ averaged across all samples could in principle be distinguished from zero with high statistical confidence (putting aside the question of how \emph{hard} the verification would be). To actually simulate the RCS experiments performed in this paper or to verify $F_{\rm XEB}$ of the samples generated by H2, one would want to perform at least $S$ such amplitude contractions for each of the $C$ circuits. While recent advances have shown that `multi-contraction' for a single circuit can be performed with cost sublinear in $S$~\cite{pan2021simulating,Kalachev2021multi,liu2024verifying}, having to draw many samples still presents a significant computational increase. Moreover, in all RCS data that we report \cite{rcs_github} we use a relatively small number of shots per circuit ($S=20$), so the sublinearity of such methods provides only a marginal benefit. We also note that when targeting $F_{\rm XEB}<1$ in classical RCS, a generic speedup of $1/F_{\rm XEB}$ is available to brute-force TN methods~\cite{markov2018quantum}. Given the large fidelity values expected even for the deepest circuits implemented in this paper ($F \geq 0.1$ at $d=24$, see \sref{sec:exp_results}), exploitation of hardware imperfections to perform classical RCS of our circuits at suitably reduced fidelity appears to offer no significant speedup.

\subsection{\label{sec:DMRG} The cost of approximate tensor network simulations}

One promising method for approximately simulating quantum circuits is to utilize a tensor-network-based ansatz suitable for capturing states with limited bipartite entanglement. At sufficiently early times, when bipartite entanglement is limited, this method can be exact even for system sizes well beyond the reach of full statevector simulation.  However, as the quantum state progresses through the circuit and becomes more entangled, it eventually needs to be compressed to maintain an ansatz constrained by available resources (in both memory and run time).  This compression inevitably causes the state to lose fidelity with respect to the exact state, reminiscent of the way gate errors in a noisy quantum circuit cause a state to lose fidelity as the depth of a quantum circuit increases. The less noisy a quantum computer's gates are, and the more entanglement-per-gate that computer can generate, the more difficult it becomes for such methods to compete with the physically achieved fidelities for highly-entangled states.

In Refs.\,\cite{PhysRevX.10.041038,PRXQuantum.4.020304}, the authors have argued that in the case of a matrix-product-state (MPS) ansatz and a random circuit, the overall fidelity $F_{\rm MPS}$ of the calculation is well approximated by simply accumulating the loss of fidelity in every compression step, and is therefore readily accessible during the calculation despite the unavailability (in general) of the exact state.  From the estimated simulation fidelity one can ascribe an error per 2Q gate $\varepsilon_{\rm MPS}$ via the relation $F_{\rm MPS}=(1-\varepsilon_{\rm MPS})^{\rm (\#~of~2Q~gates)}$, which can then be compared to the effective error per 2Q gate on quantum hardware, $\varepsilon$. Assuming that fidelity is well approximated by $F_{\rm XEB}$ for both the data output by a quantum computer and for this classical simulation method \footnote{This point has been the source of some debate in the literature, see the appendices for further discussion.}, quantum advantage in cross-entropy benchmarking is only possible if feasible classical resources cannot achieve $\varepsilon_{\rm MPS}\lesssim \varepsilon$. Below we will show that the high gate fidelities and connectivity of H2 make simulations based on MPS extremely challenging, and likely infeasible at the scale and fidelity of the circuits run in this work.  It would be interesting to investigate to what extent approximate simulations could be improved by using a more general TN ansatz than the MPS considered here. However, fast (with circuit depth) suppression of \emph{any} low-entanglement partitions is guaranteed by the way we construct our circuits.  Ultimately, H2 can produce states that are near-maximally entangled with respect to all possible partitions while maintaining high state fidelities, and sampling from such states should pose substantial challenges for existing compression-based TN methods.

All MPS results reported in this section were obtained with a density-matrix renormalization group (DMRG) algorithm similar to that described in Ref.~\cite{PRXQuantum.4.020304}.  In particular, we approximate the amplitude of a particular output bit string of a circuit using a closed-simulation approach, e.g. we evolve one MPS forward from the initial bit string to the middle of the circuit, evolve another backward from the final bit string to the middle circuit, and then compute their overlap. We only simulate depth-20 random circuits, and use intermediate MPS results after only $d/2<10$ forward and backward layers to estimate the achievable fidelity for circuits of all depths $d<20$. Given the extremely non-local geometry of our circuits, the use of a blocked MPS (in which many qubits are grouped together into a single tensor) can substantially reduce the number of long-range (inter-block) gates applied during the time-evolution if the blocks are chosen in reference to the specific random geometry in question. We explored a variety of blocking strategies using $b$ blocks of $s=56/b$ qubits per block (denoted $b\times[s]$), in each case making an assignment of qubits into blocks on a circuit-by-circuit basis by attempting to minimize the number of inter-block gates using the open source hypergraph partitioning package KaHyPar \cite{doi:10.1137/1.9781611974317.5,doi:10.1137/1.9781611974768.3}.

\begin{figure}[!t]
\centering
\includegraphics[width=0.98\columnwidth]{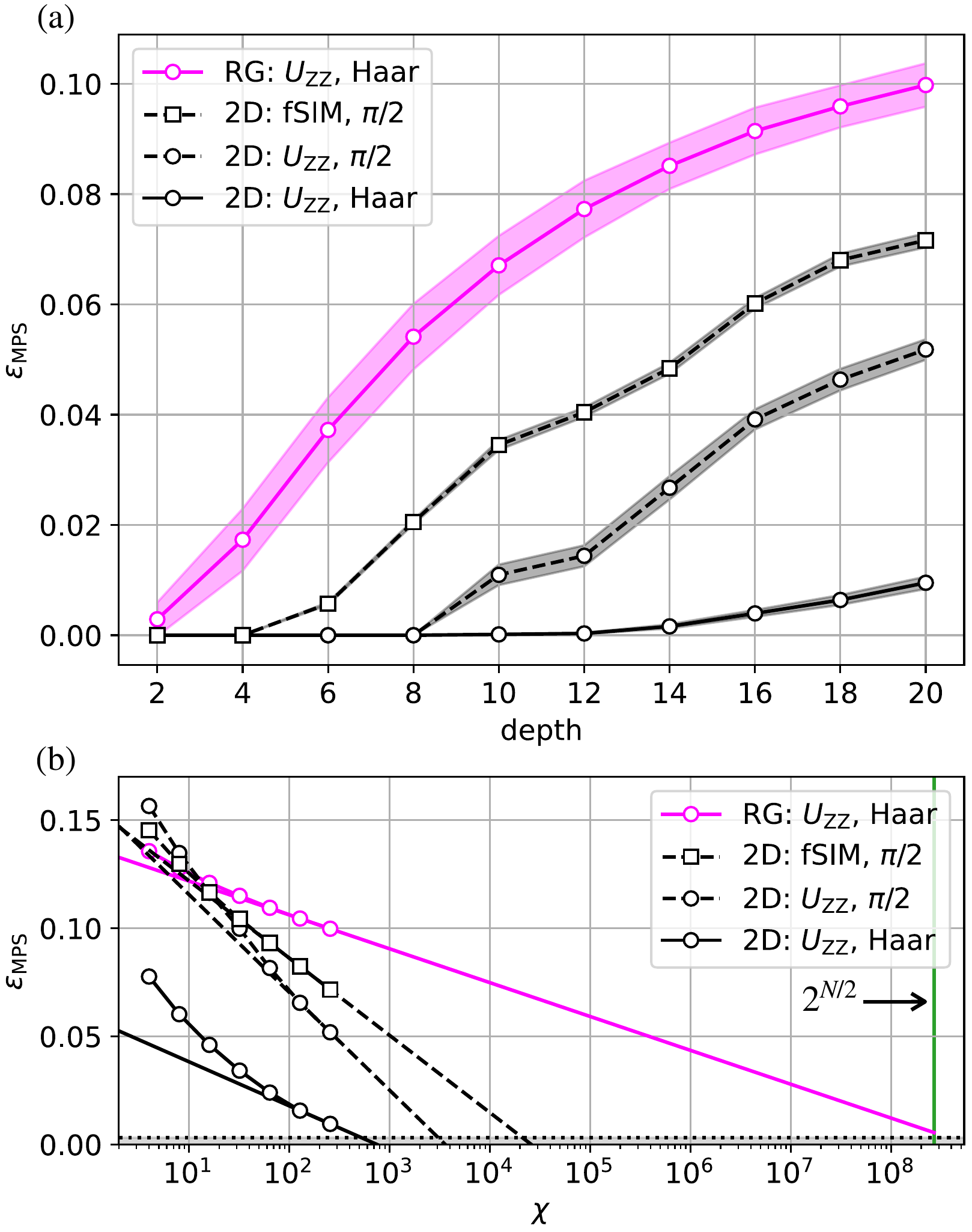}
\caption{(a) Achievable error per gate $\varepsilon_{\rm MPS}$ as a function of circuit depth using DMRG for circuits with 2D and random geometries and a variety of 1Q and 2Q gate sets. All curves use a bond dimension $\chi=256$, and shaded regions show standard deviation of the error per gate across 100 (20) circuit randomizations for random (2D) circuits. (b) Error per gate at depth 20 as a function of bond dimension. Linear extrapolation of $\varepsilon_{\rm MPS}$ to the experimental value of $\varepsilon \approx 3.2\times 10^{-3}$ (dotted horizontal line in this figure, see \sref{sec:exp_results}) from the two largest bond dimensions suggests that by depth $20$, DMRG cannot simulate random geometry circuits at the fidelities achieved on $H2$ without employing an essentially exact representation of the full statevector (MPS bond-dimension $\chi=2^{N/2=28}$, green vertical line). \label{fig:mps_epg} }
\end{figure}
\begin{figure}[!t]
\centering
\includegraphics[width=0.98\columnwidth]{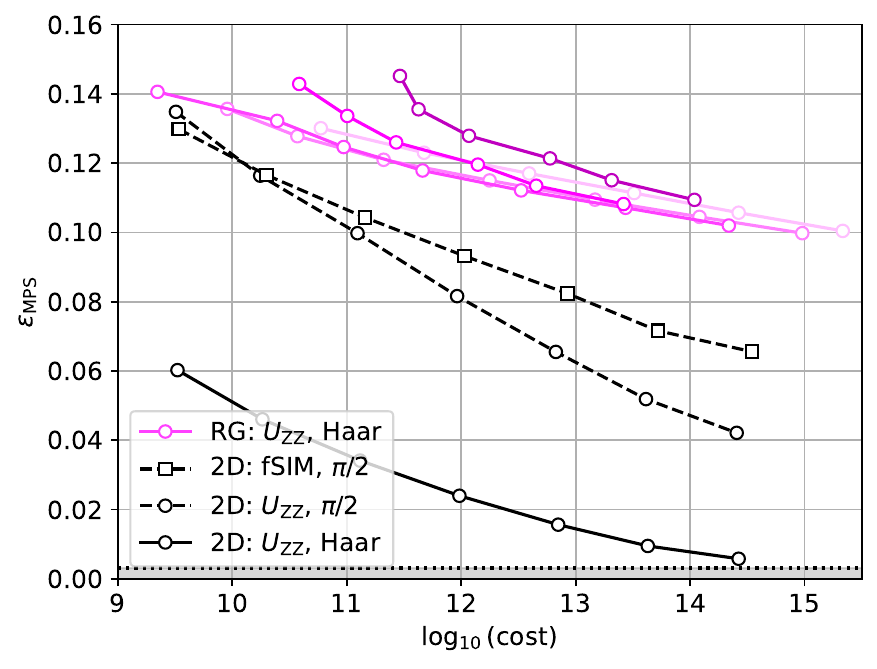}
\caption{Error per gate ($\varepsilon_{\rm MPS}$) versus computational cost (in FLOPs) for both 2D and RG circuits. The results for RG (2D) circuits are again averaged over 100 (20) random circuits. All 2D circuits employ a $7\times[8]$ blocking strategy, while a variety of blocking strategies (each a different shade of pink, see main text) are attempted for the RG circuits. Cost on the $x$-axis is approximate, and estimates the number of floating-point operations required for DMRG using various block sizes as the MPS bond dimension is increased (including the cost of applying all gates to the MPS and all MPS overlap calculations encountered in DMRG, but not the cost of QR decompositions).
\label{fig:epg_cost} }
\end{figure}

 Figure \ref{fig:mps_epg}(a) shows how the error per gate grows as a function of circuit depth for both 2D (black curves) and RG (magenta curve) circuits and a variety of 1Q and 2Q gate sets. The label ``fSIM'' (``$\pi/2$'') denotes the 2Q (1Q) gate set of Ref.\ \cite{arute2019}, while ``$U_{ZZ}$'' (``Haar'') denotes the 2Q (1Q) gate set used for the circuits run on H2. For all 2D simulations we used a $7\times[8]$ blocking strategy on a grid similar to Google's Sycamore chip (though with a slightly different aspect ratio of 7 columns and 8 rows to accommodate $56$ qubits, making each column an MPS tensor), and chose the same gating pattern as that used for the ``quantum supremacy'' circuits in Ref.\ \cite{arute2019}.  For consistency the RG circuit results are also plotted with a $7\times[8]$ blocking strategy, though we show in \fref{fig:epg_cost} that results for random circuits are only weakly dependent on blocking strategy. Comparing the solid black and magenta curves, one can see that with the same gate sets RG circuits generate bipartite entanglement (heralded by the growing error rate at fixed bond dimension) much faster than 2D circuits.  While utilizing 1Q and 2Q gates fine-tuned to generate maximum entanglement also can hinder the performance of DMRG in 2D, overall the impact of geometry is more severe. We note that the circuits run here could be made to entangle marginally faster by utilizing the $\pi/2$-labeled gate set in \fref{fig:mps_epg}, see also the discussion in Appendix~\ref{sup:mps_bound}.  On the other hand, the 2Q fSIM gate utilized in Ref.~\cite{arute2019} \emph{does not} make RG circuits harder to simulate because it can be decomposed as SWAP$\times U_{ZZ}(\theta)$; all SWAP gates in a given layer can be compiled into permutations of the qubit labels in future layers, and results in a translation between different RG circuits in the same circuit ensemble considered. Figure \ref{fig:mps_epg}(b) shows how the error per gate at depth $20$ decreases when increasing the bond-dimension $\chi$ of the MPS, which controls how much entanglement the MPS can capture. From a linear fit to the largest two bond dimensions ($\chi=128$ and $\chi=256$), one can see that by depth $20$ the random circuits appear to require close to the maximum possible bond dimension of $2^{N/2}$ to achieve an error per gate comparable to the experiment [$\varepsilon(N=56)\approx3.2\times 10^{-3}$, black dotted line], at which point the MPS ansatz becomes exact.

In \fref{fig:epg_cost}, we compare the scaling of $\varepsilon_{\rm MPS}$ for RG circuits as a function of total computational cost using various blocking choices. Various shades of pink in \fref{fig:epg_cost} correspond to blocking choices $4\times[14],~7\times[8],~ 14\times[4],~28\times[2]$, and $56\times[1]$ (ordered from light to dark). While the blocking choice significantly impacts the computational cost at fixed bond dimension, overall we find that different blocking choices are all roughly comparable from the standpoint of achievable $\varepsilon_{\rm MPS}$ at fixed computational cost.  We also plot $\varepsilon_{\rm MPS}$ for 2D circuits of this geometry and using the same 1Q and 2Q gate set modifications as in \fref{fig:mps_epg}.  Once again, while the use of gate sets tailored to maximize the growth of entanglement can substantially increase the error per gate at fixed cost in 2D, the impact of random geometry on the cost to achieve a fixed error-per gate is more severe.

\subsection{\label{sec:tts} Time requirements for verifiable RCS}

Random circuit sampling~---~and any related statements about its asymptotic hardness, either classical or quantum~---~can be formulated purely in reference to the task of drawing samples from a particular distribution.  There is a related but conceptually independent question regarding the sample efficiency of \emph{verifying} random circuit sampling (or any sampling task): How many samples must be provided to offer compelling evidence of having sampled from the claimed distribution? If closeness to the ideal distribution is judged by $F_{\rm XEB}$, then the specific question becomes: If someone claims to be able to sample from a distribution with a particular value of $F_{\rm XEB}$, how many samples must they provide in order for us to be confident that they can in fact do so \footnote{It is important to note that actually computing $F_{\rm XEB}$ classically from the quantum data is expected to be infeasible classically in the regime that RCS is classically hard, so the specific notion of verifiability discussed here applies to the ability to verify the data \emph{in principle}, and not in practice.}?  Since the trivial algorithm of guessing random bitstrings produces $F_{\rm XEB}\approx 0$, the experimental task of RCS has often been implicitly defined as generating sufficiently many samples such that $F_{\rm XEB}$~---~were it ever to be computed from the provided samples~---~would be greater than zero with high statistical confidence.

Since the statistical uncertainty on $F_{\rm XEB}$ scales inversely with the square root of the number of samples taken, the total required quantum run time of a verifiable RCS experiment must scale as
\begin{align}
\label{eq:Tq}
T_{\rm q}\sim \tau_{\rm q}/F_{\rm XEB}^2 = \tau_{\rm q}/(1-\varepsilon)^{N d}\approx \tau_q \times e^{\varepsilon N d},
\end{align}
where $\tau_q$ is the quantum run time of a single shot of one circuit. This equation makes it clear that a very fast quantum computer (small $\tau_{\rm q}$) may produce enough data to witness even small statistical deviations from complete noise/randomness ($F_{\rm XEB}\ll 1$) in a reasonable amount of time. On the other hand, increased fidelity and connectivity greatly enhance the deviations of a quantum computer's output from random, thereby reducing the amount of data required and potentially compensating for a slower clock speed.  While the discussion of quantum run time for verifiable RCS may seem academic, it is important to note that a similar trade off between fidelity and clock speed can be identified in most efforts to stretch the power of noisy quantum computers via error mitigation techniques \cite{temme2017error}. Ultimately, these methods amount to extracting a weak signal encoding the noiseless result of a calculation or simulation from a background of noise in a similar spirit to RCS. While individual circuit run times on H2 are slow compared to superconducting devices (about $1$ second per sample for the circuits in this paper), the high fidelity and connectivity offset this deficit, enabling convincing RCS demonstrations with comparable run times to other state-of-the-art demonstrations.
\begin{figure}[!t]
\centering
\includegraphics[width=0.98\columnwidth]{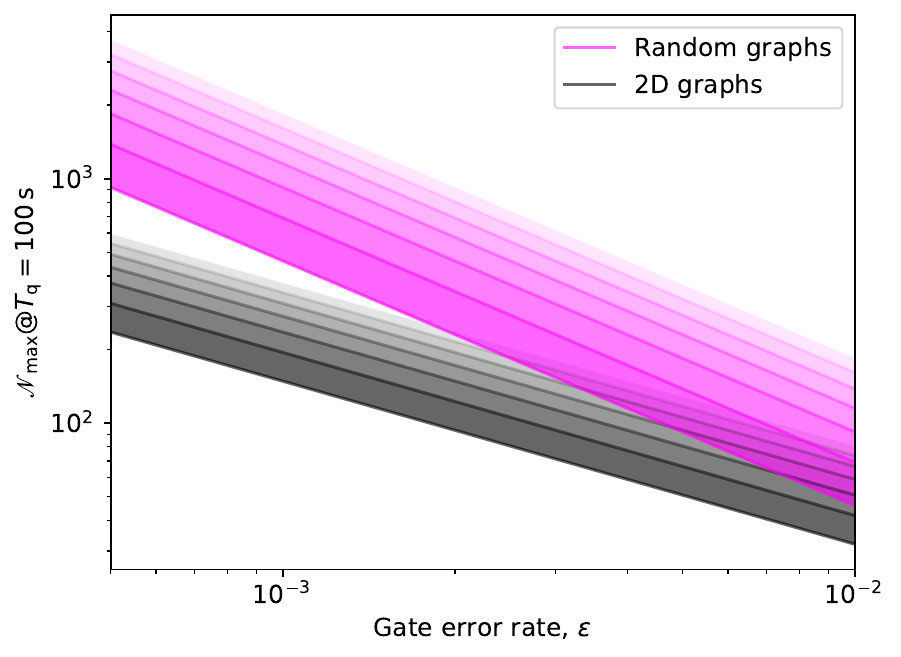}
\caption{Maximum effective qubit number $\mathscr{N}_{\rm max}$ achievable with 100 seconds of data as a function of the 2Q gate error rate. The black (magenta) solid curve at the bottom of each region shows results for 2D (random) geometries assuming a circuit execution time of $\tau_{\rm q}=1\,{\rm s}$. Each progressively fainter line above the lowest corresponds to another order-of-magnitude reduction in the assumed circuit time $\tau_{\rm q}$, with the upmost dashed line corresponding to $\tau_{\rm q}=1\,\mu$s for both geometries.}
\label{fig:tts}
\end{figure}

Equation \eqref{eq:Tq} implicitly establishes a region of circuit volumes in the $N$-$d$ plane within which a quantum computer can extract verifiable RCS data given a fixed amount of total run time $T_{q}$,
\begin{align}
\label{eq:Nd_region}
d \leq d_{\rm max}(N)\equiv \frac{\log(T_{\rm q}/\tau_{\rm q})}{\varepsilon N}.
\end{align}
By computing $\mathscr{N}_{d,N}$ for circuits of a particular geometry and maximizing it over $d\leq d_{\rm max}(N)$, we can determine the maximum effective qubit number accessible to a given quantum computer in a fixed total amount of run time $T_{\rm q}$, denoted $\mathscr{N}_{\rm max}(\varepsilon,\tau_q,T_q)$, which in turn determines the maximum classical difficulty of simulating the data output by such an experiment. For this purpose we compute simplified models of the contraction costs for 2D and RG circuits in order to extrapolate contraction costs beyond the range displayed in \fref{fig:d_v_n}, with parameters fitted to the contraction cost estimates obtained from cotengra for $N\leq 110$ (see Appendix~\ref{sup:simp_cost}). From these models we can compare effective qubit numbers achievable in 2D and random geometries using the same random gate set over a broad range of assumptions about the circuit time $\tau_{\rm q}$ and the 2Q gate error rate $\varepsilon$, as shown in \fref{fig:tts} for a fixed total quantum run time of $T_{\rm q}=100\,{\rm s}$ .  In this figure, the bottom boundary (darkest solid line) of each shaded region corresponds to an assumed circuit time of 1 second, and each progressively lighter line above represents a decrease of $\tau_{q}$ by one order of magnitude (with the top most line corresponding to a circuit time of $1\,\mu{\rm s}$).  The fast saturation of classical simulation cost with depth for RG circuits translates into significant enhancements in the achievable effective qubit number (i.e.\ classical simulation costs) at fixed gate error and circuit time. We note that the estimated effective qubit numbers in \fref{fig:tts} depend explicitly on the achievable error per gate $\varepsilon$. To what scale the QCCD architecture can be pushed while maintaining completely arbitrary connectivity at some fixed small $\varepsilon$ is a technical question that will only be answered by technical progress. While it is true that at \emph{some} system size, memory errors incurred during the transport required to implement RG circuits (which will scale with $N$ in an architecture dependent way) will exceed the 2Q gate error rates. However, we believe the scale and current performance of H2 is far from the limits of what is achievable with current technologies, especially considering recent results on junction transport \cite{PhysRevLett.130.173202,delaney2024scalable}.

\section{The fidelity of random-geometry circuits on H2} \label{sec:exp_results}

\begin{figure*}[!!t]
\centering
\includegraphics[width=1.0\textwidth]{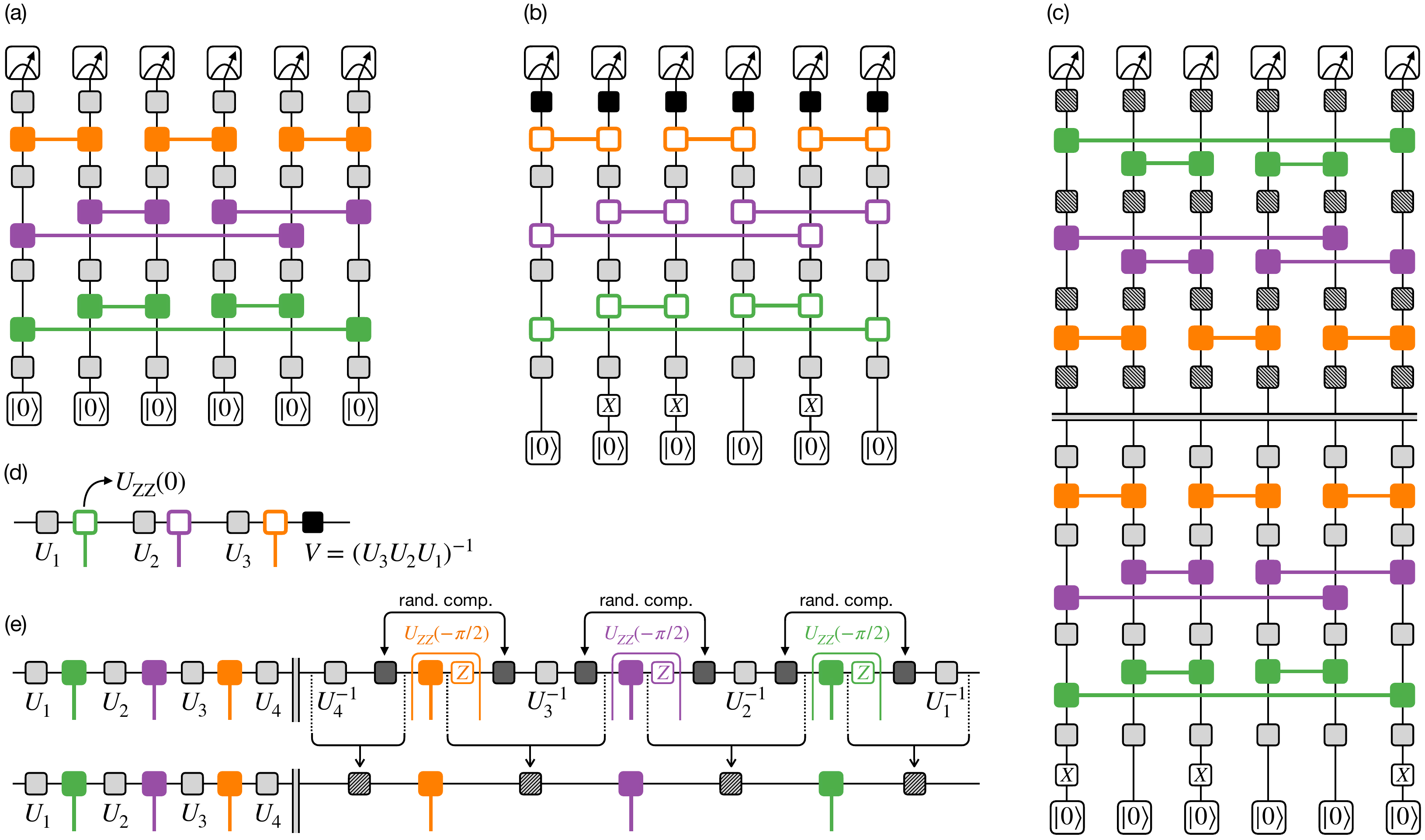}
\caption{
Illustrative examples of the circuits implemented experimentally in this paper. (a) A depth-3 RCS circuit on 6 qubits (see \fref{fig:circuit} for a description of the notation). (b) Transport 1Q RB version of the same circuit. 2Q gates with unfilled boxes denote the use of a $U_{ZZ}(0)$ gate, which induces the same transport and cooling operations as the $U_{ZZ}(\pi/2)$ but does not apply a 2Q gate. The circuit is initialized in a random bit string, in this case 011010, and the 1Q gates at the end (solid black squares) invert the cumulative action of all prior 1Q gates in order to return to this initial state, as shown in (d) for a single qubit. (c) Mirrored version of the same circuit, here initialized in the random bit string 101001. On the reversed part of the circuit all 1Q gates are inverted, $U_{ZZ}(-\pi/2)$ gates are created from $U_{ZZ}(\pi/2)$ gates by appending suitable 1Q $Z$ rotations, and randomized compiling is used to twirl any potentially coherent errors in the 2Q gates into incoherent errors (avoiding any potential cancellation of coherent errors between the forward and backward halves of the circuit).  All 1Q gates on the reverse half are then compiled down to the gates represented as hashed squares in (c), as shown in (e) for a single qubit.
\label{fig:exp_circuits}}
\end{figure*}
\begin{figure*}[!t]
\centering
\subfloat[\label{fig:n56dscan}]{\includegraphics[scale=0.438]{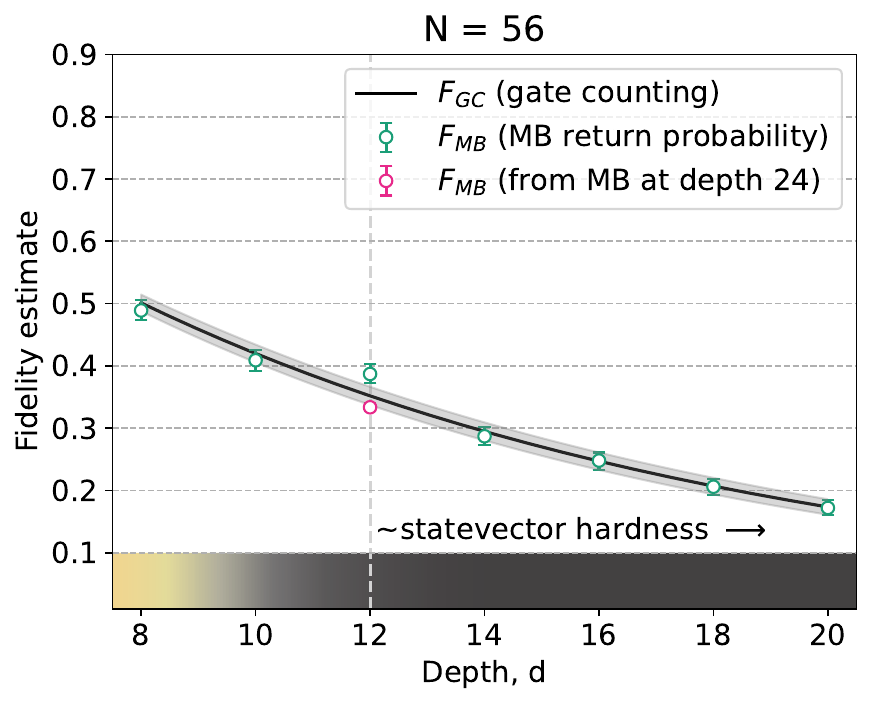}}
\subfloat[\label{fig:nscan}]{\includegraphics[scale=0.438]{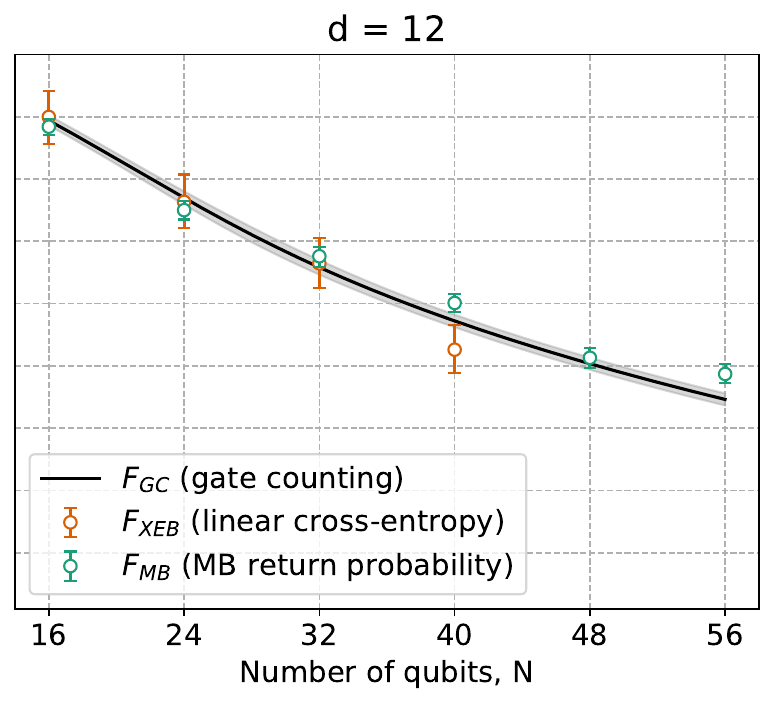}}
\subfloat[\label{fig:n40dscan}]{\includegraphics[scale=0.438]{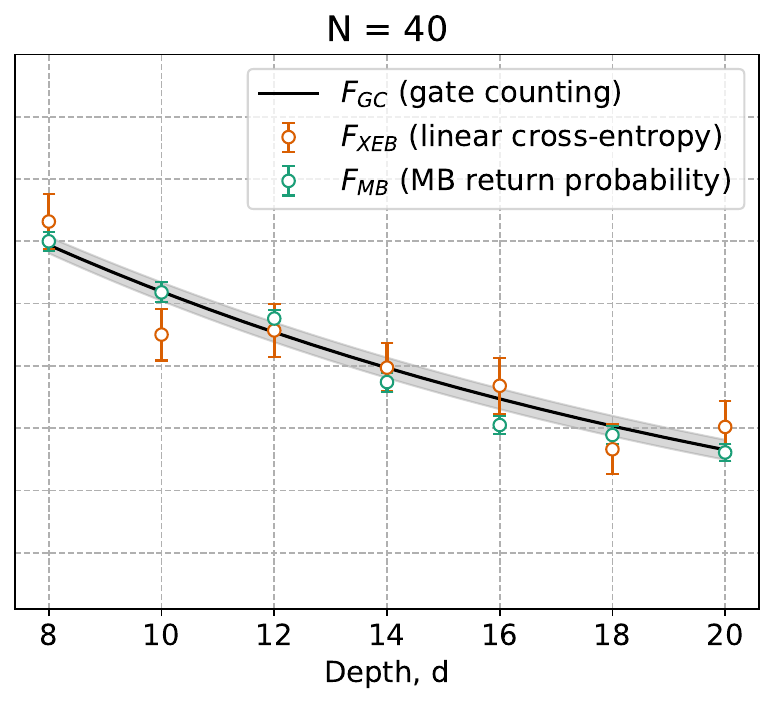}}
\caption{Fidelity estimation from H2 data at a variety of qubit numbers and circuit depths. (a) $F_{\rm MB}$ (MB return probability, green) at $N=56$ as a function of depth. We also plot a fidelity estimate at depth $12$ (pink) inferred from the MB return probability at depth 24 from mirror circuits that were constructed from the exact depth-12 RCS circuits. The color gradient corresponds to the complexity density from Fig.~\ref{fig:d_v_n}. (b) $F_{\rm XEB}$ (linear cross-entropy, orange) and $F_{\rm MB}$ (green) as a function of qubit number $N$ at fixed depth $d=12$. (c) $F_{\rm XEB}$ (orange) and $F_{\rm MB}$ (green) as a function of depth at $N=40$, the largest of the $N$ we took data at that was classically verifiable with our computational resources. All uncertainties plotted represent $1\sigma$ confidence intervals on the data. The gray shaded regions represent $1\sigma$ confidence intervals on the gate-counting model arising from propagation of uncertainties on the component operation fidelities. \label{fig:data} }
\end{figure*}

We have carried out RCS experiments on RG circuits at a variety of qubit numbers $N$ and depths $d$, and all samples obtained from these experiments are reported in Ref.~\cite{rcs_github}. A small example RCS circuit ($N=6$, $d=3$) is shown in \fref{fig:exp_circuits}(a). The RCS experiments were randomly interleaved with other diagnostic experiments meant to validate their fidelity.  Our primary tool for fidelity estimation is a mirror-benchmarking (MB) technique similar to that described in Refs.~\cite{mayer2021theory,proctor2022measuring}, which we carry out at all qubit numbers and circuit depths reported in \fref{fig:data}. Mirror benchmarking provides a fidelity estimate for a depth-$d$ circuit by mirroring a depth-$d/2$ circuit such that the output state is identical to the input state, as in \fref{fig:exp_circuits}(c). Randomized compiling \cite{Wallman2016} is used on the mirrored half of the circuit to prevent the potential cancellation of coherent errors. Bias in our SPAM fidelity is removed by preparing the initial state in a random bitstring, such that the ``correct'' final measurement is an unbiased combination of $0$ and $1$ outcomes. The fraction of samples in which the qubits all return to the expected state, referred to as the MB return probability and denoted $F_{\rm MB}$, serves as an estimate of the overall fidelity of the output state for a depth-$d$ circuit. For all RCS circuits at $N\leq 40$ we also directly compute the linear cross-entropy $F_{\rm XEB}$ from the obtained samples, which serves as a complimentary estimate of the fidelity.  For all higher qubit numbers~---~most notably all data at $N=56$~---~we still report samples from the RCS experiments in Ref.~\cite{rcs_github} but do not compute $F_{\rm XEB}$. While it should be computable for $N=48$ with considerable effort, and at $N=56$ for the shallowest depths, it does not appear possible to compute $F_{\rm XEB}$ for the $N=56$ circuits at $d\gtrsim 12$ with any realistic amount of classical resources (see \sref{sec:slicing}).

We also employ a simple gate-counting estimate based on measured fidelities of the component operations in the circuit, referred to as the ``digital error model" in Ref.~\cite{arute2019}. The largest contribution to the overall circuit infidelity is 2Q gate errors. The 2Q gate average infidelity $\epsilon_{\rm 2Q}$ reported in Tab.~\ref{tab:RB_avgs} is inferred from 2Q RB experiments interleaved with the rest of the data (see Appendix~\ref{sec:2Q_RB}). The next leading source of circuit infidelity is memory error, i.e.\ the cumulative effect of errors that happen to qubits when not doing 2Q gates. This memory error has several origins including reference clock and magnetic field drifts, imperfections in spatial phase tracking and other calibration errors or drifts, leakage out of the qubit manifold, and transport failures; see Ref.~\cite{moses2023race} for further discussion. Since memory error is closely related to the transport time, its magnitude varies with circuit geometry and qubit number. Therefore, to estimate the contribution of memory error we run a separate transport 1Q RB experiment at each qubit number [see \fref{fig:exp_circuits}(b), Appendix~\ref{supp:t1qrb} and Tab.~\ref{tab:transport_1qrb}]. The geometry of the transport 1Q RB circuits at low depths is taken to exactly match the same circuits used in the RCS experiments. By making the replacement $U_{ZZ}(\pi/2)\rightarrow U_{ZZ}(0)$ for all 2Q gates, the transport 1Q RB circuits replicate nearly identical transport steps, cooling, 1Q gates, and overall timing as the RCS circuits without ever applying 2Q gates. In the absence of 2Q gates, the ideal circuit prepares each qubit in a known state. From the decay (with circuit depth) of the probability that each qubit ends in the correct state, we can extract an estimate of the aggregate non-2Q gate contributions to the error per qubit per layer, denoted $\epsilon_{\text{mem}} (N)$ \footnote{We generated random $d$-regular graphs up to $d = \min(30,N - 1)$. This covered all values of $N,d$ run for both MB and XEB. For the deeper transport 1Q RB sequences, circuit layers were repeated cyclically after reaching the maximum $d=30$.}.

Combining the memory error with the estimate of 2Q gate error from randomized benchmarking, the effective process/entanglement infidelity per 2Q gate is
\begin{align}
\varepsilon(N) = \frac54 \epsilon_{2Q} + 2\times \frac32 \epsilon_{\text{mem}} (N),
\end{align}
where the prefactors incorporate the conversion between average and process fidelities~\cite{Nielsen2002} as well as the fact that there are two qubits in each 2Q gate \footnote{Strictly speaking, leakage errors should be subtracted out from the average fidelity data prior to conversion to process fidelities, and then added back in as a process error.  Estimate of our leakage rates from both 2Q RB and transport 1Q RB suggest an aggregate leakage probability of $7(2)\times 10^{-4}$ per 2Q gate. As this rate is small compared to the quoted process error per 2Q gate of $3.2(1)\times 10^{-3}$, we expect such corrections to be very small.}. Note that we do \emph{not} include a contribution from $\epsilon_{1{\rm Q}}$ listed in Tab.\,\ref{tab:RB_avgs}, since 1Q gate errors are already accounted for in $\eps_{\rm mem}(N)$.
A reasonable model for the overall fidelity of the circuit then simply counts the number of 2Q gates and state preparation and measurement operations and takes the corresponding fidelities to the appropriate power:
\begin{align}
F_{\rm GC}(N,d) = \bigl(1 - \varepsilon(N) \bigr)^{Nd / 2} \bigl(1-p_{\text{SPAM}}\bigr)^N. \label{eq:gc}
\end{align}
The SPAM error $p_{\text{SPAM}}$ is measured from an independent SPAM experiment, and reported in Tab.~\ref{tab:RB_avgs}. As demonstrated in Appendix~\ref{sec:XEB_v_fidelity}, this gate-counting model for the circuit fidelity tends to systematically underestimate both the true circuit fidelity as well as the fidelity estimates $F_{\text{XEB}}$ and $F_{\text{MB}}$ provided by the linear cross-entropy and the average return probability of mirror circuits, respectively. Consequently, following the analysis of Appendix~\ref{sec:XEB_v_fidelity}, the effective depth in the gate-counting model is shifted by approximately $1.12$ for direct comparison with $F_{\rm XEB}$ and $F_{\rm MB}$.

We report our experimental results in Fig.~\ref{fig:data}. The set of all circuits for which we report data in Fig.~\ref{fig:data}(a,b) were submitted to the H2 quantum computer in a random order so as to fairly average across any drift in computer performance, using default settings accessible to any user of the device.  The set of $N=40$ circuits for which we verify the linear cross-entropy $F_{\rm XEB}$ in Fig.~\ref{fig:data}(c) were submitted (and executed) immediately afterwards, also in a random order. In Fig.~\ref{fig:data}(a), where we expect that $F_{\rm XEB}$ is not computable except with substantial effort at the lowest depths, we report MB estimates of the fidelity achieved by H2 at $56$ qubits on RG circuits out to depth $20$. The estimated fidelities are roughly two orders of magnitude larger than those reported in prior RCS demonstrations on classically difficult circuits \cite{arute2019,PhysRevLett.127.180501,ZHU2022240,morvan2023phase}. We also execute MB circuits at depth $24$ that are mirrored versions of the exact depth-$12$ RCS circuits to provide another consistent estimate of the fidelity at depth $12$ for $N=56$. This procedure is similar to the one proposed in Ref.~\cite{Proctor22} but without the randomly compiled reference circuit and with SPAM errors added back in to better match expected circuit performance.

In Fig.~\ref{fig:data}(b) we show that $F_{\rm XEB}$ computed up to $N=40$ agrees well with $F_{\rm MB}$ and the gate-counting model across a range of $N$ at depth 12, where we expect that the cost of simulating the circuits is roughly equal to the statevector simulation cost, c.f. Sec.~\ref{sec:TNcost}. Further verification that $F_{\rm XEB}$ is well estimated by $F_{\rm MB}$ is provided by taking data at $N=40$ across a wide range of circuit depths in Fig.~\ref{fig:data}(c), and computing $F_{\rm XEB}$ from the obtained samples on the Perlmutter supercomputer. All MB and RCS experiments were executed with 50 circuit randomizations and 20 shots per randomization at each depth, with the exception of the depth-$24$ experiment which was executed with 50 circuit randomizations and $100$ shots per randomization. All experimental data is provided at \cite{rcs_github}, including data from random circuits for which we could not compute $F_{\rm XEB}$.

\section{Outlook}

Demonstrating a convincing computational advantage of quantum computers over classical computers is widely regarded as a key milestone on the path towards broadly useful quantum computation, and much progress towards this goal has been made in recent years. From a formal point of view, one would ideally like for such a demonstration to be based around a problem for which an exponential asymptotic advantage can be claimed as the size of the problem (and the computer used to tackle it) grows.  Since the original proposal for RCS as a task for demonstrating exponential quantum advantage \cite{boixo_circuit_sampling}, substantial progress has been made in understanding its asymptotic difficulty both with and without circuit noise. For example, compelling evidence now exists that \emph{noiseless} RCS requires exponential (in the qubit number) run time for a classical computer \cite{bouland2018,bouland2022,Movassagh2023}.

In the absence of quantum error correction, the asymptotics of \emph{noisy} RCS (i.e.\ the task of sampling from the output distribution that results from running a quantum circuit with a fixed level of noise per gate) are relevant but remain comparatively less well understood. Notable recent developments include a proof that under certain assumptions about the noise structure and circuit depth, RCS is asymptotically easy for classical computers if the error-rate per gate is held constant as the system size is scaled up \cite{10.1145/3564246.3585234}. We emphasize that \emph{none} of these results (hardness in the absence of noise or easiness at fixed noise rate) is conclusive regarding how the difficulty of RCS grows over a particular range of system sizes given a particular error rate. Easiness results suggest that at sufficiently large noise rates, RCS may be classically feasible at all system sizes. But it is natural to expect that as error rates are reduced, the system sizes over which the asymptotic difficulty of noiseless RCS controls the empirical hardness will grow \footnote{It is worth noting that the classical approach to RCS used to show $N$-asymptotic easiness at fixed noise rate in Ref.~\cite{10.1145/3564246.3585234} scales exponentially with the inverse of the gate error rate, leaving open the possibility that the window of system sizes controlled by noiseless asymptotics may be quite large even at modest gate error rates.}.

Ultimately, which set of results control the behavior of classical RCS in the range of system sizes and error rates accessible to experiments is unclear, but one thing is clear: The larger one can make a quantum computer while continuing to push down error rates (for example, in a manner that maintains overall circuit fidelities while increasing $N$), the more likely it is that we will find tasks at which non-error-corrected quantum computers dramatically outperform the best classical algorithms. The high gate fidelities and arbitrary connectivity afforded by the trapped-ion QCCD architecture have enabled RCS to be carried out in a computationally challenging regime and at completely unprecedented fidelities, leaving open considerable room to scale such demonstrations up \emph{even without} further progress in reducing gate error rates. We are confident that H2 is by no means near the boundary of how far such demonstrations can be pushed using the QCCD architecture. For example, increasing system sizes while also reducing memory errors, circuit times, and 2Q error rates all appear achievable with current technologies by moving towards natively 2D trapping architectures \cite{PhysRevLett.130.173202,delaney2024scalable}, and switching to qubit ions (such as $^{137}{\rm Ba}^{+}$) that afford better SPAM fidelities than $^{171}{\rm Yb}^{+}$ \cite{an2022high} and admit visible wavelength laser-based 2Q gates with more favorable error budgets. Given that RG circuits require such low depths to become hard to simulate (and a fixed fraction of qubits contribute to the worst-case hardness of simulation even as $N$ is scaled up at \emph{constant} depth), near-term scaling progress in the QCCD architecture should enable the faithful execution of quantum circuits whose simulation lies well beyond the reach of any conceivable classical calculation.

\section*{Acknowledgments}
We acknowledge the entire Quantinuum team for their many contributions towards successful operation of the H2 quantum computer with $56$ qubits, and Honeywell for fabricating the trap used in this experiment. We thank Bill Fefferman, Miles Stoudenmire, Xavier Waintel, Scott Aaronson, Garnet Chan, Stefanos Kourtis, Michael Gullans, and Dominik Hangleiter for helpful discussions.
M.L., P.N., R.S. and M.P. wish to thank Daniel Pinto, Lori Beer, Andrew J.\ Lang, George Sherman, Stephen Winer, Jennifer Lavoie, and Emily Mullins for their support, as well as the research scientists at the Global Technology Applied Research center of JPMorgan Chase for insightful discussions.  
M.L., Y.A., and D.L. acknowledge support from the Office of Science, U.S. Department of Energy, under contract DE-AC02-06CH11357 at Argonne National Laboratory and the U.S. Department of Energy, Office of Science, National Quantum Information Science Research Centers.
This research used supporting resources at the Argonne Leadership Computing Facilities. The Argonne Leadership Computing Facility at Argonne National Laboratory is supported by the Office of Science of the U.S. DOE under Contract No. DE-AC02-06CH11357. 
This research used resources of the National Energy Research Scientific Computing Center, a DOE Office of Science User Facility supported by the Office of Science of the U.S. Department of Energy under Contract No.~{DE-AC02-05CH11231} using NERSC award NERSC~{DDR-ERCAP0029825}. Updates to author affiliations after completion of this work: C. Volin, \emph{Oxford Ionics, Oxford OX5 1PF, United Kingdom}; J. Colina, \emph{Los Alamos National Laboratory, Los Alamos, New Mexico 87545, USA}.

\section*{Data availability}

All data presented, along with OpenQASM descriptions of the circuits from which the data was measured, are available at \cite{rcs_github}.

\clearpage
\newpage

\section*{Appendices}
\beginappendix

\section{Primitive operation benchmarks at $N=56$\vspace{0.5em}\label{sec:primitive_benchmarks}}

\subsection{State-preparation and measurement (SPAM) experiment \vspace{0.5em}}
State-preparation and measurement (SPAM) errors are measured with two experiments preparing all eight gate-zone qubits in the $| 0 \rangle$ (and $| 1 \rangle$) state and measuring how often the incorrect `1' (and `0') outputs are returned. The associated probabilities (the ratio between observed wrong outcomes and the number of trials) are denoted $p(1|0)$ and $p(0|1)$, respectively. Due to the nature of the photon counting measurement \cite{PhysRevA.76.052314} there is an imbalance between probabilities $p(1|0)=3.8(7)\times 10^{-4}$ and $p(0|1)=26(2)\times 10^{-4}$. The average $[p(1|0)+p(0|1)]/2$ is reported in Tab.\,\ref{tab:RB_avgs} as the SPAM error $p_{\textrm{SPAM}}=1.47(9)\times 10^{-3}$. Uncertainty is calculated from the standard binomial confidence interval.

\subsection{1Q randomized benchmarking (1Q RB) \vspace{0.5em}}
\begin{figure}[!h]
\centering
\includegraphics[width=\columnwidth]{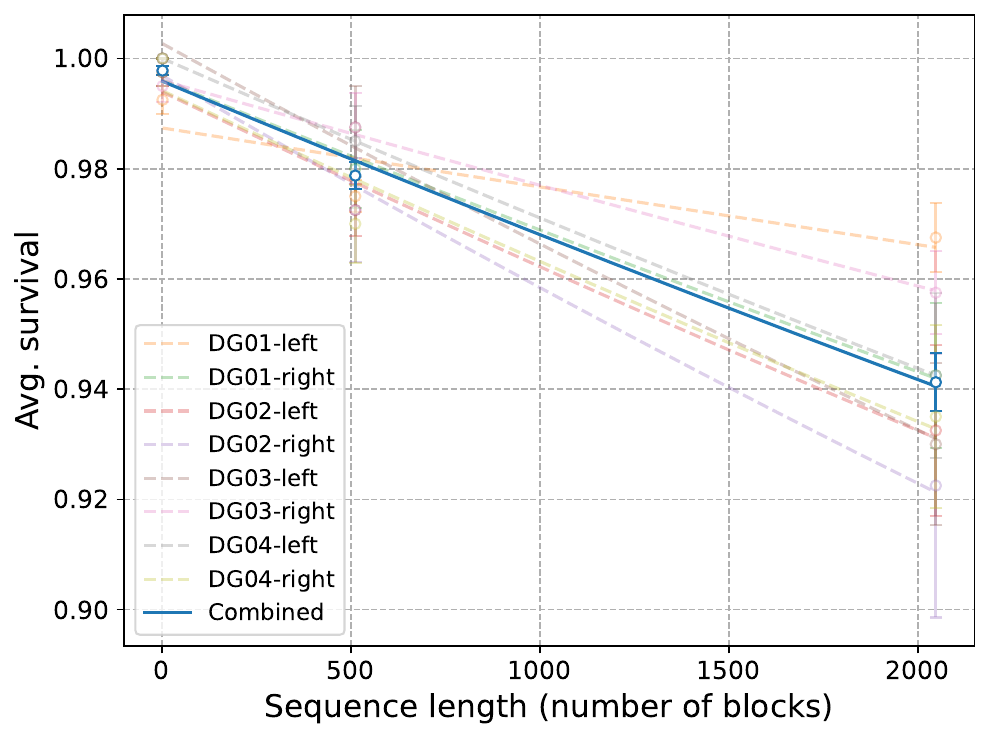}
\caption{1Q RB decay curve for all four gate zones (labeled DG01-DG04 in the legend) each with two qubits (dashed lines, the two qubits labeled as ``left'' and ``right'') and combined estimate averaged over all eight gate-zone qubits (solid line). \label{fig:SQ_RB} }
\end{figure}

1Q randomized benchmarking (1Q RB) was performed on the eight gate-zone qubits with standard Clifford randomization~\cite{Magesan12}. We used sequence lengths [2, 512, 2048] each with 4 randomizations and 100 shots per randomization as shown in Fig.~\ref{fig:SQ_RB}. The RB decay rate $\lambda_{1Q}$ is estimated by fitting the survival probability from each sequence length $p(m)$ with the first-order RB decay equation $p(m) = A \lambda_{1Q}^m + 1/2$~\cite{Magesan12}, where the asymptote is fixed to $1/2$ by randomizing the final output state~\cite{Harper19}. The average 1Q gate infidelity is calculated by combining the results over all eight gate-zone qubits and converting the RB decay rate to an average infidelity per 1Q Clifford to give $\epsilon_{1Q} = (1-\lambda_{1Q})/2 = 2.9(4)\times 10^{-5}$. Uncertainty is calculated from a semi-parametric bootstrap resampling and shown with a one-sigma standard deviation~\cite{Meier06, EfroTibs1993}. 

\subsection{2Q randomized benchmarking (2Q RB) \vspace{0.5em}}
\label{sec:2Q_RB}
\begin{figure}[!h]
\centering
\includegraphics[width=\columnwidth]{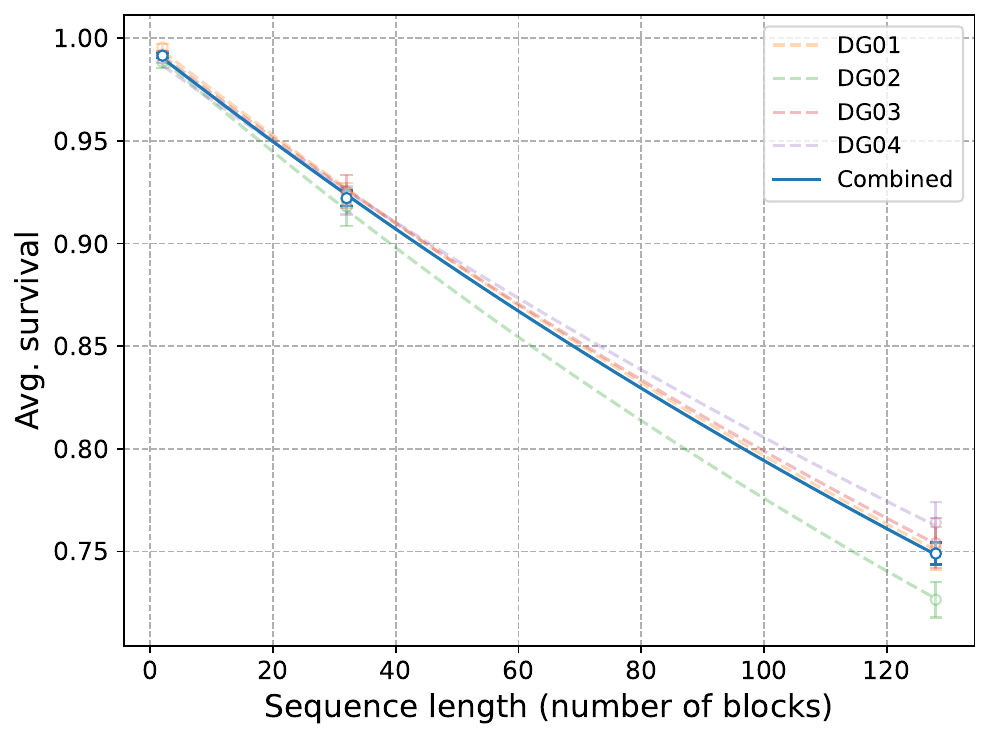}
\caption{2Q RB decay curve for all four gate zones (labeled DG01-DG04 in the legend, dashed lines) and combined estimate averaged over all four gate zones (solid line). \label{fig:TQ_RB} }
\end{figure}

2Q randomized benchmarking (2Q RB) was performed on the eight gate-zone qubits with standard Clifford randomization~\cite{Magesan12}. We used sequence lengths [2, 32, 128] each with 4 randomizations and 100 shots per randomization. 2Q RB was repeated seven total times interspersed over the entire data collection and the results were combined and analyzed together as shown in Fig.~\ref{fig:TQ_RB}. The RB decay rate $\lambda_{2Q}$ is estimated by fitting the survival probability from each sequence length $p(m)$ with the first-order RB decay equation $p(m) = A \lambda_{2Q}^m + 1/4$~\cite{Magesan12}, where the asymptote is fixed to $1/4$ by randomizing the final output state~\cite{Harper19}. The average 2Q gate infidelity was estimated by combining all four pairs of gate-zone qubit's 2Q RB data and scaling the average 2Q Clifford infidelity by the average number of $U_{ZZ}(\pi/2)$ gates per 2Q Clifford (1.5) to give $\epsilon_{2Q} = 3(1-\lambda_{2Q}^{2/3})/4 = 1.57(5)\times 10^{-3}$. Uncertainty is calculated from a semi-parametric bootstrap resampling and shown with a one-sigma standard deviation~\cite{Meier06, EfroTibs1993}. 

\begin{figure*}[!t]
\centering
\subfloat[\label{fig:tsqrb}]{\includegraphics[scale=0.438]{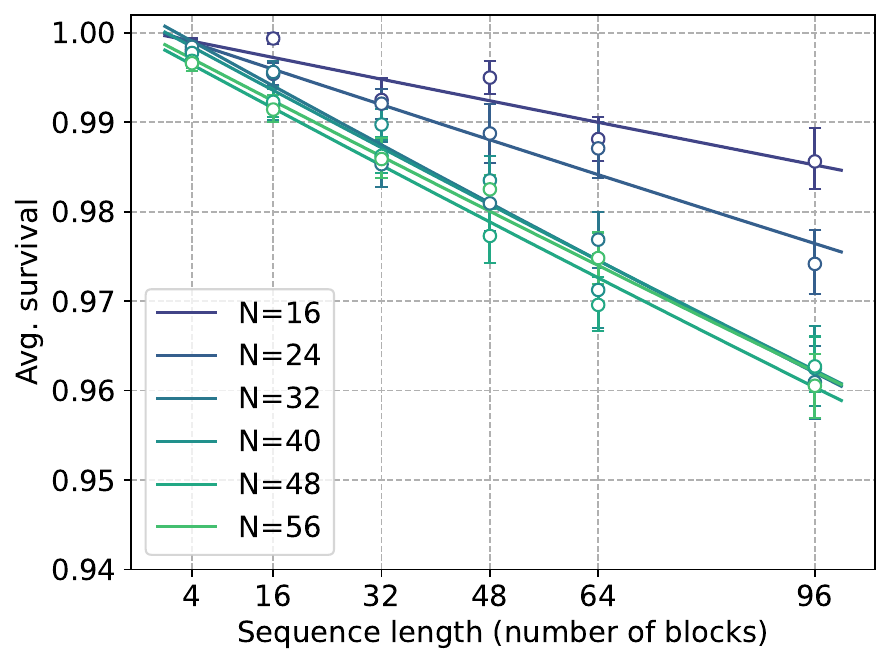}}
\subfloat[\label{fig:tsqrb_n40}]{\includegraphics[scale=0.438]{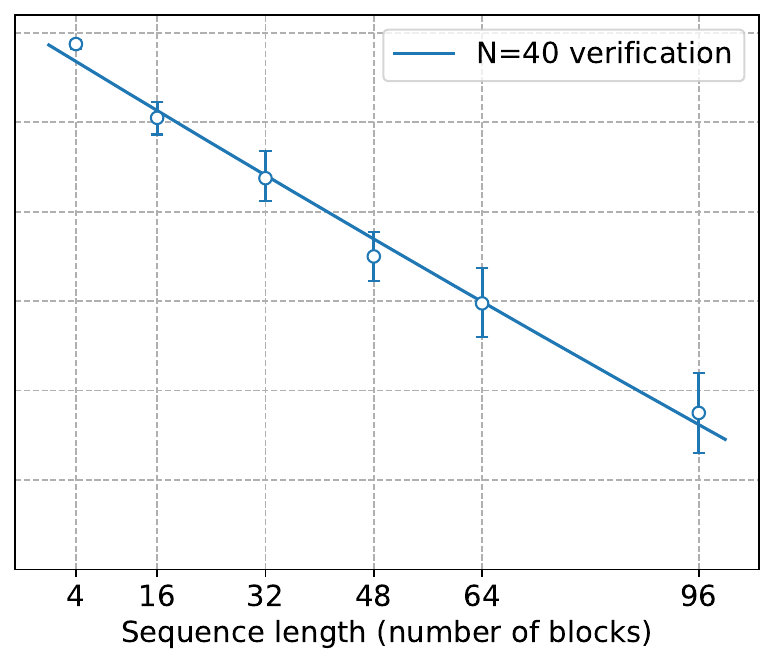}}
\subfloat[\label{fig:mem_err_N}]{\includegraphics[scale=0.438]{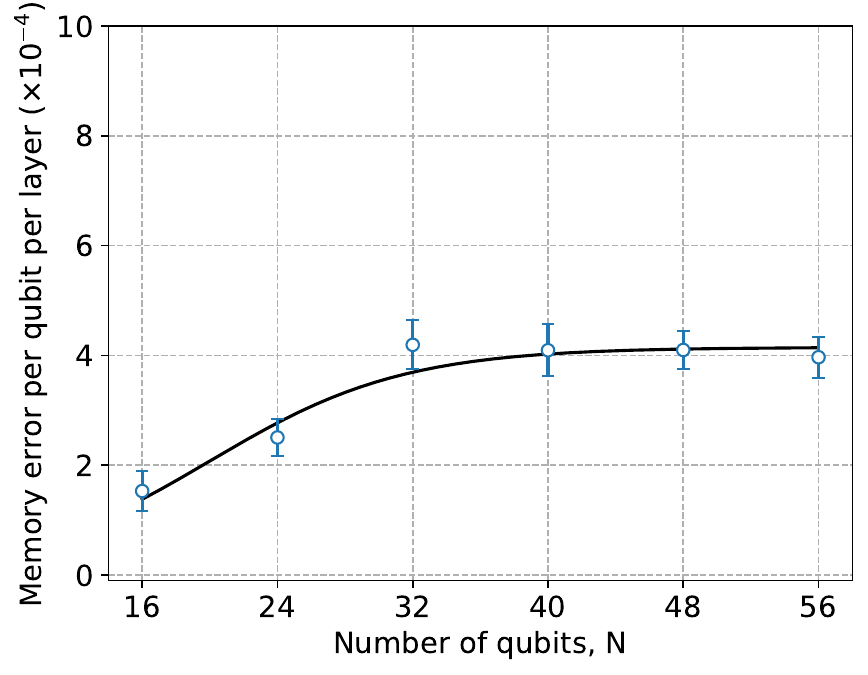}}
\caption{(a) Transport 1Q RB data and decay curves for each qubit number $N$ at which the fidelity is estimated in the main text. (b) Transport 1Q RB data and decay curve for the $N=40$ verification data set collected subsequent to the rest of the data. (c) Memory error per qubit per layer extracted from the decay curves in (a), along with the logistic function model described in the text.}\label{fig:transport_sqrb_combined} 
\end{figure*}

\subsection{Transport one-qubit randomized benchmarking (Transport 1Q RB) \label{supp:t1qrb}\vspace{0.5em}}
Transport 1Q randomized benchmarking (transport 1Q RB) was performed on up to all 56 qubits with random 1Q unitaries selected from SU(2). In transport 1Q RB, rounds of Haar-random 1Q random unitaries are interleaved with random transport operations. The random transport operations are selected to mimic the circuits used in the main text and are implemented with a dummy 2Q gate that forces the ion transport and cooling  but does not apply the lasers that generate the 2Q gate. We used sequence lengths $[4,16,32,48,64,96]$ each with 10 randomizations and 10 shots per randomization, at each $N \in [16,24,32,40,48,56]$. The data for each qubit was accumulated and analyzed as a 1Q RB experiment (as detailed above) performed in parallel across all $N$ qubits, as displayed in Fig.~\ref{fig:tsqrb}, allowing us to extract an error per qubit that accounts for memory errors accrued during the random transport as well as errors from the 1Q gates. A second transport 1Q RB experiment was also executed at $N=40$ (see Fig.~\ref{fig:tsqrb_n40}), interleaved with the verification set of $N=40$ RCS circuits whose results are displayed in Fig.~\ref{fig:data}c, since these circuits were executed subsequent to the rest. In Tab.~\ref{tab:transport_1qrb} we report the resulting estimate of memory error per qubit per layer extracted from each experiment. Uncertainty is calculated from a non-parametric bootstrap resampling with $1000$ resamples and is shown with a one-sigma standard deviation.

\begin{table}[]
\begin{ruledtabular}
\begin{tabular}{lcc}
Number of qubits  &$\epsilon_{\text{mem}}$ ($\times 10^{-4}$) & $\varepsilon(N)$ \:($\times 10^{-3}$)   \\[.3em]  \hline \\[-.6em]
$N=16$  & 1.5(4) & 2.4(2)   \\[.1em] 
$N=24$   & 2.5(4) & 2.8(2)     \\[.1em] 
$N=32$   & 4.2(4) & 3.1(2)  \\[.1em] 
$N=40$   & 4.1(5) & 3.2(2)   \\[.1em] 
$N=40$ (verification set)   & 4.6(5) & 3.4(1) \\[.1em] 
$N=48$ & 4.1(4)& 3.2(1)   \\[.1em] 
$N=56$ & 4.0(4)& 3.2(1)  \\[.1em] 
\end{tabular}
\end{ruledtabular}
\caption{Memory error per qubit per layer extracted from a separate transport 1Q RB experiment at each fixed qubit number $N$, and the resulting effective process infidelity per 2Q gate $\varepsilon(N)$.}
\label{tab:transport_1qrb}
\end{table}

\begin{figure}[!t]
\centering
\includegraphics[width=\columnwidth]{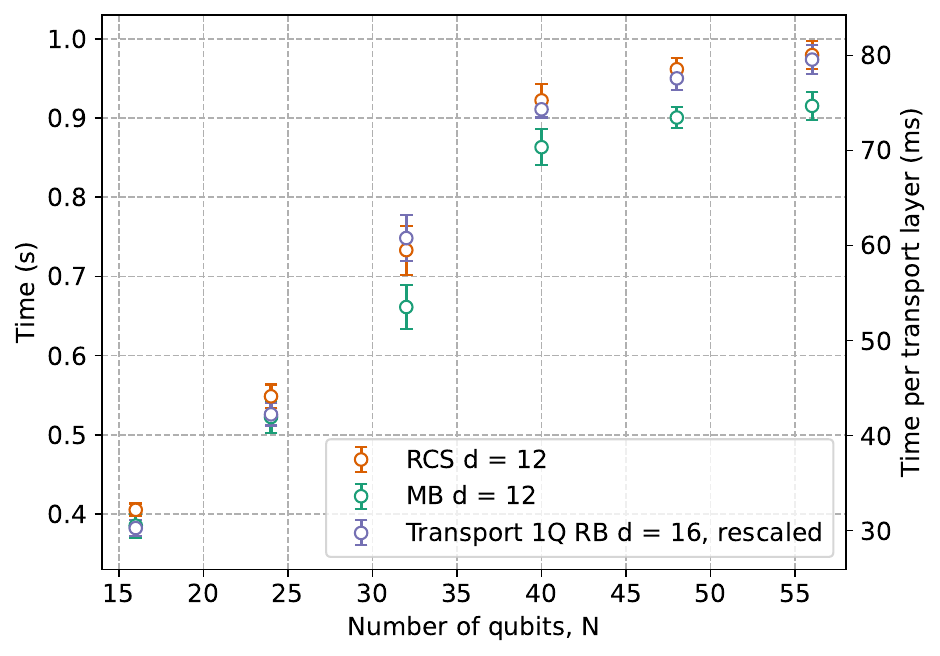}
\caption{Average time per shot on the H2 quantum computer for each type of circuit at $d=12$ as a function of number of qubits, $N$. Since there is no transport 1Q RB data at sequence length $12$, we take the transport 1Q RB data from sequence length $16$ and rescale times linearly to length $12$ (see Fig.~\ref{fig:timing} for evidence that the time per layer of transport 1Q RB sequences scales nearly perfectly linearly with depth).
\label{fig:timingNrange} }
\end{figure}

In order to produce estimates of the fidelity from gate counting of circuits at intermediate qubit numbers in Fig.~\ref{fig:data}b, we require a model for the memory error at all possible qubit numbers. As the qubit number is increased towards the maximum of $56$ supported by the device, the time required for a full round of transport operations increases and eventually plateaus (since there are always $56$ qubits in the trap, once a sufficient fraction of the qubits are involved in a given program there is no longer a substantial increase in transport time with further increase in qubit number; see Fig.~\ref{fig:timingNrange}).
We emphasize that the time reported for sorting at $N = 32$ is larger than that reported in Ref.~\cite{moses2023race} for $N=32$, because the time required for $N=32$ now includes transport operations on all $56$ qubits ($24$ of which are loaded in the trap but not involved in the circuit). As an empirical model for the resulting memory error as a function of $N$ (interpolating between zero error as $N\to 0$ and a maximum level of error as $N\to 56$) we employ a logistic (sigmoid) function $L$ of the form
\begin{align}
    L(N) = \frac{A}{1+e^{-k(N-N_0)}},
\end{align}
representing interpolation between extreme values of $0$ and $A$, centered at $N_0$, with slope governed by $k$. A nonlinear least-squares fit of $L(N)$ to the measured memory error as a function of $N$ yields the fit parameters $A = 4.1(2) \times 10^{-4}$, $N_0 = 20(2)$, $k = 0.18(5)$, and is depicted in Fig.~\ref{fig:mem_err_N}. Propagating these fit values and their uncertainties allows estimates of the memory error per qubit per layer at all $N$, continuously interpolating between $N=16$ and $N=56$. In combination with the other component benchmarks reported in Tab.~\ref{tab:RB_avgs}, the gate-counting model in Eq.~\eqref{eq:gc} then supplies the estimates of the fidelity as a function of $N$ reported as $F_{\rm GC}$ in Fig.~\ref{fig:data}b. 

\begin{figure}[!t]
\centering
\includegraphics[width=1.1\columnwidth]{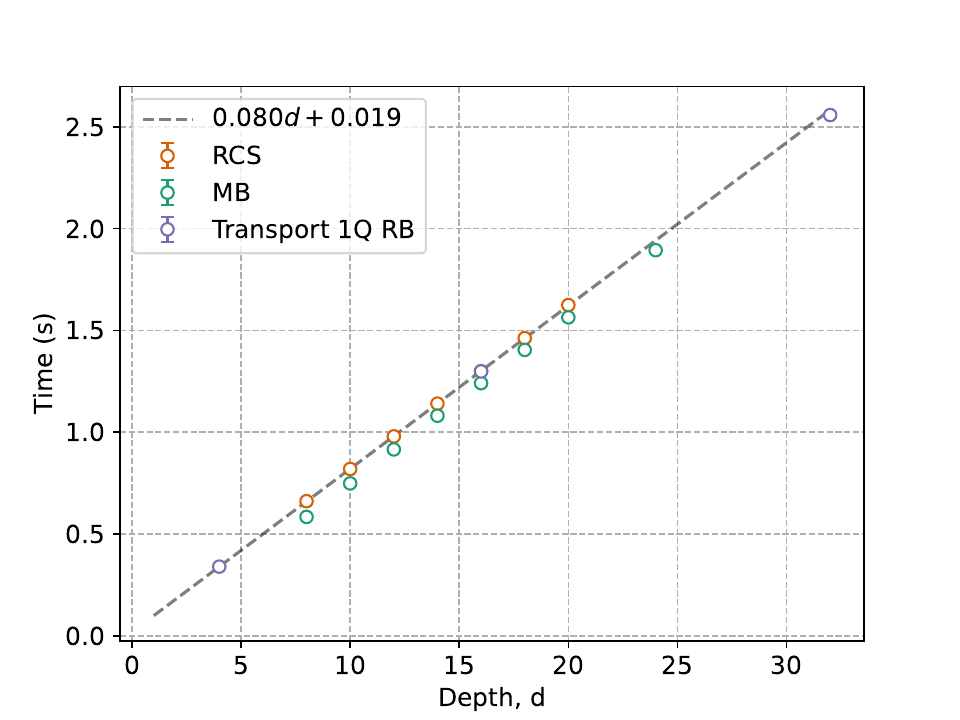}
\caption{Average time per shot on the H2 quantum computer for each type of circuit at $N=56$ as a function of depth. For the transport 1Q RB (and RCS) data, the execution time is essentially exactly linear: fitting all depths out to $96$ of the transport 1Q RB data yields the model: $\text{time} = (0.080d + 0.019) \text{ s.}$, as depicted. Note that the range of the $x$ axis is restricted to $32$ in the plot to make it easier to visualize the MB and RCS data. Uncertainties are smaller than point size.\label{fig:timing} }
\end{figure}

\begin{figure}[!t]
\centering
\includegraphics[width=1.1\columnwidth]{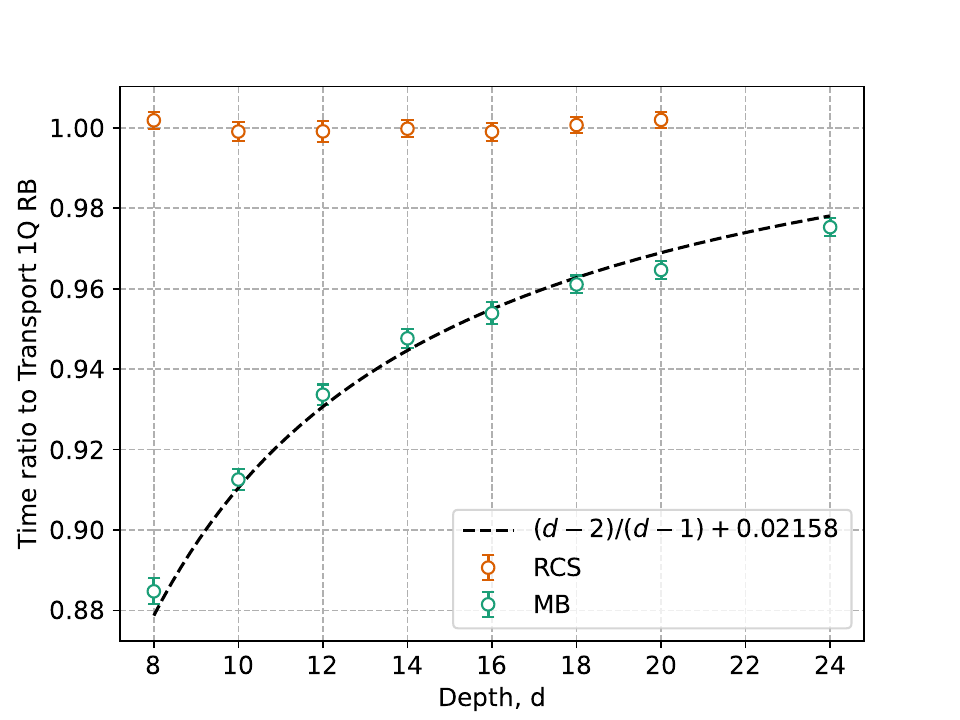}
\caption{The ratios of average shot time at each depth for $N=56$ MB and RCS circuits to the expected time of transport 1Q RB circuits by interpolating the linear model in Fig.~\ref{fig:timing}.\label{fig:timing_ratios} }
\end{figure}

It is interesting to examine to what extent shot time on the H2 quantum computer depends on circuit structure. In Fig.~\ref{fig:timing} we provide evidence that for RCS and transport 1Q RB circuits, which have identical gate topologies but different gate angles, the average time per shot at $N=56$ is essentially identical and nearly perfectly linear with circuit depth. Mirror benchmarking circuits, on the other hand, are expected to have one fewer layers of transport compared to RCS and transport 1Q RB circuits, since there should be no transport required to execute the first layer in the backwards half of the circuit. In Fig.~\ref{fig:timing_ratios} we investigate the extent to which this is true by comparing the shot times of RCS and MB circuits to the times of the corresponding transport 1Q RB circuits at the same depth, computed by interpolating the linear model in Fig.~\ref{fig:timing}. RCS circuits demonstrate essentially identical timings to the corresponding transport 1Q RB circuits. While MB circuits roughly demonstrate the expected $(d-2)/(d-1)$ scaling, they also have a constant multiplicative overhead of approximately $2.2\%$ compared to transport 1Q RB that may be due to differences in the required transport to execute the mirror circuit.

\begin{figure*}[!!t]
\centering
\includegraphics[width=0.98\textwidth]{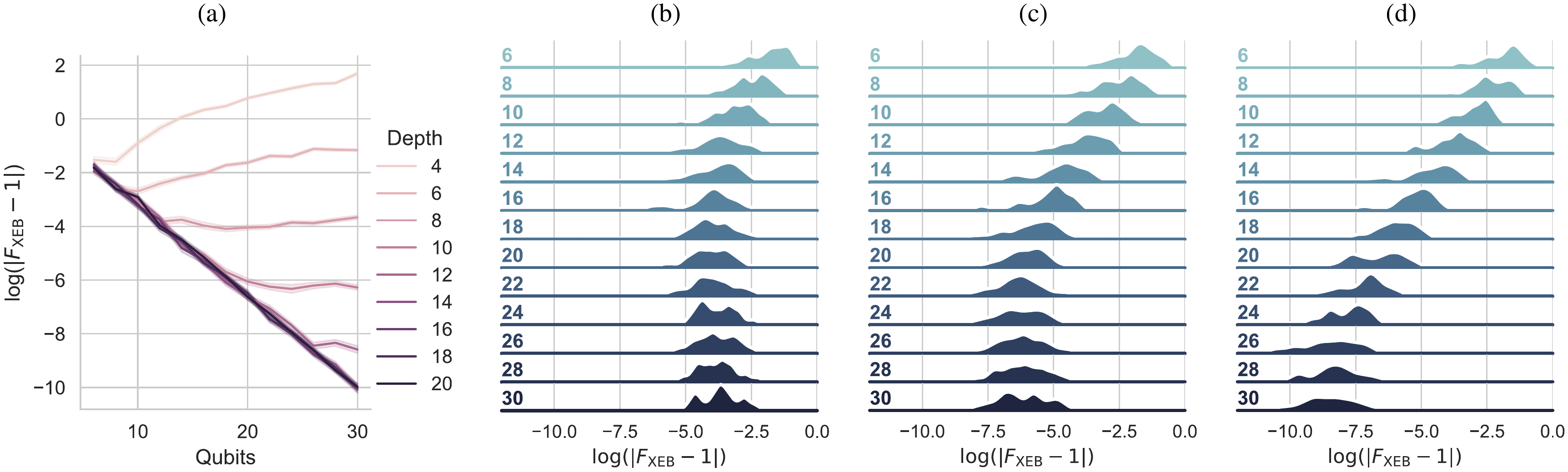}
\caption{(a) Deviations of $F_{\rm XEB}$ from unity in the absence of noise as a function of qubit number $N$ for RG circuits of various depths.  Each curve shows the median over 100 circuits, and shaded regions represent one standard error. (b-d) Distributions of $|F_{\rm XEB}-1|$ over circuits for various $N$ at (b) $d=8$, (c) $d=10$, and (d) $d=12$. The histograms are smoothed probability densities of $\log{|F_{\rm XEB}-1|}$ over 100 random circuits at the stated value of $N$ (different $N$ offset vertically, labeled on the left of each plot) and $d$.  Small values indicate that the distribution of probabilities at the output of a given circuit is well converged (in its second moment) to the Porter-Thomas distribution. Over the limited range of $N$ studied, it is clear that the vast majority of the distribution lies at values well below $1\%$ for $d\geq10$ and $N\geq 16$.
\label{fig:xeb_flucs}}
\end{figure*}
\begin{figure}[!!t]
\centering
\includegraphics[width=0.98\columnwidth]{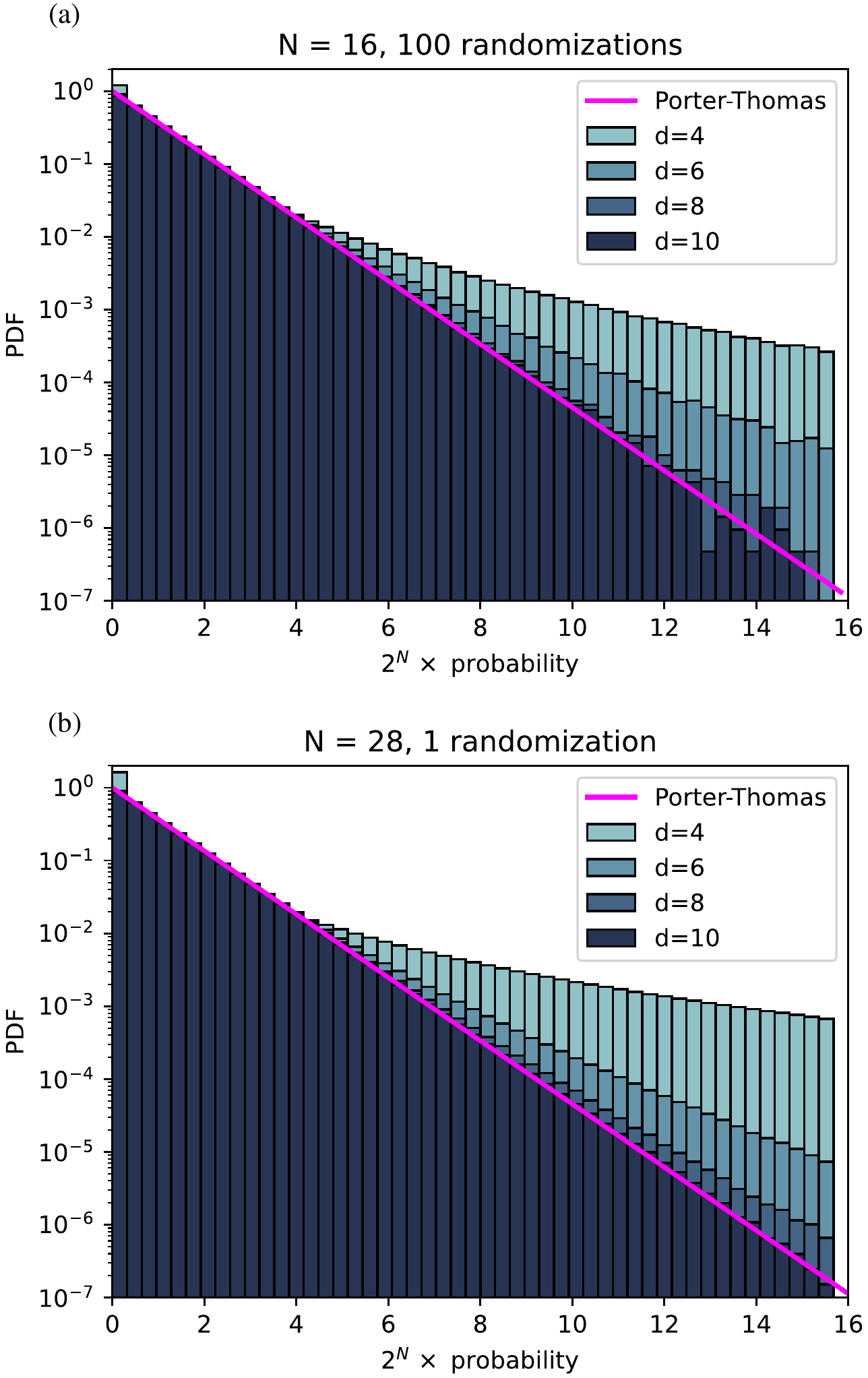}
\caption{Histograms of output probabilities for circuits on (a) $N=16$ qubits and (b) $N=24$ qubits, both at various depths. All probabilities (horizontal axis) are rescaled by $2^N$, and the vertical axis is the normalized probability density function (PDF) over rescaled probabilities. By depth $10$, the distribution is well-converged to Porter-Thomas upon averaging over a $100$ circuits for $N=16$, and self-averages to Porter-Thomas for a single circuit by $N=28$.
\label{fig:PT_hists}}
\end{figure}

\section{Comparison of various fidelity estimators\vspace{0.5em}\label{sec:XEB_v_fidelity}}

It is generally expected that the fidelity of a sufficiently deep random circuit will be roughly equal to the probability that no constituent operation has failed; we refer to this fidelity estimation strategy as gate counting, and label the fidelity estimated in this manner $F_{\rm GC}$.  To estimate the actual fidelity of a circuit experimentally (in the regime where it is not computable even assuming an accurate noise model is known), we use the mirror benchmarking technique outlined in Ref.~\cite{mayer2021theory} adapted to the random circuits run here.  In particular, to assess the fidelity of a depth-$d$ random graph (RG) circuit we take a depth-$d/2$ RG circuit (assuming $d$ is even) and then append to it the same circuit with all gates inverted and run in the reverse order, such that the overall circuit (ideally) implements the identity map.
Randomized compiling \cite{Wallman2016} is applied to the reverse circuits to prevent the possible cancellation of coherent errors.
We choose the initial state to be a random computational basis state to remove any potential SPAM bias, measure the fraction of shots that return the (randomized) initial bit string, and denote this fraction $F_{\rm MB}$.  Strictly speaking, the decay of $F_{\rm MB}$ with depth provides an estimate of the variance of the Pauli fidelities of the per-layer error channel of the circuit.
From this estimate, the per-layer fidelity can be bounded. However, as the numerics below confirm, $F_{\rm MB}$ provides an excellent estimate of the overall circuit fidelity even under a highly realistic local noise model.

For circuits deep enough that the output distribution of the ideal circuit is well-approximated by the Porter-Thomas distribution, we also expect the linear cross-entropy fidelity $F_{\rm XEB}$ to agree with both of the above estimates.  In this section, we analyze to what extent
\begin{align}
F\approx F_{\rm MB}\approx F_{\rm XEB}
\end{align}
is valid for the circuits in the manuscript.  First, we assess the convergence to Porter-Thomas of the output distribution of ideal circuits as a function of depth and system size. At the depths where such convergence is adequate, we then run numerical simulations with a detailed noise model of our machine at a variety of moderate system sizes, checking that all three quantities agree in a manner that is very nearly system-size independent when the noise is scaled in such a manner that the probability of a gate failure per layer of 2Q gates is held fixed.

\subsection{Convergence of RG circuits to Porter-Thomas without noise}

For insufficient depths, the probabilities of bitstrings at the output of RG circuits will not be distributed according to the Porter-Thomas distribution even in the absence of circuit noise, and linear XEB fails to be a good fidelity estimator.  However, \fref{fig:xeb_flucs} shows that the convergence to Porter-Thomas in the second moment (equivalent to $|F_{\rm XEB}-1|\rightarrow 0$ in the absence of noise) happens very quickly as the depth is increased, with the bulk of the distribution over $100$ random circuits showing fluctuations well below the $1\%$ level for $d\geq 10$. In \fref{fig:PT_hists} we show the full distribution of binned output probabilities at depths $d\in\{4,6,8,10\}$. Even for $N=16$ (the smallest number of qubits for which data is presented in the paper), upon averaging the distribution over 100 circuits we find that the convergence to Porter-Thomas is excellent over the entire distribution by depth $10$ (\fref{fig:PT_hists}(a)). By $N=28$ we find good convergence at depth $10$ even for a single random circuit (\fref{fig:PT_hists}(b)).

\begin{figure*}[!t]
\centering
\subfloat[\label{fig:F_comp}]{\includegraphics[scale=0.57]{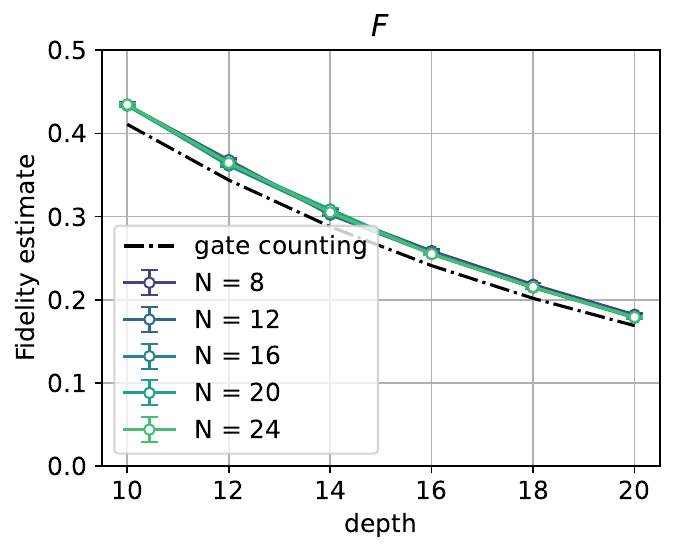}}
\subfloat[\label{fig:F_MB_comp}]{\includegraphics[scale=0.57]{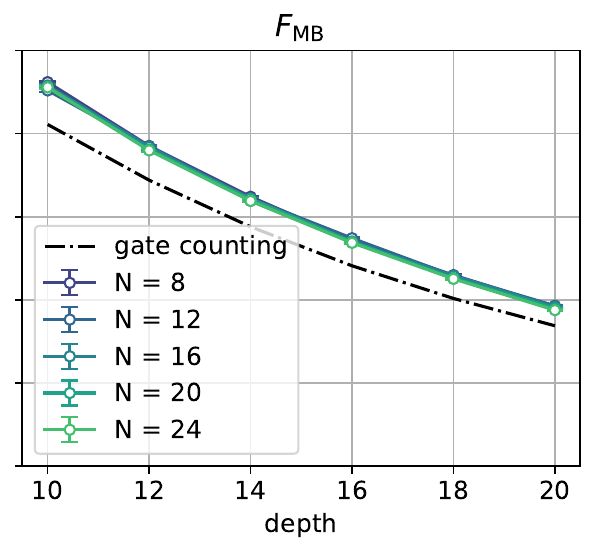}}
\subfloat[\label{fig:F_XEB_comp}]{\includegraphics[scale=0.57]{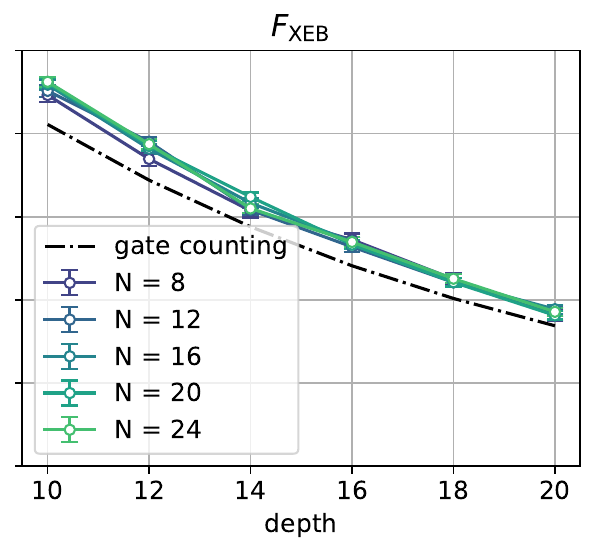}}
\caption{Fidelity collapse when scanning $N$. The three fidelity estimators plotted are (a) True fidelity of the output state with respect to the ideal state; (b) Mirror benchmarking fidelity, with the fidelity at depth $d$ inferred from the return probability of a mirrored depth-$d/2$ circuit; (c) $F_{\rm XEB}$, the linear cross-entropy benchmarking fidelity.  All simulations are randomized over 100 random circuits, and all error bars (generally too small to resolve) are $1\sigma$ confidence intervals from bootstrap resampling.
\label{fig:fids_comp_N}}
\end{figure*}

\subsection{Numerical studies with a detailed noise model}
The simulations reported in this section utilize a noise model consisting of 2Q stochastic Pauli channels attached to each $U_{\rm ZZ}(\pi/2)$ gate and a memory error attached to each qubit during the idling time between sequential 2Q gates that are applied to it.  The stochastic Pauli channel is inferred from cycle benchmarking data~\cite{Erhard2019}, taken on H1-1 and rescaled according to the measured 2Q RB fidelity on H2 such that its average infidelity is the measured $\epsilon_{2Q}=1.57\times 10^{-3}$. The memory error per qubit per layer is inferred from transport 1Q RB data measured at $N=56$ (see Appendix~\ref{sec:primitive_benchmarks}), and the associated average infidelity is $\epsilon_{\rm mem}(56)=4.0\times 10^{-4}$. While the transport 1Q RB data does not distinguish between different types of memory errors, we assume the per-qubit-per-layer memory error is entirely attributable to coherent dephasing; while we expect memory errors to be only partially coherent, we make this assumption because it seems most likely to challenge the assumptions underlying the various fidelity estimates. Finally, we rescale all noise rates by a factor of $56/N$ for simulations performed on $N$ qubits, such that the error rate per layer of 2Q gates is held fixed to the $N=56$ value as we change $N$. The gate-counting estimate works by converting these error rates into process infidelities, and then multiplicatively composing the process fidelities of the circuit components as 
\begin{align}
F_{\rm GC}=\Big(1-\varepsilon(N)\Big)^{Nd/2},
\end{align}
where here
\begin{align}
\varepsilon(N) = \frac{56}{N}\bigg(\frac54 \epsilon_{2Q} + 2\times \frac32 \epsilon_{\text{mem}} (56)\bigg)
\end{align}
is the rescaled effective process infidelity per 2Q gate.
Figure \ref{fig:fids_comp_N}(a) shows calculations of the actual circuit fidelity compared to this gate-counting estimate for various system sizes and over a range of circuit depths.  The fidelities are obtained by quantum trajectories simulations, and we average over 100 random circuits taking 1024 trajectories for each.  Error bars represent $1\sigma$ confidence intervals obtained by bootstrap resampling of the numerical data (1000 resamples).  Figure \ref{fig:fids_comp_N}(b,c) show simulations of $F_{\rm MB}$ and $F_{\rm XEB}$ under the same conditions (quantum trajectories simulations averaging over 100 circuits with 1024 shots per circuit, error bars from bootstrap resampling).  An important conclusion of the numerical data in \fref{fig:fids_comp_N} is that the data collapses well for all $N$ simulated, strongly suggesting that the observed similarity between the various fidelity estimates should hold at $N=56$.

\begin{figure}[!t]
\centering
\includegraphics[width=0.98\columnwidth]{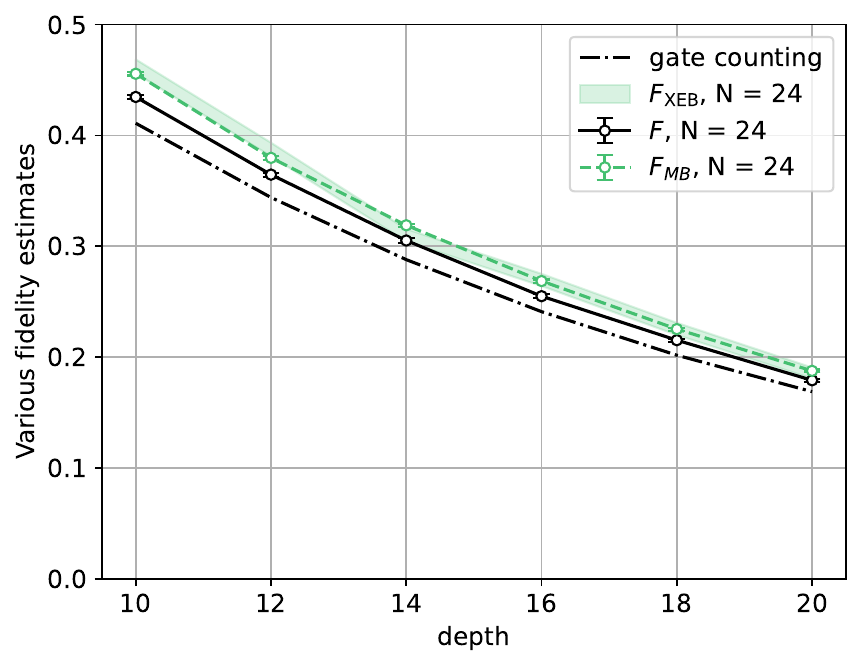}
\caption{Comparing various fidelities at $N=24$, which was the largest system size we performed noisy simulations for. Error bars (for $F$ and $F_{\rm MB}$) and the shaded region (for $F_{\rm XEB}$) are $1\sigma$ confidence intervals obtained by bootstrap resampling.
\label{fig:fids_comp_N24}}
\end{figure}

Figure \ref{fig:fids_comp_N24}  shows comparisons of $F$, $F_{\rm GC}$, $F_{\rm MB}$, and $F_{\rm XEB}$ at the largest simulated system size, $N=24$. All three fidelity estimates agree reasonably well with the computed $F$, though there are small systematic differences between the various estimates that have a known physical origin \cite{gao2021limitations}.  The gate-counting argument assumes that if there are one or more errors in the circuit, the output state will be orthogonal to the ideal one. While this assumption is approximately true for very deep circuits, errors near the beginning of the circuit (when the state is only weakly entangled) do not effectively orthogonalize the state, and therefore gate counting underestimates the true fidelity.  Both $F_{\rm MB}$ and $F_{\rm XEB}$ also reflect this imperfect state orthogonalization, but now due to errors at either the beginning \emph{or} end of the circuit, causing these estimators to agree well with each other but to \emph{overestimate} the true fidelity by the same amount that gate counting underestimates it. Note that the small discrepancies between these various fidelity estimates and the true fidelity are generally comparable to or smaller than the error bars reported in the manuscript, and we interpret the numerical data in Figs.~(\ref{fig:fids_comp_N}, \ref{fig:fids_comp_N24}) as confirmation that all three fidelity estimates~---~$F_{\rm GC}$, $F_{\rm MB}$, and $F_{\rm XEB}$~---~are suitably accurate for our purposes. In particular, this data supports our expectation that $F_{\rm XEB}$ of the data taken from $56$-qubit circuits (and available at \cite{rcs_github}) should agree well with the mirror benchmarking return probabilities for those circuits, despite the fact that we have not computed the cross-entropy of that data ourselves (and we do not know how it could be computed except possibly at the lowest depths with significant effort).  However, because $F_{\rm GC}$ underestimates the true fidelity while $F_{\rm MB/XEB}$ overestimates it, the cumulative affect is a resolvable (at the level of our error bars) discrepancy between $F_{\rm GC}$ and $F_{\rm MB/ XEB}$.  Since the discrepancy arises due to the inadequate discounting of errors near the early and late-time boundaries of the circuit by $F_{\rm GC}$, we fit the gate-counting fidelity at a shifted depth $d\rightarrow d-\delta$ to the $F_{\rm MB}$ numerics at $N=24$ to obtain a depth shift of $\delta \approx  1.12$. We use this inferred depth shift to report the gate-counting fidelities in \fref{fig:data} of the manuscript.

\section{Quantifying uncertainties in the experimental data\label{sec:uncertainties}}

To compute the uncertainties associated to measurements of $F_{\rm XEB}$ and $F_{\rm MB}$, we employ non-parametric bootstrapping~\cite{EfroTibs1993}. An experimental data set is a collection of observed bitstrings $\{x_j\}=S$, and for all $F_{\rm XEB/MB}$ data reported in \fref{fig:data}, each data point corresponds to an experiment in which $50$ random circuits are sampled $20$ times each, for a total of $1000$ samples per experiment.
For a given outcome $x_j$, we let $f_j = 2^N P(x_j)$ (with $P(x_j)$ the probability of $x_j$ for the ideal circuit)
in the case of $F_{\rm XEB}$,
or $f_j\in\{0,1\}$ with ``1'' or ``0'' indicating a ``right'' or ``wrong'' value in the case of $F_{\rm MB}$.
We let $f$ be the expectation value of $f_j$,
and $\hat f$ its estimate obtained by averaging $f_j$ over the experimental outcomes: $\hat f=\frac{1}{N_s}\sum_j f_j$, where $N_s$ is the total number of samples. In the non-parametric bootstrap procedure, one creates a large number $r$ of synthetic data sets $\{\mathcal{S}_{1},\mathcal{S}_2,\dots,\mathcal{S}_{r}\}$, each one obtained by resampling the experimental data with replacement.  The quantity of interest is then averaged over the data in each synthetic data set to produce a distribution of estimates $\{\hat f_1,\hat f_2,...\hat f_r\}$. A $1\sigma$ confidence interval $[f_{-1\sigma},f_{+1\sigma}]$ is assigned to this distribution according to the appropriate $1\sigma$ quantiles, and a confidence interval for $f$ is then constructed by reflecting the bootstrapped confidence interval about twice the empirically observed mean $\hat{f}$,
\begin{align}
{\rm CI} = [2\hat{f} - f_{+1\sigma}, 2\hat{f} - f_{-1\sigma}].
\end{align}
This reflection is intended to remove bias in the construction of the bootstrap interval arising from the fact that the mean of the bootstrap distribution may not match the empirically observed mean.

We consider two methods for resampling from the observed data set. One possible method is to pool together all of the samples associated to a particular experiment and resample from this combined pool, which we call the \emph{aggregate resampling} method. Another, which we call the \emph{double resampling} method, is to first resample from the pool of circuits associated to a particular experiment, and then to further resample from the observed samples for each of the resampled circuits. In general, as we provide evidence for below, the double resampling method results in a more conservative (wider) confidence interval. To maintain consistency with previously reported methods for randomized benchmarking \cite{moses2023race}, we choose the double resampling method to quantify uncertainties of component benchmarks. However, which confidence interval is most appropriate depends on the extent to which the underlying distribution being sampled from for each circuit fluctuates from circuit to circuit.  To understand which is more appropriate in the context of estimating $F_{\rm MB}$ and $F_{\rm XEB}$, we first assess to what extent these quantities vary from circuit-to-circuit given a fixed error model.

\begin{figure}[!t]
\centering
\includegraphics[width=0.98\columnwidth]{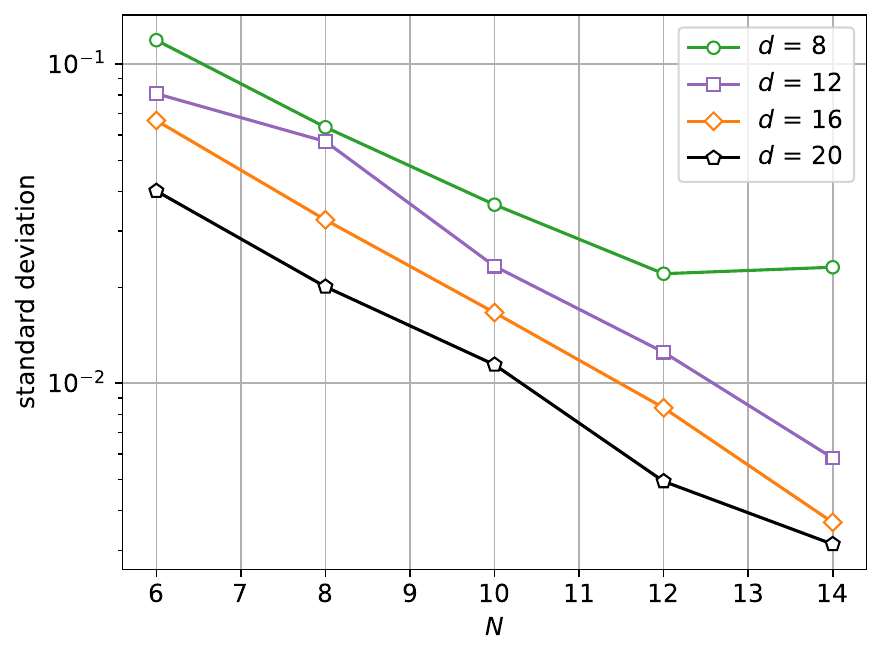}
\caption{Standard deviation of $F_{\rm XEB}$ computed over $100$ random circuits, and plotted versus system size for various depths. For this calculation, $F_{\rm XEB}$ is computed for each circuit by full density matrix simulation using a depolarizing noise model for the 2Q gates (with depolarizing parameter at each $N$ rescaled to have the same error-per-layer as the measured value for H2 at $N=56$). The standard deviation falls below $0.01$ as $N$ is increased for all but the shallowest ($d=8$) circuits, which saturate at a slightly higher value with increasing $N$. 
\label{fig:dm_flucs}}
\end{figure}

From numerics on small system sizes, we believe the fluctuations in both quantities from circuit to circuit are well below the $1\%$ level for the circuits run in this paper ($d\geq 8$ at $N=56$ and $d\geq 12$ at all $N\geq 16$). In the case of $F_{\rm XEB}$, evidence for this statement is provided in \fref{fig:dm_flucs}.  There, we perform density matrix simulations of noisy circuits at small systems sizes ($N\leq 14$), and observe that the standard deviation of $F_{\rm XEB}$ over circuits falls below $10^{-2}$ for all depths $d\geq 12$. A more relevant source of fluctuations from circuit to circuit comes from drift in machine performance. Considering that all circuits for all experiments are randomly interleaved in time, while all shots for a single circuit are taken in rapid succession, it is reasonable to model drifts in machine performance by assigning each circuit a random but fixed value of $\varepsilon$, drawn from a distribution around its experimentally measured average. For $F_{\rm XEB}$, we model a single circuit as sampling from a model probability distribution that is a convex mixture of the ideal Porter-Thomas distribution (with probability $F(\varepsilon)$) and the uniform distribution (with probability $1-F(\varepsilon)$).  For $F_{\rm MB}$ we model each circuit as a Bernoulli trial (biased coin flip) with a probability $F(\varepsilon)$ of returning the correct bit string and probability $1-F(\varepsilon)$ of a wrong bitstring. In both cases the circuit fidelity $F$ depends on the per 2Q gate process infidelity $\varepsilon$, which we take to be a random variable that is Gaussian-distributed about its experimental mean value $\varepsilon_{\rm exp}\approx3.2\times10^{-3}$ with standard deviation $\mu\times \varepsilon_{\rm exp}$. Diagnostic and calibration data from the time period during which the data reported in \sref{sec:exp_results} was taken suggests a conservative upper bound $\mu\lesssim 0.2$. The set of all probability distributions over the collection of outcomes obtainable in this way is referred to as the \textit{experimental model}. 

\begin{figure}[!t]
\centering
\includegraphics[width=1.0\columnwidth]{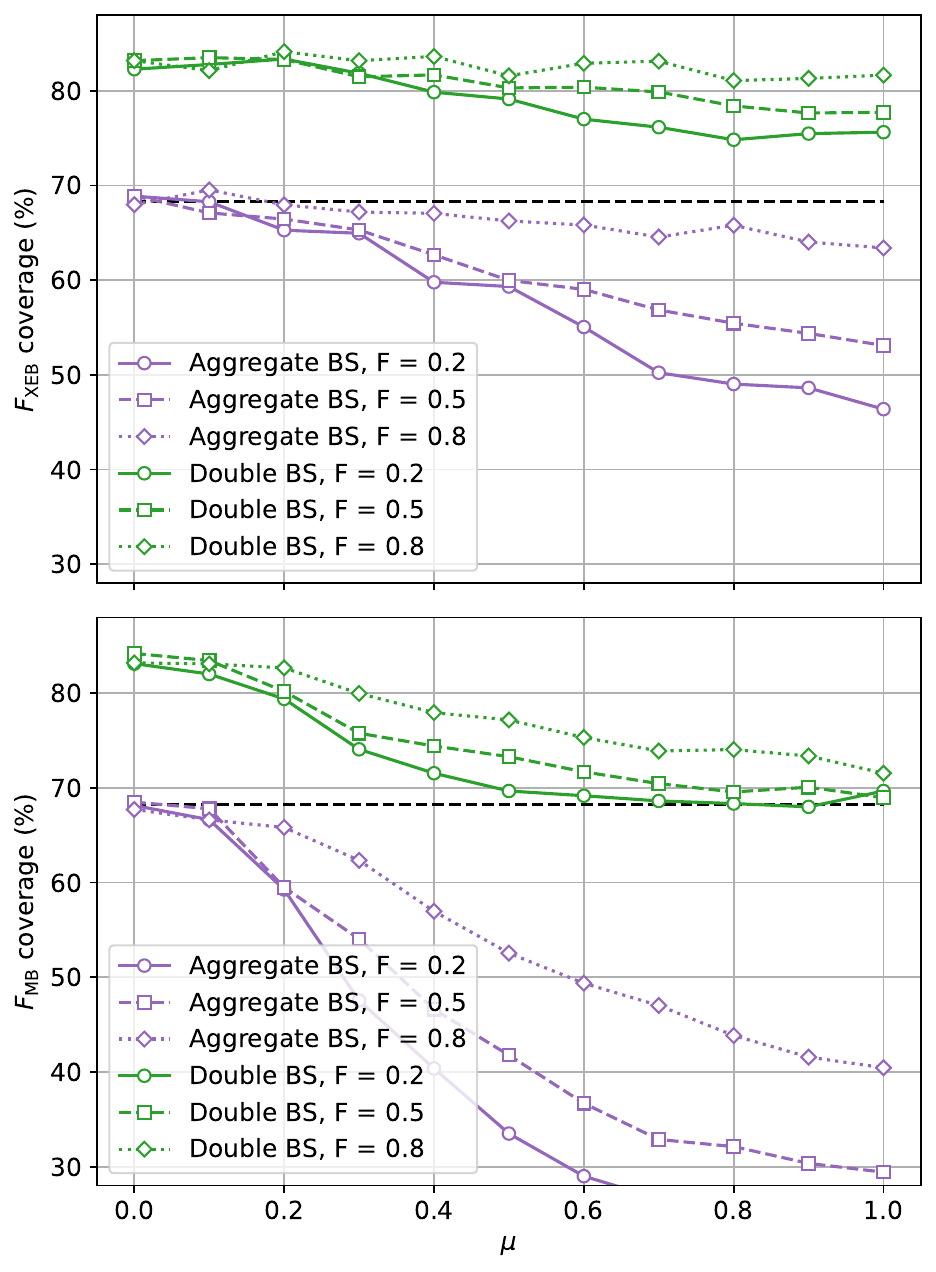}
\caption{Coverage analysis for aggregate bootstrapping and double-bootstrapping applied to simulated experiments for both $F_{\rm xEB}$ (upper) and $F_{\rm MB}$ (lower).  Double bootstrapping generally over-covers the data, most severely in the small $\mu$ limit where circuit-to-circuit fluctuations in the underlying distribution are small. Aggregate resampling generally provides an appropriate confidence interval for $\mu<0.2$ for $F_{\rm XEB}$ (and $\mu<0.1$ for $F_{\rm MB}$), but can under-cover the data in either case at large $\mu$ (especially for small values of the circuit fidelity $F$.)
\label{fig:coverage}}
\end{figure}

We use this experimental model to perform a coverage analysis by simulating $N_{E}=4000$ experiments, each with $50$ circuits and $20$ shots per circuit.  For each data set we perform both aggregate bootstrapping and double bootstrapping using $r=4000$ resamples. We then compute bootstrapped confidence intervals for each of the $N_E$ simulated experiments from the synthetic bootstrap data and the empirical mean from that experiment, and compute the fraction of experiments for which the true value (estimated by the mean across all experiments) is contained within that confidence interval, referring to that fraction as the \emph{coverage}.  A proper $1\sigma$ confidence interval must provide coverage $\ge68.27\%$ for all distributions in the experimental model, and this level is denoted by the black dashed line in \fref{fig:coverage}.  In general, for the expected $\lesssim 20\%$ fluctuations in $\varepsilon$ ($\mu\lesssim0.2$) we see that aggregate resampling provides a confidence interval for $F_{\rm XEB}$ with the required coverage, while double resampling provides excessive coverage (resulting in an overly conservative confidence interval). For $F_{\rm MB}$ the coverage of aggregate resampling falls off faster, requiring $\mu\lesssim 0.1$ to be adequate.
In either case, for extremely large fluctuations in $\varepsilon$ ($\mu=1$) we see that aggregate resampling fails to provide adequate coverage (and thus underestimates the size of the correct $1\sigma$ confidence interval), while the coverage of double resampling remains adequate.  Since we expect the actual circuit fluctuations in $\varepsilon$ to be $\lesssim 20\%$, we believe that aggregate resampling is likely more appropriate for assigning confidence intervals to our $F_{\rm XEB}$ data, though it may slightly underestimate the appropriate confidence intervals for $F_{\rm MB}$. Given that the confidence intervals are larger for $F_{\rm XEB}$, and in the interest of consistency, we opted to use the aggregate resampling method to report \emph{all} $1\sigma$ error bars on both $F_{\rm MB}$ and $F_{\rm XEB}$ in the manuscript.

\begin{figure}[!t]
\centering
\includegraphics[width=1.1\columnwidth]{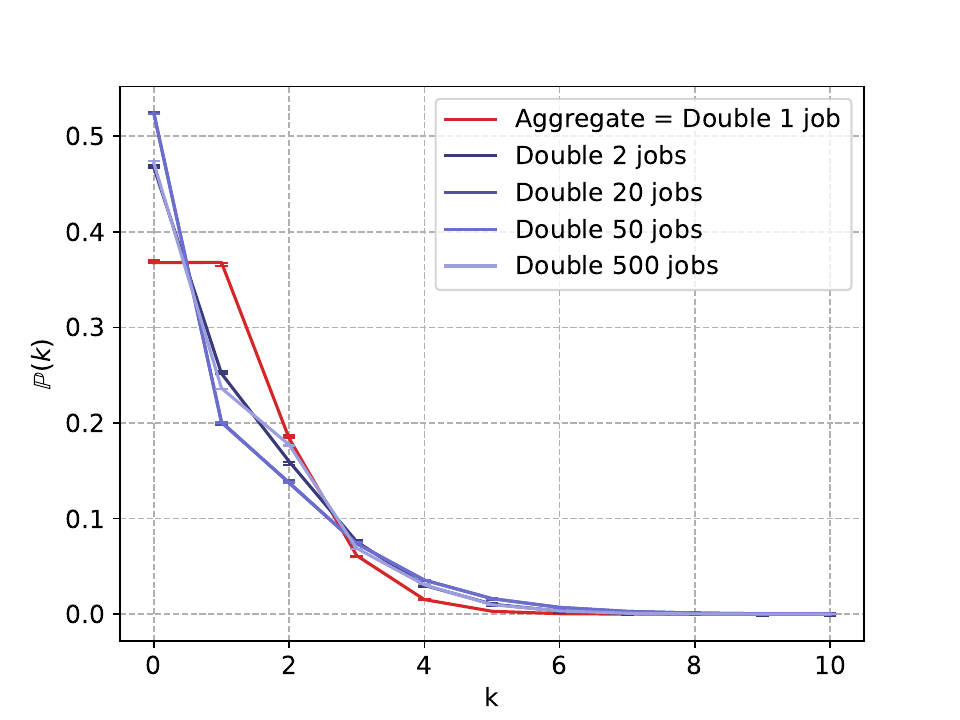}
\caption{
Probability distributions $\mathbb{P}(k)$ of resampling a single labeled shot $k$ times, depending on the aggregate or double resampling methods with $1000$ shots binned into a variable number $N_j$ of jobs and $1000/N_j$ shots per each job. In all cases except the $N_j=1$ and $N_j=1000$ limits where the aggregate and double resampling methods are equivalent, the double resampling distributions contain significantly higher weight at $k = 0 $ and $k \geq 3$. Numerically modeled points with error bars are overlaid on the solid line representing the analytic expressions in Eqs.~\eqref{eq:aggregate}, \eqref{eq:double}.
\label{fig:resampling_probs}}
\end{figure}

The larger variance estimate provided by the double resampling method can be explained analytically as follows. Consider the probability of any particular individual shot appearing $k$ times when $N_s$ shots are resampled. In the aggregate resampling method, this probability is binomially distributed:
\begin{align}
\mathbb{P}_{\text{agg}}(k) = \left(\frac{1}{N_s}\right)^{k}\left(1-\frac{1}{N_s}\right)^{N_s- k} {N_s \choose k}. \label{eq:aggregate}
\end{align}
In the double resampling method, the probability that a particular labeled shot appears $k$ times depends on how many times its corresponding job is resampled. If there are $N_j$ total jobs, with $N_{\text{per}}$ shots per job and $N_s = N_j N_{\text{per}}$,
\begin{align}
    \mathbb{P}_{\text{doub}}(k) = \sum_{j=0}^{N_j} \left(\frac{1}{N_j}\right)^{j}\left(1-\frac{1}{N_j}\right)^{N_j- j} {N_j \choose j}\: \mathbb{P}(k \, |\, j ), \label{eq:double}
\end{align}
where $\mathbb{P}(k \, |\, j )$ is the probability that a given labeled shot appears $k$ times given that its corresponding job is resampled $j$ times. To compute $\mathbb{P}(k \, |\, j )$, one must consider all of the ways that the $k$ appearances can be spread over the $j$ job resamples. Note that the $k$ appearances may be restricted to only a subset of the $j$ resamples of size $n_b$ ``bins", as long as $k \leq n_b N_{\text{per}}$. Therefore, we compute $\mathbb{P}(k \, |\, j )$ by summing over the set of partitions of $k$ of fixed size $n_b$, $P_{k,n_b} = \{(p_1, \ldots, p_{n_b}) \in (\mathbb{Z}^{+})^{n_b}\,|\, \sum_i p_i = k\}$ as follows
\begin{align}
\mathbb{P}(k \, |\, j ) &= \sum_{n_b = 0}^j  \biggl[{j \choose n_b} \left(1 - \frac{1}{N_{\text{per}}} \right)^{(j-n_b)N_{\text{per}}} \nonumber \\&\qquad\qquad \qquad \qquad\times \sum_{P_{k,n_b}} \mathbb{P}(k\,|\,P_{k,n_b}) \biggr].
\end{align}
The first term in the exterior sum counts the number of ways of choosing the $n_b$ bins from the $j$ job resamples. The second term is the standard binomial term accounting for the remaining $j-n_b$ jobs, each with $N_{\text{per}}$ shots, where the labeled shot does not appear. The interior sum is over $\mathbb{P}(k\,|\,P_{k,n_b})$, the probability of the $k$ appearances being distributed amongst bins according to each possible partition $P_{k,n_b}$. Note that the set of all such partitions includes ``equivalent" partitions where the same set of $p_i$ are permuted over the different bins $n_b$, which are counted separately in the sum. This final conditional probability is simply the product of binomial distributions over the $n_b$ bins:
\begin{align}
\mathbb{P}(k\,|\,P_{k,n_b}) = \prod_{i=1}^{n_b} \left(\frac{1}{N_{\text{per}}} \right)^{p_i}\left(1-\frac{1}{N_{\text{per}}} \right)^{N_{\text{per}} - p_i} {N_\text{per} \choose p_i}.
\end{align}
The distributions $\mathbb{P}_{\text{agg}}(k)$ and $\mathbb{P}_{\text{doub}}(k)$ are depicted in Fig.~\ref{fig:resampling_probs} along with data averaged across $100$ instances of numerical resampling experiments in which $1000$ resamples were drawn from $1000$ shots by either method and different job sizes. For all job sizes except $N_j=1$ and $N_j=1000$ where the double resampling method reduces to aggregate resampling, $\mathbb{P}_{\text{doub}}(k)$ contains significantly higher weight at $k = 0$ and $k \geq 3$. One interpretation of this result is that in a given resampling experiment using the double resampling method, the data tend to ``clump", favoring multiple samples of the same shots and excluding other shots entirely, owing to the fact that each resampling experiment has a chance to exclude some jobs entirely. The higher variance estimate provided by the double resampling method can consequently can be viewed as an instance of Sheppard's correction \cite{sheppard} in which binned data tend to exhibit slightly higher variance estimates; a tendency to replace individual shots by multiple occurrences of the same shot is similar to replacing all data in a bin with the midpoint of the bin.

\section{Construction of circuits with random geometries\vspace{0.5em}\label{sec:rand_circs}}

Considering each qubit as a node in a graph, a set of edges $\mathcal{E}$ connecting those nodes is said to be mutually disjoint if no two edges in $\mathcal{E}$ contain the same node. If we associate a 2Q gate to each edge in $\mathcal{E}$, those gates have mutually non-overlapping support and can all be executed (in principle) in parallel.  In a slight abuse of notation we use $\mathcal{E}$ to also denote the set of 2Q gates associated to the edges in $\mathcal{E}$, and we refer to $\mathcal{E}$ as a single 2Q-layer of a circuit.  Each gate in $\mathcal{E}$ will be the H2 quantum computer's native perfect entangler $U_{\rm ZZ}(\pi/2)=e^{-i(\pi/4)Z\otimes Z}$ (see Appendix~\ref{sec:gate_choice} for further discussion regarding this gate choice). Similarly, let $\mathcal{S}$ denote a set of 1Q gates assigned to each qubit; we refer to $\mathcal{S}$ as a single 1Q-layer of a circuit. Each 1Q gate $S\in \mathcal{S}$ is chosen to be a Haar-random SU$(2)$ gate, which can be decomposed as $S=R_{z}(\psi){\rm U1q}(\theta,\phi)$.  Here
\begin{align}
{\rm U1q}(\theta,\phi)=e^{-i(\theta/2)(\hat{X}\cos\phi+\hat{Y}\sin\phi)}
\end{align}
is our native 1Q gate, and $R_{z}(\psi)$ is a $z$-rotation that is performed in software (i.e.\ by pushing the $z$-rotation through future 1Q gates and modifying their phases accordingly).  A random depth-$d$ circuit is then defined by interleaving such layers as

\begin{align}
C= \mathcal{S}_d\mathcal{E}_d\mathcal{S}_{d-1}\mathcal{E}_{d-1}\dots\mathcal{S}_1\mathcal{E}_1\mathcal{S}_0.
\end{align}

A powerful feature of QCCD architecture is the flexibility to choose which pairs of qubits are gated in a given 2Q layer: Regardless of the physical locations of the qubits after the previous layer, any set $\mathcal{E}$ can contain arbitrary pairs and the associated gates will be executed directly (e.g. without logical SWAP gates) by physically rearranging the ions such that the qubits associated to each 2Q gate in $\mathcal{E}$ are located next to each other.  In this work we consider the ensemble of random circuits in which the sets $\mathcal{E}_1,\dots,\mathcal{E}_d$ are obtained via the following procedure:
\begin{enumerate}
\item{Select a random $d$-regular graph $\mathcal{G}$ on $N$ vertices.}
\item{Find a proper edge coloring of $\mathcal{G}$ with $d$ colors.}
\item{Assign a unique index $j=1,\dots,d$ to each color in the edge coloring, and put each edge with color $j$ into the set $\mathcal{E}_j$.}
\end{enumerate}
A proper edge coloring is defined as an assignment of a color to each edge in a graph such that no two edges of the same color impinge on the same node, and hence each $\mathcal{E}_j$ is a proper (mutually disjoint) layer of 2Q gates (see \fref{fig:circ_tn}(a) for an example of a proper edge coloring of a 3-regular graph on 6 nodes using 3 colors). By Vizing's theorem, $\mathcal{G}$ is guaranteed to have a proper edge coloring using a minimum number of colors equal to either $d$ (in which case $\mathcal{G}$ is referred to as class I) or $d+1$ (in which case $\mathcal{G}$ is referred to as class II). Note that in step 2 of the above procedure we implicitly assume that the sampled graph is class I.  While this assumption is not guaranteed to be true, it is known that random regular graphs of degree $d\geq 3$ are asymptotically almost always (i.e. with probability approaching 1 for $N\rightarrow\infty$) class I \cite{Janson1995,wormald1997}, and all $d$-regular graphs with even $N$ and $d\geq 2\lceil N/4 \rceil-1$ are provably class I \cite{2014arXiv1401.4159C}.  Empirically, we are able to find proper $d$-colorings of all but a very small subset of randomly sampled graphs in the ranges of $(N,d)$ that we have considered.  In the rare event that no proper $d$-coloring is found in step 2, we reject the graph and return to step 1.

\section{Bounds on contraction difficulty of random-geometry circuits\vspace{0.5em}\label{sup:bounds}}

\begin{figure*}[!t]
\centering
\includegraphics[width=2.05\columnwidth]{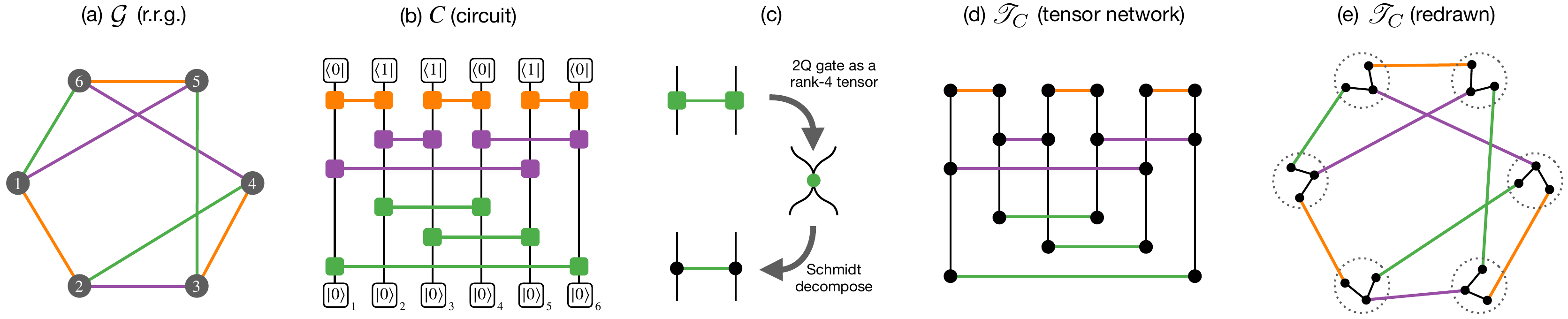}
\caption{(a) To construct a depth-$3$ circuit on $6$ qubits, we first select a random 3-regular graph on $n$ vertices, denoted $\mathcal{G}$, and assign to it a proper edge coloring using $3$ colors.  (b) $\mathcal{G}$ induces the circuit $C$ shown by associating each vertex (edge) in $\mathcal{G}$ to a qubit (2Q gate) in $C$; each 2Q-gate layer of $C$ is in correspondence with a given color in the coloring of $\mathcal{G}$. Qubits are assumed (without loss of generality since random 1Q gates are applied immediately after initialization) to be initialized in the $\ket{0}$ state, and layers of $1Q$ gates in $C$ have been absorbed into the 2Q gates for clarity.  The probability that the circuit outputs a given product state (here the state $\ket{011010}$) can be represented as a tensor network as shown in (c) by associating each 2Q gate to a rank-4 tensor. (d) Upon Schmidt-decomposing those gates and absorbing the rank-1 tensors corresponding to input and output states into adjacent 2Q gate tensors, we obtain the tensor network $\mathscr{T}_C$. (e) An equivalent redrawing of $\mathscr{T}_C$ reveals its similarity to the underlying graph $\mathcal{G}$ from which the circuit was generated.
\label{fig:circ_tn}}
\end{figure*}

To any unitary quantum circuit $C$ one can associate a tensor network $\mathscr{T}_C$ in which gates are complex-valued tensors and wires (qubits) are the indices/legs of those tensors. In this representation, 1Q  gates become rank-2 tensors (one incoming and one outgoing qubit wire) and 2Q gates become rank-4 tensors (two incoming and two outgoing qubit wires).  The initial state $\ket{\psi_j}$ of each input qubit $j$ is a rank-1 tensor, and the output of the tensor network (uncontracted legs) represents the quantum state at the output of the circuit, $\ket{\Psi}=C\bigotimes_j\ket{\psi_j}$. The overlap between $\ket{\Psi}$ and a particular tensor-product of single qubit states $\bigotimes\ket{\varphi_j}$ is then obtained by contracting each output leg of the tensor network with a rank-1 tensor corresponding to the 1Q state $\ket{\varphi_j}$ associated to that qubit.  As an example, in \fref{fig:circ_tn}(b) we show the depth $d=3$ quantum circuit $C$ induced by the $3$-regular graph $\mathcal{G}$ shown in \fref{fig:circ_tn}(a), as described in Appendix~\ref{sec:rand_circs}. The tensor network associated to this circuit, denoted $\mathscr{T}_C$, is shown in \fref{fig:circ_tn}(d). Note that we have decomposed each rank-4 gate tensor into a tensor product of two rank-3 tensors as shown in \fref{fig:circ_tn}(c).  For the $U_{ZZ}(\pi/2)$ gates used here, this decomposition has Schmidt-rank two. Therefore, both the colored lines (originating from gates) and the black lines (originating from qubit wires) have dimension $2$, and we need not distinguish between them when determining the time or memory cost of tensor contraction.  We have also absorbed the rank-1 input and output tensors into the nearest gate tensor.   Figure \ref{fig:circ_tn}(e) shows how the tensor network $\mathscr{T}_C$ is structurally similar to the graph $\mathcal{G}$ from which the circuit $C$ was generated.

\subsection{Lower bound}
To arrive at the TN $\mathscr{T}_C$ in the form shown in \fref{fig:circ_tn}(e) we have already decomposed each rank-4 2Q gate tensor into two rank-3 tensors. However, we could also consider the contraction cost of the original circuit TN ($\mathscr{T}_C^{0}$, not shown in \fref{fig:circ_tn}). The TN $\mathscr{T}_C$ cannot be harder to contract than $\mathscr{T}_C^{0}$, because one viable contraction path for $\mathscr{T}_C$ consists of first grouping each gate back to a single rank-4 tensor, thereby achieving $\mathscr{T}_C^0$ as an intermediate tensor (and this process occurs at much lower cost than future operations in contracting $\mathscr{T}_C^0$). Therefore, the contraction cost of $\mathscr{T}_C$ lower bounds the contraction cost of the original circuit TN $\mathscr{T}_C^{0}$, and we only need to lower bound the cost of contracting $\mathscr{T}_C$ in the form shown in \fref{fig:circ_tn}(e).

Below we will prove that for $d\geq 3$ and with probability (over the distribution of circuits considered) approaching 1 as $N\rightarrow\infty$, the contraction cost for our circuits obeys
\begin{align}
\mathscr{C}_d=\lim_{N\rightarrow \infty}\frac{\log_2({\rm cost})}{N}\geq c
\label{eq:lower_bound}
\end{align}
for some constant $0<c<1$. This result should be compared to the observation that $\mathscr{C}_d=0$ in \textit{any} local and finite-dimensional geometry.  The underlying methodology is related to the work of Markov and Shi (see Theorem 1.7 in Ref.~\cite{markov2008simulating}), however the proof is more direct and requires very little graph-theoretic machinery; ultimately, it is based on the recognition that for circuits with gates drawn from graphs with good expansion properties, TN contraction requires forming intermediate tensors that either cut an extensive number of qubit wires, gates, or both.

\begin{figure}[!t]
\centering
\includegraphics[width=0.732\columnwidth]{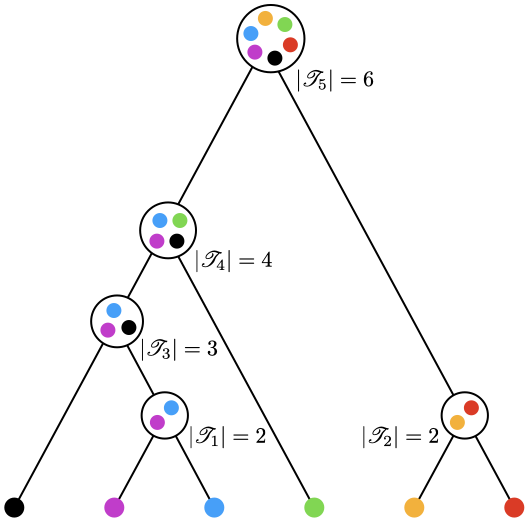}
\caption{Representation of the contraction of a tensor network $\mathscr{T}$ into a scalar as a rooted binary tree flowing from bottom (leaves) to top (root) (here we assume that $\mathscr{T}$ has no open legs so that it can be contracted to a scalar, as in the case of a TN representing the probability amplitude of a given state at the output of a quantum circuit). Here $\mathscr{T}$ is chosen to have 6 tensors, which are represented by the leaves of the tree (each is given a unique color), while every other vertex represents a merged tensor formed by contracting any legs shared by its two children. The root represents the final (scalar) value of the contraction.
\label{fig:contraction_tree}}
\end{figure}

$\mathscr{T}_C$ is composed of $N d$ original tensors, and its contraction into a scalar (a single rank-0 tensor whose value is the amplitude $\bra{\Phi}C\ket{\Psi}$ for a particular output state $\ket{\Phi}$) proceeds as a sequence of individual contractions that can be represented as a rooted binary tree (see \fref{fig:contraction_tree} for an example), with the original tensors as leaves and the final scalar as the root. Each vertex in the tree corresponds to a merged tensor during the contraction process, and its children (the two tensors merged to make it) are connected to it by edges. Each vertex can be labeled by its descendants, e.g. which of the leaves (original tensors) have been absorbed into the associated tensor, and those leaves are shown explicitly inside the vertices in \fref{fig:contraction_tree}. The tensor obtained by the $j^{\rm th}$ contraction is denoted $\mathscr{T}_j$, and we define its size $|\mathscr{T}_j|$ as its number of descendants (i.e. the number of original tensors that merged to form it). We refer to the number of legs of the tensor $\mathscr{T}_j$ as its \emph{rank} $r_j$. Because all edges of $\mathscr{T}_C$ have dimension 2 (due to our gates being rank-2 entanglers), the contraction cost (measured in FLOPs) associated to a particular contraction tree is lower-bounded as
\begin{align}
{\rm cost}\geq \max_j2^{r_j}.
\end{align}
The cost of contracting $\mathscr{T}_C$ is obtained by minimizing the contraction cost over all possible contraction trees.  Note that there must be a smallest value of $j$ such that $|\mathscr{T}_{j}|\geq Nd/3$ (i.e., as one proceeds with the contraction, there must be a first time when two merged tensors result in a final tensor of size at least $Nd/3$), and we call this intermediate tensor $\mathscr{T}_{\rm min}$. Since every other tensor has (by assumption) size less than $Nd/3$ when the $j^{\rm th}$ contraction happens, we also know that $|\mathscr{T}_{\rm min}|\leq 2Nd/3$. Our goal here is to ensure that both the set of original tensors contained in $\mathscr{T}_{\rm min}$ and the set of all other tensors are \emph{both} extensive, which will ultimately be used to ensure that an associated subset of the vertices of $\mathcal{G}$ has a large edge boundary (which requires that the subset and its complement are both large).  We could have used any number of tensors $k\times Nd$ with $0<k<1/2$ (rather than our choice $k=1/3$) when defining $\mathscr{T}_{\rm min}$ ($1/2$ is problematic since then $\mathscr{T}_{\rm min}$ could contain almost every original tensor).

\begin{figure}[!t]
\centering
\includegraphics[width=0.7\columnwidth]{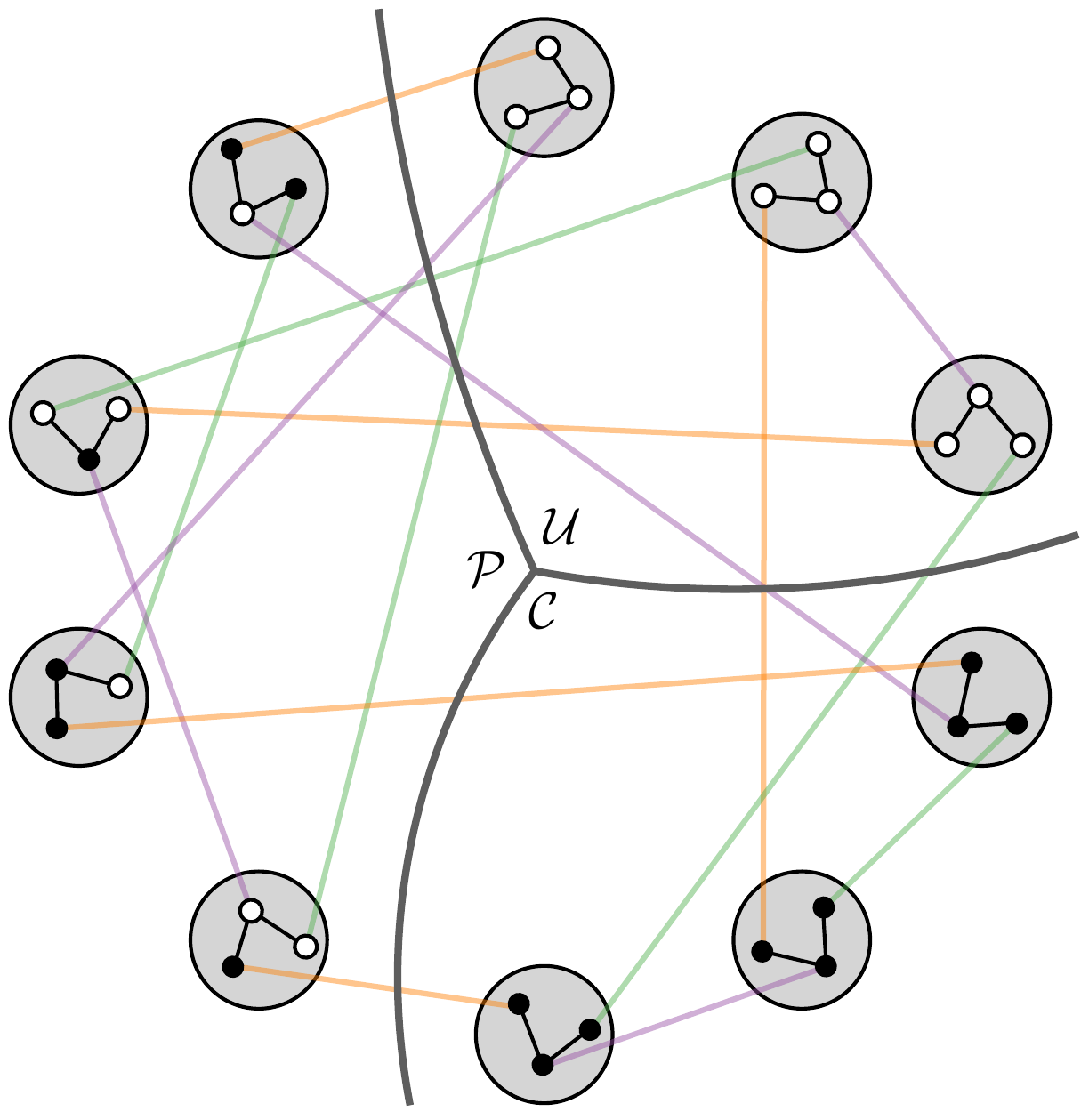}
\caption{Example tensor network for a depth-3 random circuit on 10 qubits, with a possible intermediate tensor in the contraction process denoted as the collection of black tensors (e.g. they will all be contracted together into a single tensor during the contraction process).  There is always an intermediate tensor built by contracting at least $1/3$ and at most $2/3$ of the original tensors, and in this example we are depicting an intermediate tensor built by contracting exactly half $(Nd/2=15)$ the original tensors. Each grey disk, which we refer to as a bag, is associated with a particular qubit and contains all tensors having legs originating from that qubit's wire. Those bags can be subdivided into sets $\mathcal{C}$, $\mathcal{U}$, and $\mathcal{P}$, according to whether they are fully contracted, completely uncontracted, or partially contracted.
\label{fig:contraction_bound}}
\end{figure}

For simplicity of illustration let's assume that $|\mathscr{T}_{\rm min}|=(Nd/2)$, i.e. $\mathscr{T}_{\rm min}$ is composed of exactly half of the primitive tensors. Figure \ref{fig:contraction_bound} shows an example of $\mathscr{T}_C$ for a 10-qubit depth-3 circuit in which we have labeled the $Nd/2$ original tensors forming $\mathscr{T}_{\rm min}$ as filled circles ($\sbullet[1.0]$), and all of the original tensors not merged into $\mathscr{T}_{\rm min}$ as open circles ($\circ$).  To each qubit in the circuit we can associate a \emph{bag} of tensors belonging to the gates that act on that qubit; in the depth-3 example of \fref{fig:contraction_bound} each bag (grey disk) contains exactly 3 tensors.  The set of all bags can be decomposed as $\mathcal{C}\cup\mathcal{U}\cup\mathcal{P}$, such that $\mathcal{C}$ contains the bags for which all constituent tensors have been contracted into $\mathscr{T}_{\rm min}$, $\mathcal{U}$ contains the bags for which none of the constituent tensors have been contracted into $\mathscr{T}_{\rm min}$, and $\mathcal{P}$ contains the bags for which some but not all constituent tensors have been contracted into $\mathscr{T}_{\rm min}$.
 
It is helpful to consider the strategy of statevector simulation, in which all gates of a given color are absorbed sequentially, causing all bags to be in $\mathcal{P}$.  More generally, each bag of type $\mathcal{P}$ is associated with at least one qubit-wire protruding from the intermediate tensor $\mathscr{T}_{\rm min}$, and choosing the set $\mathcal{P}$ to be large corresponds to a predominantly space-like slicing of the network (e.g. created by slicing in a way that divides tensors at later times from those at earlier times), suffering a cost that is exponential in the number of qubits.  A productive contraction strategy for sufficiently shallow spatially local lattices is to avoid bags of type $\mathcal{P}$ entirely, corresponding to making time-like cuts through the tensor network such that only gate wires protrude from $\mathscr{T}_{\rm min}$.  Judiciously choosing the set $\mathcal{C}$ (to have a small edge boundary) can greatly reduce the contraction cost relative to the statevector strategy. The fundamental reason why \eref{eq:lower_bound} holds for RG circuits is that \emph{neither} strategy (nor any other strategy, as we show below) succeeds in producing an intermediate tensor $\mathscr{T}_{\rm  min}$ with a number of legs that is sublinear in $N$. The time-like strategy can be seen to fail by considering the extreme limit in which $\mathcal{P}$ is empty: The number of gate wires exiting $\mathscr{T}_{\rm min}$ is then determined by the number of edges connecting bags in $\mathcal{U}$ to bags in $\mathcal{C}$, and this number is guaranteed to scale linearly with the size of $\mathcal{C}$ whenever the underlying graph $\mathcal{G}$ is a good expander, as is the case for random regular graphs.
The above reasoning is somewhat heuristic, but it can readily be turned into a proof of \eref{eq:lower_bound}. Specifically, we need to show that with probability (over the considered class of graphs considered) approaching 1 for large $N$, no contraction strategy can yield $\mathscr{T}_{\rm min}$ with rank $r_{\rm  min}< cN$ for some constant $c$.

If we consider $\mathscr{T}_C$ as a graph (tensors $\rightarrow$ nodes and legs $\rightarrow$ edges) and denote the set of black nodes in \fref{fig:contraction_bound} as $V$, the rank of $\mathscr{T}_{\rm min}$ is the size of the edge-boundary of $V$, denoted by $|\partial V|$, and \eref{eq:lower_bound} will be true as long as $|\partial V|\geq c N$. The edge boundary $\partial V$ contains all colored edges that originate from $\mathcal{C}$ and  terminate on vertices in $\mathcal{U}$ or unfilled vertices in $\mathcal{P}$. The number of such edges is lower-bounded by ${\rm max}(0,|\partial\mathcal{C}|-d|\mathcal{P}|)$. Since every bag in $\mathcal{P}$ also contributes at least one black edge to $\partial V$ (not already counted in the colored edges considered above) we conclude that
\begin{align}
|\partial V|\geq |\mathcal{P}|+{\rm max}(0,|\partial\mathcal{C}|-d|\mathcal{P}|).
\label{eq:boundary_bound}
\end{align}
The first and second terms in \eref{eq:boundary_bound} capture the costs of attempting to contract the tensor network in a time-like (Schr\"odinger) or space-like direction, respectively. The edge boundary of $\mathcal{C}$ is constrained relative to its size by the edge isoperimetric number of $\mathcal{G}$, defined as
\begin{align}
I(\mathcal{G})=\min_U|\partial U|/|U|,
\label{eq:isoperimetric}
\end{align}
with $U$ ranging over all subsets of nodes in $\mathcal{G}$ containing at most half of the vertices in $\mathcal{G}$.  Either $\mathcal{C}$ or its complement satisfies the constraint of having at most half of the vertices in $\mathcal{G}$. The complement of $\mathcal{C}$ is lower-bounded in size by $N/3$ (because $|V|<2Nd/3$) and therefore
\begin{align}
\label{eq:C_boundary}
|\partial C|\geq I(\mathcal{G})\min(|\mathcal{C}|,N/3).
\end{align}

The notion that random regular graphs are (with high probability) good expanders can be formalized in the following theorem by Bollob\'as \cite{BOLLOBAS1988241} regarding the isoperimetric number of $\mathcal{G}$:

\

\noindent\textbf{Good edge expansion of random d-regular graphs}: \textit{Consider the distribution of $d$-regular graphs on $N$ nodes sampled uniformly at random, and define $\eta(d)=2\sqrt{(\log 2)/d}$.  For all $d\geq 3$, and with probability approaching $1$ as $N\rightarrow\infty$, the isoperimetric number $I(\mathcal{G})$ of a graph $\mathcal{G}$ sampled from this distribution is bounded below as:}
\begin{align}
\label{eq:iso_bound}
I(\mathcal{G})\geq \frac{d}{2}\big(1-\eta(d)\big).
\end{align}

\

\noindent Combining \eref{eq:iso_bound} with Eqs.\ (\ref{eq:boundary_bound}, \ref{eq:C_boundary}) yields the lower bound
\begin{align}
\label{eq:delV}
|\partial V|&\geq |\mathcal{P}|+d\!\times\! {\rm max}\bigg(0,\frac{\min(|\mathcal{C}|,N/3)[1-\eta(d)]}{2}-|\mathcal{P}|\bigg).
\end{align}
Recall that by the definition of $\mathscr{T}_{\rm min}$, the size of $V$ is constrained to satisfy
\begin{align}
\frac{dN}{3}\leq |V|\leq \frac{2dN}{3}.
\end{align}
Since each bag in $\mathcal{C}$ contributes exactly $d$ vertices to $V$ while each bag in $\mathcal{P}$ contributes at most $d-1$, we conclude that
\begin{align}
|\mathcal{C}|\geq\frac{N}{3}-|\mathcal{P}|,
\end{align}
from which we can further bound $|\partial V|$ as
\begin{align}
|\partial V|&\geq |\mathcal{P}|+d\times  {\rm max}\bigg(0,\frac{[N/3-|\mathcal{P}|][1-\eta(d)]}{2}-|\mathcal{P}|\bigg).
\end{align}
Upon varying $|\mathcal{P}|$, the r.h.s.\ is minimized (for $d\geq 3$) when the second argument of the ${\rm max()}$ function vanishes, ultimately giving
\begin{align}
|\partial V|&\geq \frac{N[1-\eta(d)]}{9}.
\end{align}
Note that we have not attempted to make the bound even reasonably tight, and we expect it is possible to derive a bound in which the r.h.s.\ tends to $N$ as $d\rightarrow\infty$.

\subsection{Upper bound} 
\begin{figure}[!t]
\centering
\includegraphics[width=1.0\columnwidth]{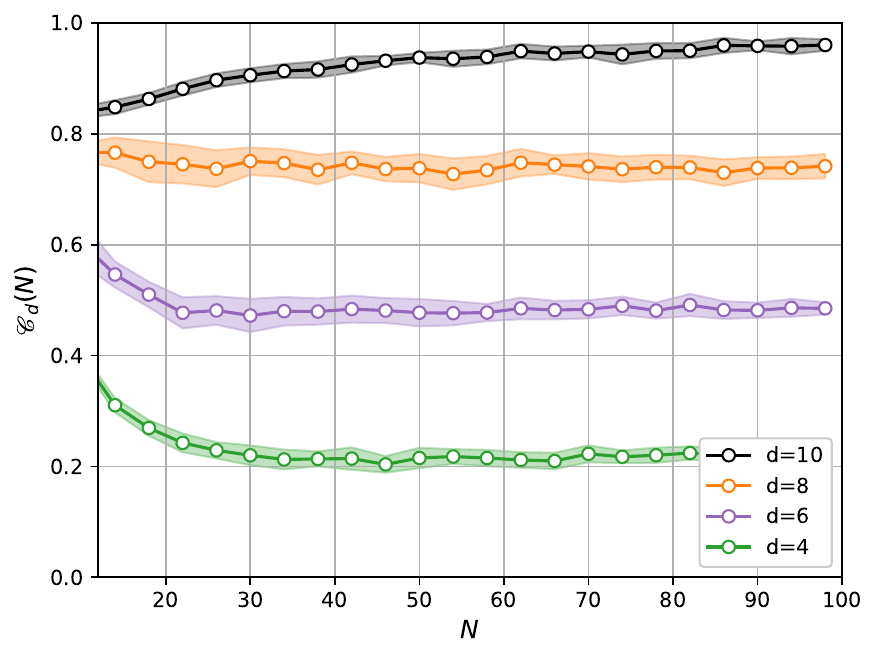}
\caption{Complexity density as a function of $N$ for circuit depths $d=4,6,8,10$, showing saturation to a constant as $N\rightarrow\infty$.  Each point is the mean complexity density over 20 random circuits (shaded regions represent one standard-deviation ).
\label{fig:fix_d_scan_n}}
\end{figure}
The previous section demonstrates that the asymptotic contraction complexity density $\mathscr{C}_d$ is finite, but does not address how the density approaches the worst-case upper bound from statevector simulation ($\mathscr{C}_d\rightarrow 1$) as the depth increases. Fig.~\ref{fig:fix_d_scan_n} shows numerical evidence that the complexity density saturates, and also that the saturated value quickly approaches 1 as $d$ increases. It is natural to ask precisely \emph{how} $\mathscr{C}_d\rightarrow 1$ as $d\rightarrow\infty$: Does $1-\mathscr{C}_d$ vanish smoothly as $d\rightarrow\infty$, and if so does it decay polynomially or exponentially? Or does it vanish abruptly at some critical value $d=d_c$?

Here we show that $1-\mathscr{C}_d$ cannot vanish abruptly by proving that individual circuit instances obey the upper bound
\begin{align}
\label{eq:upper_bound}
\mathscr{C}_d\leq 1-1/2^d.
\end{align}
This bound leaves open the possibility that $1-\mathscr{C}_d$ vanishes exponentially as $d\rightarrow\infty$; we find this scenario likely but have no proof of it. The proof of \eref{eq:upper_bound} is by analysis of an explicit contraction ordering inspired by the qubit reuse algorithm described in Ref. \cite{decross2022qubit}.

\begin{figure}[!t]
\centering
\includegraphics[width=0.8\columnwidth]{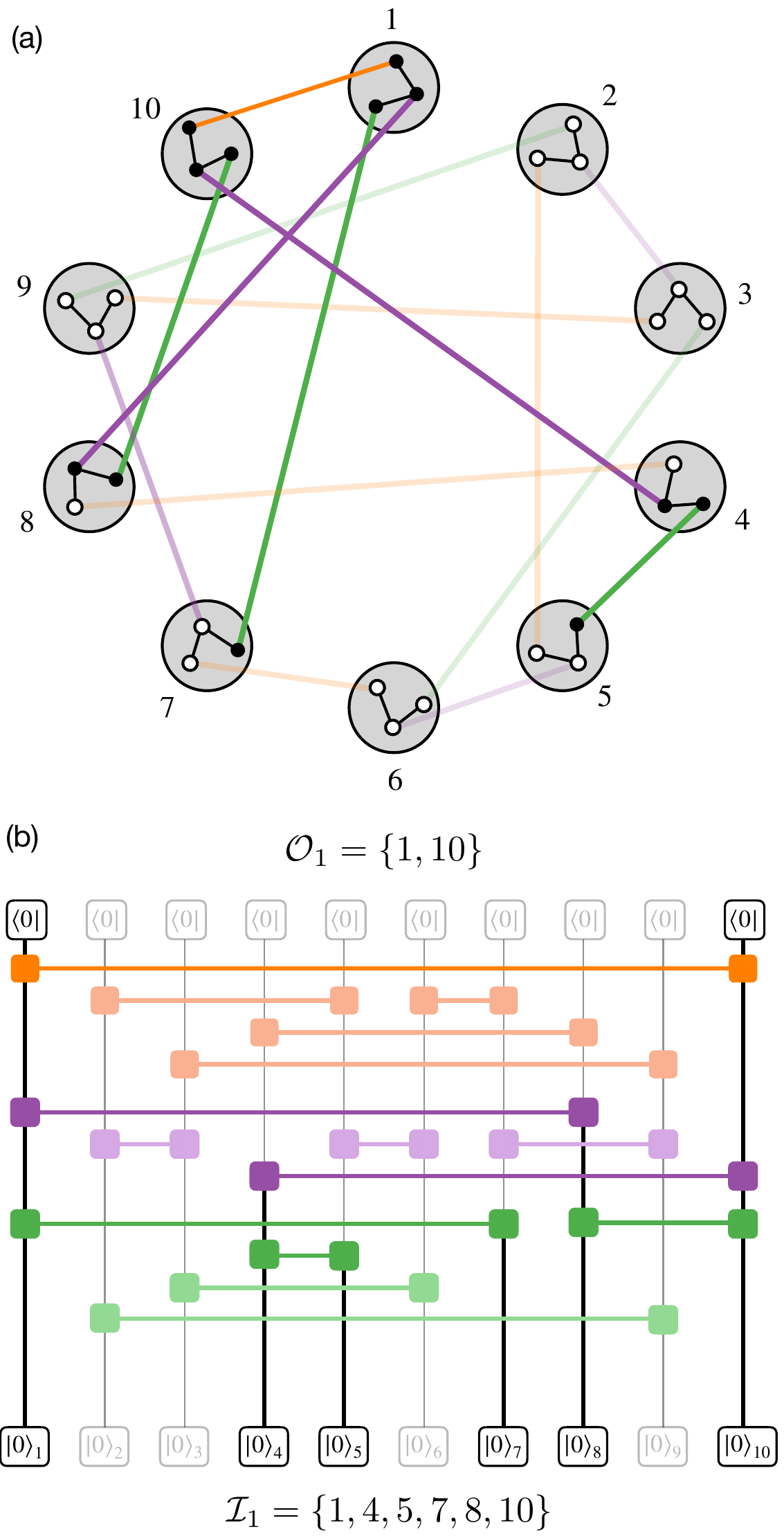}
\caption{Light-cone based contraction ordering yielding the upper bound in \eref{eq:upper_bound}.
\label{fig:light_cone}}
\end{figure}

\begin{figure*}[!t]
\centering
\includegraphics[width=2.0\columnwidth]{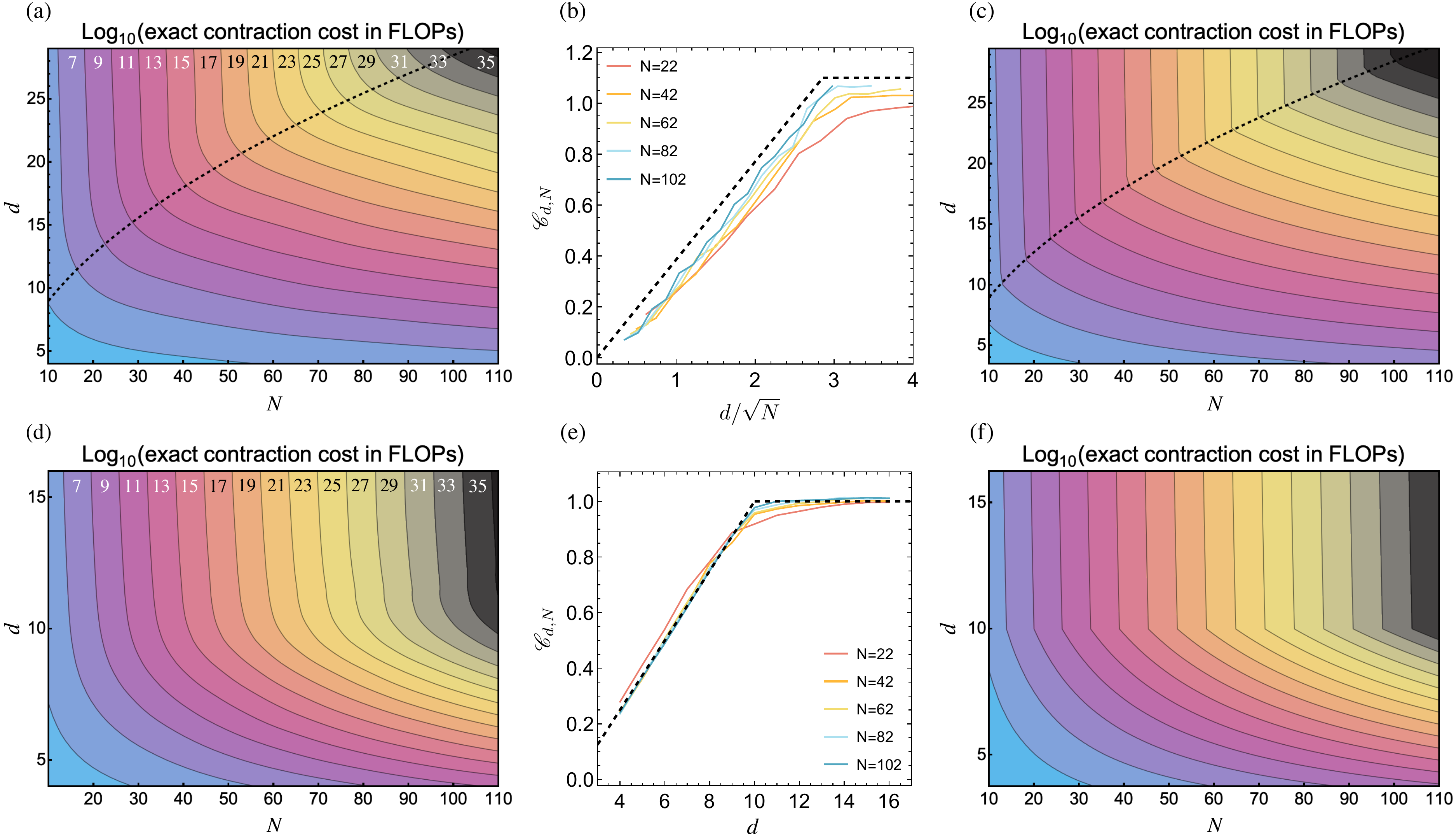}
\caption{Models of the exact contraction cost for both 2D (top) and RG (bottom) circuits.  (a,d) show the exact contraction cost estimated using cotengra for 2D and RG circuits, respectively.  (b,e) Show slices from those plots at various $N$, showing collapse to the simple models described in the text at large $N$.  (c,f) Show plots of the cost inferred from those simple models with parameters fit from panels (b,e), showing good qualitative agreement with the results in panels (a,d).
\label{fig:tts_model} }
\end{figure*}

Pick any pair of qubits $j_1,k_1$ that are gated together in the final layer of gates and contract all tensors corresponding to gates in those qubits' past causal cone. A representative example of such a causal cone is shown for a $10$ qubit depth-$3$ circuit in \fref{fig:light_cone}.  Denote the set of output qubits by $\mathcal{O}_1=\{j_1,k_1\}$ and the collection of input qubits in the past causal cone of $\mathcal{O}_1$ by $\mathcal{I}_1$. It is easy to see that these sets satisfy
\begin{align}
|\mathcal{O}_1|&=2\\
|\mathcal{I}_1|&\leq \min(2^d,N)
\end{align}
This contraction can be performed in a time-like direction with at most $|\mathcal{I}_1|$ legs protruding from any intermediate tensor (corresponding to unterminated wires of the input qubits in $\mathcal{I}_1$). Note that the final tensor formed this way has at most $|\mathcal{I}_1|-2$ open legs, since we close the two outputs associated with qubit $j_1$ and $k_1$. Now we pick a new pair of output qubits $j_2,k_2\notin\mathcal{O}_1$ that are gated together in the final layer, define $\mathcal{O}_2=\{j_2,k_2\}\cup\mathcal{O}_1$, and define $\mathcal{I}_2$ to be the set of all qubits in the past casual cone of $\mathcal{O}_2$.  Since $\mathcal{I}_2\backslash\mathcal{I}_1$ has at most $2^d$ qubits, continuing the contraction process in a time-like direction with all gates in $\mathcal{I}_2$ that have not already been contracted cannot add more than $2^d$ legs to the existing tensor. Therefore
\begin{align}
|\mathcal{O}_2|&=4\\
|\mathcal{I}_2|&\leq \min(N,2\times 2^{d}),
\end{align}
and no intermediate tensor can have more than $|\mathcal{I}_2|-|\mathcal{O}_1|$ legs (the ``$-|\mathcal{O}_1|$'' contribution because $2$ indices corresponding to the outputs at qubits in $\mathcal{O}_1$ have been fixed). Iterating this process, we find that at stage $\ell$
\begin{align}
|\mathcal{O}_\ell|&=2\times \ell\\
|\mathcal{I}_\ell|&\leq \min(N,\ell\times 2^{d}),
\end{align}
and no intermediate tensor can have had more than $|\mathcal{I_\ell}|-|\mathcal{O}_{\ell-1}|$ legs. At some point in this process the size of $\mathcal{I}_{\ell}$ will be equal to $N$, but that cannot happen unless $\ell\geq N/2^d$. At this stage, the largest number of legs in any intermediate tensor is upper bounded by
\begin{align}
|\mathcal{I}_{\ell}|-|\mathcal{O}_{\ell-1}|&\leq N-2(\ell-1)\\
&\leq N-2(N/2^d-1)\\
&\leq N(1-1/2^{d})+2,
\end{align}
giving
\begin{align}
\mathscr{C}_{d,N}\leq 1-1/2^d+2/N.
\end{align}
Ignoring the latter term in the limit $N\rightarrow\infty$ yields \eref{eq:upper_bound}.

\section{Simple models of exact contraction cost\label{sup:simp_cost}}

In \sref{sec:tts} of the manuscript we reported comparisons of achievable effective qubit numbers for 2D and RG quantum circuits.  Those comparisons are based on a simplified model of memory-unconstrained contraction cost that we now describe.  Figures \ref{fig:tts_model}(a) and \ref{fig:tts_model}(d) show exact contraction costs determined using cotengra for 2D circuits  and RG circuits, respectively (as described in the main text).

In 2D the complexity density $\mathscr{C}_{d,N}$ is expected to scale as $\min(1,c\times d/\sqrt{N})$, with $c$ an a priori undetermined constant. Therefore, defining $\tilde{d}=d/\sqrt{N}$, we expect a plot of $f_N(\tilde{d})\equiv\mathscr{C}_{\tilde{d}\sqrt{N},N}$ to show collapse to a universal curve (when plotted vs. $\tilde{d}$) as $N$ becomes large enough to suppress boundary effects.  Figure \ref{fig:tts_model}(b) shows exactly this behavior, and based on the largest-$N$ results we choose a (generous) envelope function $f(\tilde{d})=1.1\min(1.0,0.35 \tilde{d})$ (black dashed line in \fref{fig:tts_model}(b)). Assuming this heuristic model of the complexity density and converting back into $\log_{10}(\rm contraction~cost)$ we obtain the plot in \fref{fig:tts_model}(c), showing good qualitative agreement with the behavior in \fref{fig:tts_model}(a).

For random geometry circuits the complexity density is expected to scale independently of $N$ for large $N$, approaching $1$ for large $d$ with an a priori unknown functional form.  Empirically we see that $\mathscr{C}$ grows roughly linearly until $d\approx 10$, and after this depth it saturates extremely quickly to $1$.  Together with the observation that the cost is small and constant at $d=1,2$ \cite{markov2008simulating}, this behavior motivates us to model the cost with a simple functional form $\mathscr{C}_{d,N}\approx a\times \min(1,c (d-2))$. Evaluated with the constants $(a,c)=(1,0.125)$, this ansatz gives the black-dashed curve in \fref{fig:tts_model}(e), which agrees well with the largest $N$ calculations we performed. When converted into $\log_{10}(\rm contraction~cost)$ this ansatz yields the plot in \fref{fig:tts_model}(f), showing good qualitative agreement with the results obtained using cotengra in \fref{fig:tts_model}(d).

\section{Impact of gate choice on simulation hardness \label{sec:gate_choice}}

\begin{figure*}[!t]
\centering
\includegraphics[width=2.0\columnwidth]{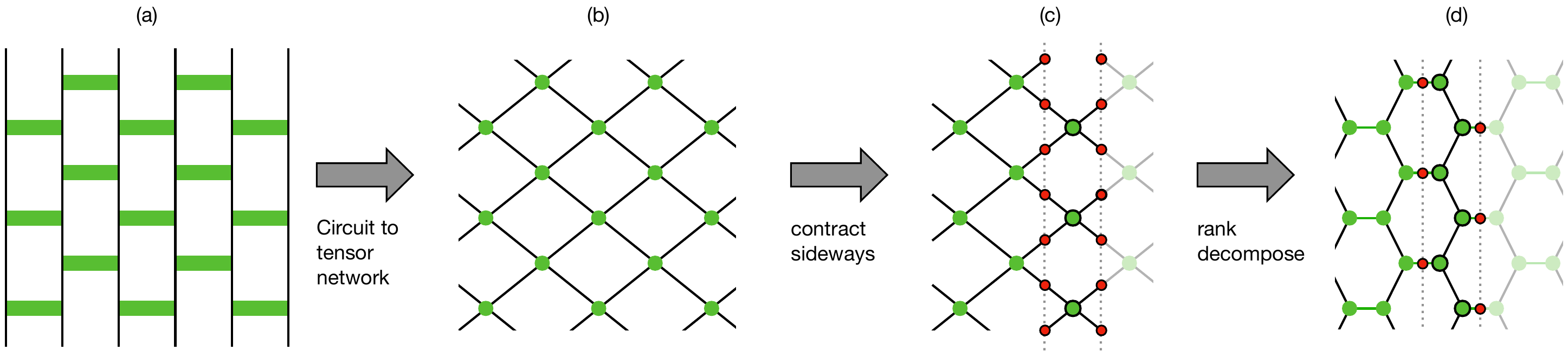}
\caption{Example of exact contraction cost dependence on gate rank for a 1D circuit. The circuit in (a) can be rewritten as the tensor network in (b) by assigning wires to legs and a rank-4 tensor to each 2Q gate. For sufficiently low depths, the tensor network in (b) can be contracted most efficiently by proceeding sideways. In (c) the dashed lines show intermediate tensors before and after application of tensors in a single vertical slice (black-outlined), and red circles show open indices both before and after application of all tensors in that slice. Each open index takes values in $\{0,1\}$ and there are $\sim d$ such indices for a depth-$d$ circuit, giving an overall contraction cost $\sim 2^d$.  The rank-4 gate tensors can always be Schmidt-decomposed into two rank-3 tensors, with those two tensors sharing a leg taking values $\{1,\dots,r_{\rm s}\}$, with $r_{\rm s}$ the Schmidt rank of the gate, as shown in (d). Sideways contraction now leads to intermediate tensors with $\sim d/2$ open indices, for a total contraction cost scaling as $\sim r_{\rm s}^{d/2}\leq 2^d$.
\label{fig:contraction_r2} }
\end{figure*}

At sufficiently high depths, optimal TN contraction is expected to proceed in a time-like direction, exhibiting computational costs approaching statevector simulation. In this limit, gate choice will have minimal impact on the cost of exact contraction.  On the other hand, at low depths the optimal contraction strategy will generally not be time-like, in which case it may benefit greatly from 2Q gates having low-rank decompositions.  A simple example is the 1D circuit shown in \fref{fig:contraction_r2}(a), which always admits the sideways contraction ordering shown in panel (c) with cost $2^d$. Decomposing the 2Q gates with Schmidt rank $r_{\rm s}$ enables a sideways contraction ordering with cost $\sim r_{\rm s}^{d/2}$.  For 2Q gates with $r_{\rm s}=4$, such as the iSWAP gate, $r_{\rm s}^{d/2} = 2^d$ and the contraction cost is the same for subfigures (c) and (d). For 2Q gates with $r_{\rm s}=2$, such as the $U_{\rm ZZ}(\pi/2)$ gate, $r_{\rm s}^{d/2} = 2^{d/2}$. Therefore, in this simple 1D example, using a rank-4 entangling gate rather than a rank-2 entangling gate effectively doubles the circuit depth.

\begin{figure}[!t]
\centering
\includegraphics[width=1.0\columnwidth]{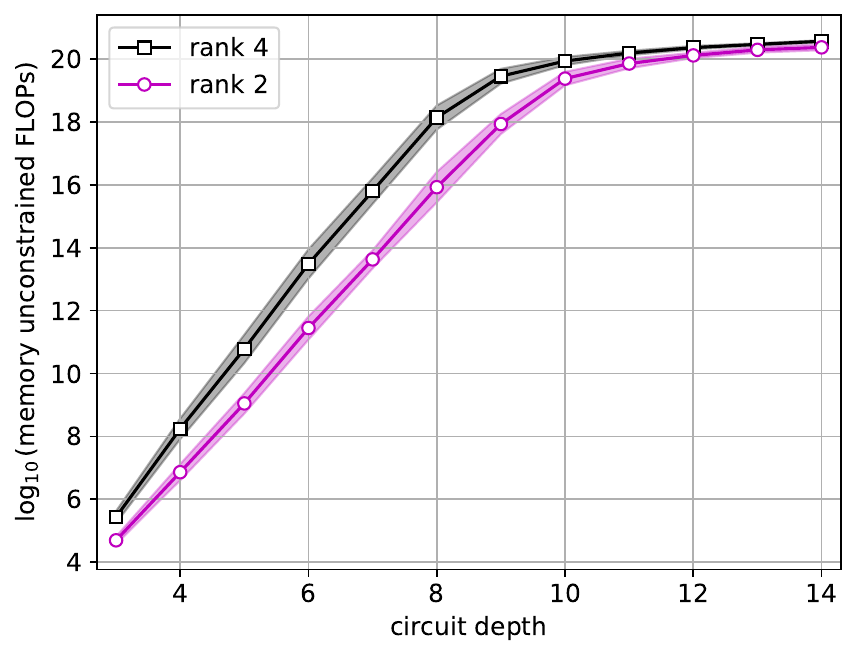}
\caption{Comparison of memory-unconstrained exact contraction costs for RG quantum circuits on $56$ qubits with rank-2 and rank-4 gates. Shaded regions represent the standard deviation on $\log_{10}({\rm FLOPs})$ computed across 20 random circuit realizations at each depth.
\label{fig:r2_v_r4} }
\end{figure}

The situation for circuits with more complicated geometries is less clear, and we have resorted to numerically studying the contraction cost dependence on gate rank by using the heuristic contraction cost optimizers provided by cotengra. Figure \ref{fig:r2_v_r4} compares the exact contraction cost of RG quantum circuits built of 2Q gates having rank-2 (circles) and rank-4 (squares) decompositions.  A natural choice of rank-4 gate would be the iSWAP gate, as it has a uniform (rank-4) Schmidt spectrum and therefore resists approximation strategies based on low-rank gate approximations. However, the iSWAP gate can be made rank-2 by factoring out a SWAP gate.  Factoring SWAP out of every 2Q gate would effectively map the original circuit on one random geometry and with rank-4 gates onto a new circuit on a \emph{different} but still random geometry with rank-2 gates, and this new circuit should have a similar exact contraction cost as the original circuit with rank-2 gates.  To avoid this simplification strategy we must use a rank-4 gate that remains rank-4 under SWAP factorization, and we choose a partial iSWAP for the numerical study in \fref{fig:r2_v_r4}. This choice is essentially arbitrary: In the space of all possible 2Q gates, all but a set of measure zero has the aforementioned property (being rank-4 and remaining rank-4 under SWAP factorization), and so nearly any 2Q gate would lead to the same exact contraction costs given by the black squares in \fref{fig:r2_v_r4}.

From \fref{fig:r2_v_r4} we see that, all else being equal, rank-4 gates should be preferred as they lead to higher contraction costs at fixed circuit depth. But all else is not equal; Quantinuum's native $U_{ZZ}(\theta)$ gate is not rank 4, and building a rank-4 gate requires at least two native gates.  In principle, it should be sufficient to append a single relatively weak entangler in a different basis to our native gate (e.g., $U_{ZZ}(\theta)\rightarrow U_{ZZ}(\theta)U_{XX}(\delta)$). However, even in the limit $\delta\rightarrow 0$ our partial entangler has a non-zero average infidelity of $\epsilon_0 \sim 4.6(6)\times 10^{-4}$ \cite{moses2023race}.  Thus, one can always achieve the same circuit fidelity using just the native 2Q gate at a depth increased by a factor of $r\equiv [\varepsilon_{2Q}+(5/4)\epsilon_0]/\varepsilon_{2Q}$.  Rescaling the depths of the rank-4 (black) curve in \fref{fig:r2_v_r4} by a factor of $r$ yields a result very similar to the rank-2 curve, suggesting that the advantage (in depth reduction at fixed contraction cost) of using a rank-4 gate is largely offset by the disadvantage (in achievable depth at fixed fidelity) due to the relatively lower rank-4 gate fidelity.  Therefore, we have opted to take the simpler route of applying a single $U_{ZZ}(\pi/2)$ to each qubit pair gated in each layer of our circuits (as described in Appendix~\ref{sec:rand_circs}).

\section{Comparison of various contraction-order optimizers \label{sup:optimizers}}

Finding truly optimal contraction orders for a generic tensor network is a hard problem, and the cost estimates for exact contraction given in the main manuscript are all based on heuristic algorithms deployed by cotengra for contraction order optimization.  To understand how the resulting FLOP costs are situated relative to other state-of-the-art TN contraction  optimization packages, we have made a variety of direct performance comparisons between cotengra, cuQuantum \cite{bayraktar2023cuquantum}, tamaki \cite{tamaki2019positive}, and multilevel \cite{ibrahim2022constructing}. State-of-the-art tensor network contraction algorithms used for simulating random circuit sampling use various techniques to achieve sublinear costs in the number of samples \cite{liu2022validating,pan2022solving,kalachev2021classical,Kalachev2021multi,huang2020classical,vincent2022jet}, including sparse states, caching, etc. Here, we focus on single amplitude contraction optimization since very few samples (20 in this paper) are produced from each unique circuit.

The most relevant cost comparison is actual contraction time, which requires executing the contraction order obtained on classical computing hardware. However, optimizers such as tamaki and multilevel do not integrate tensor network slicing directly, which means that the contraction order is not readily usable. Another metric that can be directly compared across all four optimizers considered is the minimum width of a tree decomposition obtained by the optimizer, which determines the largest intermediate tensor size required for the contraction order found.

\begin{figure}[!ht]
    \begin{center}
    \includegraphics[width=\columnwidth]{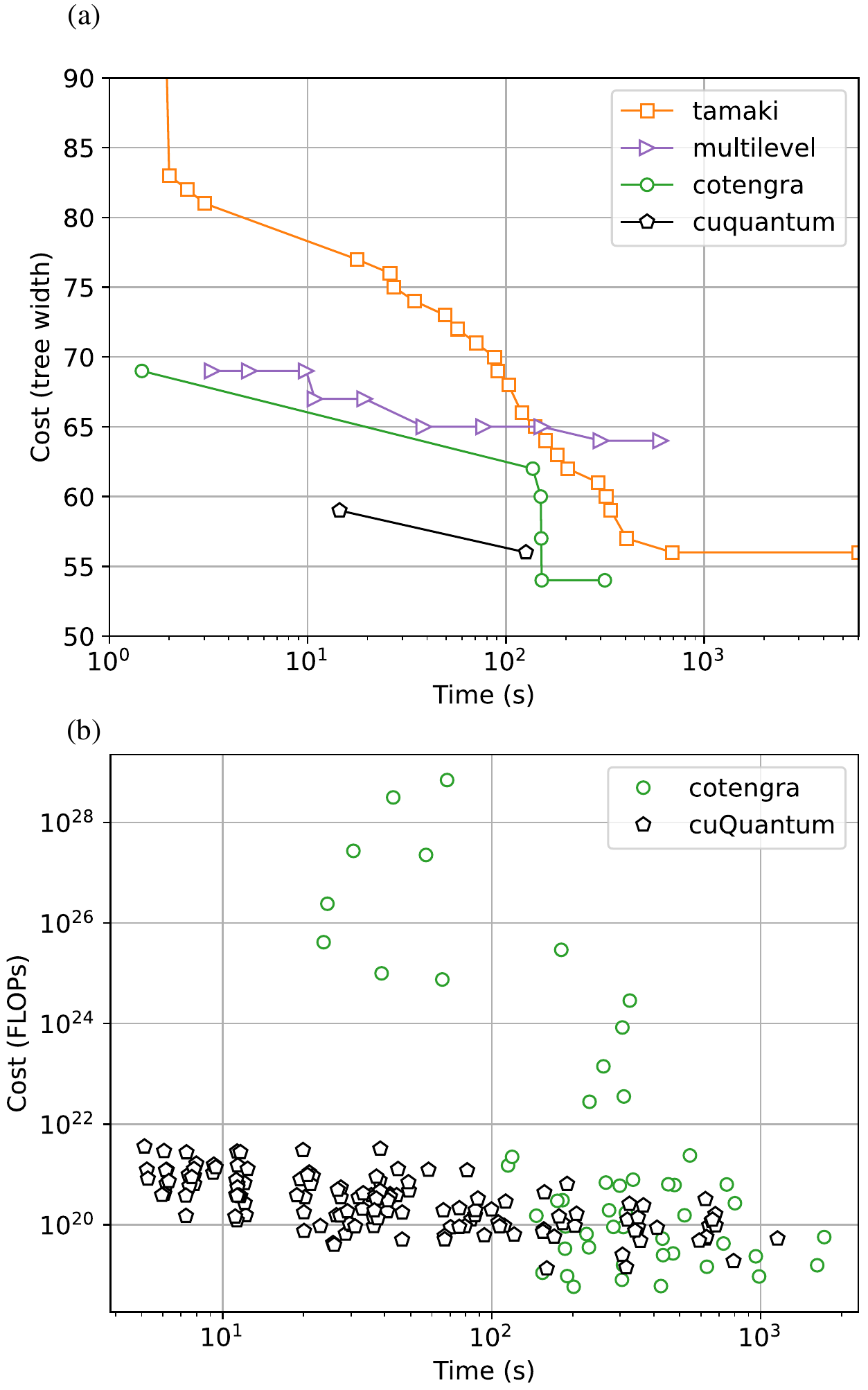}
    \end{center}
    \caption{Optimized contraction cost as a function of optimization time using a single CPU thread of an AMD EPYC ``Milan" processor. (a) Treewidth for a single quantum circuit without slicing. (b) FLOP count for 10 quantum circuits with slicing.}
    \label{fig:optimizer_benchmark}
\end{figure}

We choose a randomly generated $56$ qubit depth-$10$ RG circuit, and allow the corresponding tensor network (without simplification, since tamaki and multilevel do not support it yet) to be optimized by each optimizer.  Fig. \ref{fig:optimizer_benchmark}(a) shows the width of the lowest-width tree decomposition obtained after a given amount of optimization time, and demonstrates that cuQuantum \cite{bayraktar2023cuquantum} and cotengra \cite{Gray2021hyperoptimized, gray2024hyperoptimized} generally outperform tamaki and multilevel given fixed optimization time. Further, these optimizers supports slicing and tensor network simplification out of the box. However, we have not been able to successfully utilize simplification of circuits containing many diagonal gates (such as $U_{ZZ}$ gates) in cuQuantum. Therefore, we concentrate on comparing cuQuantum against cotengra for tensor networks without simplification.

To better compare cuQuantum and cotengra, we generate 10 random $(N=56,\,d=10)$ circuit instances to capture random fluctuations in the optimized costs (due to both the differences in circuits and randomness in the optimizer). Both optimizers readily return the contraction cost in FLOP counts, which is reasonable proxy for the actual contraction time, and the results are shown in Fig.~\ref{fig:optimizer_benchmark} (b). We observe that at very short times, cuQuantum is capable of producing good optimization results. On the other hand, while cotengra requires more time to produce good results, it is generally very competitive with cuQuantum if both are allowed $\gtrsim 100$ seconds of optimization time.

\begin{figure*}[!t]
\includegraphics[width=1.95\columnwidth]{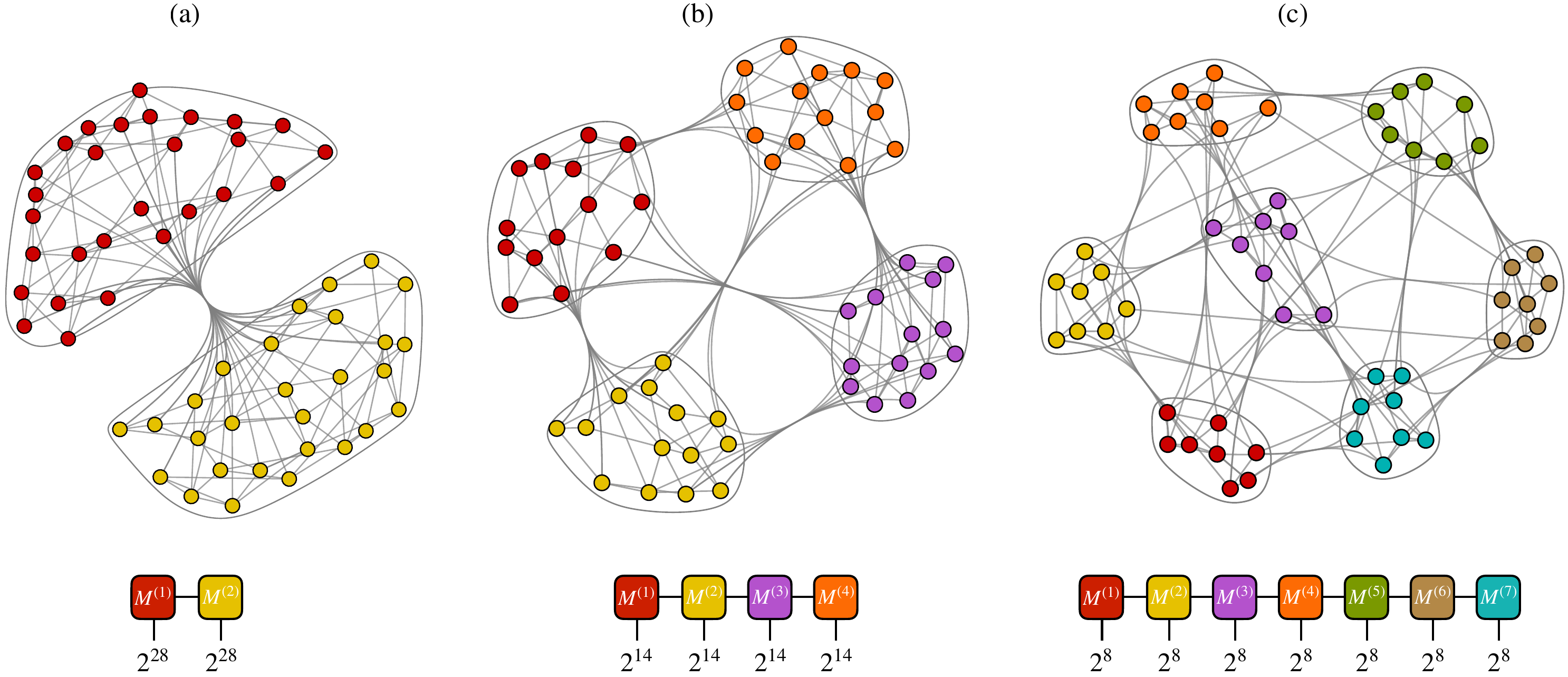}
\caption{Graph partitioning for blocked MPS simulations of a depth $5$ circuit on $56$ qubits.  The graph from which the circuit is constructed (shown without edge coloring) is partitioned into $b=2$ (a), $b=4$ (b), or $b=7$ (c) using KaHyPar. The objective is to minimize the number of inter-block edges, which ultimately (upon assigning blocks to MPS tensors) minimizes the number of gates that must be applied between MPS tensors.}
\label{fig:mps_blocking}
\end{figure*}

\section{Achievable fidelities with DMRG \label{sup:mps_bound}}

Time-evolving an approximate tensor network ansatz is a promising method to simulate quantum circuits with a limited fidelity. The most well studied tensor networks in the context of quantum circuit simulations are matrix product states (MPS) and projected entanglement pair states (PEPS). For simulating transition amplitudes in random circuit sampling, the best performing algorithm that we are aware of is the variant of time-dependent DMRG presented in Ref.~\cite{PhysRevX.10.041038, PRXQuantum.4.020304}, which is based around an approximate MPS representation of the full quantum state.  Compared to exact simulation, the difficulty of approximate simulation~---~especially in the regime of sub-maximal entanglement~---~is much more difficult to reliably assess.  This appendix elaborates on the methods used in \sref{sec:DMRG} of the manuscript, and extends them by estimating how the cost of approximate simulation grows with circuit depth for RG circuits using a combination of DMRG and the bounds on achievable MPS fidelities described in Ref.~\cite{morvan2023phase}.

An MPS representation of a quantum state can be chosen in many different ways, and we consider blocking of the qubits into $b$ blocks $[\mathcal{S}_1,\dots,\mathcal{S}_b]$ of size $[s_1,\dots,s_b]$, where each set $\mathcal{S}_j$ contains the qubits in the associated block and is assigned to a single MPS tensor. In the event that all blocks are the same size, we refer to a given blocking scheme compactly as $b\times[s]$. We denote by $\mathcal{I}_j$ a super-index containing a single binary-valued index for each qubit in the associated block $\mathcal{S}_j$. Each block can then be assigned a tensor $M^{(j)}_{\alpha_j,\mathcal{I}_j,\beta_j}$, and the full quantum state is approximated as a contraction over these tensors as

\begin{align}\label{MPS}
\ket{\Psi}=&\sum_{\mathcal{I}_1,\dots,\mathcal{I}_b}\sum_{\alpha_1=1}^{\chi}\dots\sum_{\alpha_{b-1}=1}^{\chi}M_{\mathcal{I}_1,\alpha_1}^{(1)}M_{\alpha_1,\mathcal{I}_2,\alpha_2}^{(2)}\dots \nonumber \\
&M_{\alpha_{b-1}\mathcal{I}_b}^{(b)}
\ket{\mathcal{S}_1}\otimes\cdots\otimes\ket{\mathcal{S}_b}.
\end{align}
Here $\chi$ is the bond dimension, which we assume to be the same for all tensors but in general need not be. An MPS of this form, and its relation to the blocking of qubits, is illustrated graphically for $b=2,4,7$ in \fref{fig:mps_blocking}(a,b,c).  A blocked MPS can always be expanded into a more standard single site MPS, with the blocked qubits corresponding to groups of adjacent single-site tensors. In this view, the blocking effectively allows the nominal inter-block bond dimension $\chi$ to grow larger within the blocks; in the presence of inhomogeneous entanglement structure, blocking that is congruent with such structure can plausibly improve the fidelity of an MPS at fixed total resources.  For a given blocking strategy $b\times[s]$, we assign qubits to blocks by using the open source hypergraph partitioning package  KaHyPar to minimize the number of inter-block 2Q gates.  Examples of such partitionings, and their correspondence with blocked MPS are shown in \fref{fig:mps_blocking} for $b=2,4,7$.

We say that a given quantum computer can be ``classically spoofed'' if, given a random circuit $C$, classical RCS can be achieved with comparable fidelity to that achieved by the quantum computer (e.g., as measured by the achievable XEB of samples) using a comparable amount of time per sample. One approach to spoofing would be to evolve an MPS through the circuit to approximate $\ket{\Psi}=C\ket{0}$, and then query the amplitudes of $\ket{\Psi}$ in the computational basis. While this approach has the benefit that sampling can be performed very directly, Ref.~\cite{PRXQuantum.4.020304} advocated an alternate approach in which the circuit is split into lower depth circuits, for example as $C=C_2 C_1$.  By approximating (through MPS evolution) both $\ket{\Psi_1}=C_1\ket{0}$ and $\ket{\Psi_2}=C_2^{\dagger}\ket{\bm{x}}$, computational basis state amplitudes at the output of $\mathcal{C}$ can then be computed from the overlap of those states as
\begin{align}
\label{eq:circ_split}
\bra{\bm{x}}C_2 C_1\ket{0}=\langle \Psi_2|\Psi_1\rangle,
\end{align}
and such calculations can then be used (e.g., in combination with rejection sampling) to draw samples from the output distribution of the circuit. An improved variation of this simulation scheme is to split the circuit into three pieces, as $C=C_2 C_M C_1$, computing amplitudes as expectation values $\bra{\Psi_2}C_M\ket{\Psi_1}$. However, we have found that for circuits with random geometries, having even a single layer of gates in $C_M$ increases the contraction cost dramatically, and we have only attempted simulations with the circuit decomposed as in \eref{eq:circ_split}.

\begin{figure}[!t]
\includegraphics[width=0.98\columnwidth]{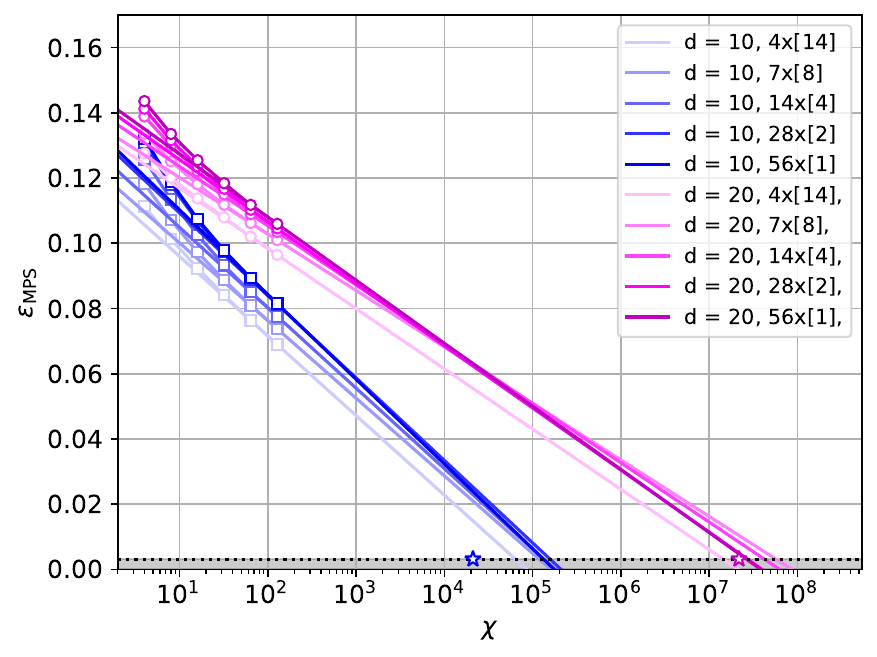}
\caption{Error per gate $\varepsilon_{\rm MPS}$ in DMRG simulations for depth-10 (blue) and depth-20 (magenta) circuits using a variety of blocking strategies.  The lines are a linear fit to the largest two bond dimensions plotted for each blocking strategy ($\chi=64,128$), and provide estimates of the bond dimension required to achieve the experimental error rates (dotted horizontal line at the bottom of the plot). The stars are lower bounds on the required bond dimension for an optimal bipartite MPS extracted from Clifford simulations (see \fref{fig:clifford_purity} and supporting text).
\label{fig:chi_extrap}}
\end{figure}

The DMRG calculations reported in the paper can be used to estimate the required bond dimension to achieve comparable circuit fidelities to those achieved in the experiment.  In particular, one can extrapolate the error per gate as a function of bond dimension to the $\varepsilon \approx 3.2\times 10^{-3}$ value measured in the experiment for various blocking choices at different depths. Figure (\ref{fig:chi_extrap}) shows such an extrapolation for circuits of depth $10$ (where the distribution of output probabilities becomes well converged to Porter-Thomas \emph{and} the exact contraction cost first becomes nearly saturated) and depth $20$ (the deepest circuits for which we report experimental samples). We find only a weak dependence of the extrapolated bond-dimension on the blocking choices, and in general find that bond dimensions in excess of $10^{5}$ ($10^7$) are likely necessary for classical spoofing of H2 via DMRG given circuits of depth $10$ $(20)$. Note that all of the blocking results lead to error-per gate curves that are concave up, and extrapolation from relatively small bond dimensions may well bias the extrapolated bond dimension down relative to the true requirements.

In Ref.~\cite{morvan2023phase}, the authors consider the bond dimension requirements to write down a bipartite MPS $\ket{\Phi_{\rm MPS}}$ that approximates an arbitrary state $\ket{\Psi}$ with fidelity $f=|\bra{\Phi_{\rm MPS}}\Psi\rangle|^2$. In particular, they show that the fidelity is bounded above as $f^2\leq \chi {\rm Tr}(\rho^2)$, where $\rho$ is the reduced density matrix of the exact state $\ket{\Psi}$ on either side of the bipartition. Since the computation of amplitudes by evolving two MPSs toward the middle of the circuit achieves an effective simulation fidelity $F\sim f^2$ (being diminished by the loss of fidelity of both the forward and backward evolutions), they conclude that the simulation fidelity $F$ achievable at fixed bond dimension is bounded above as $
F\leq \chi\,{\rm Tr}(\rho^2)$.

\begin{figure}[!t]
\includegraphics[width=0.98\columnwidth]{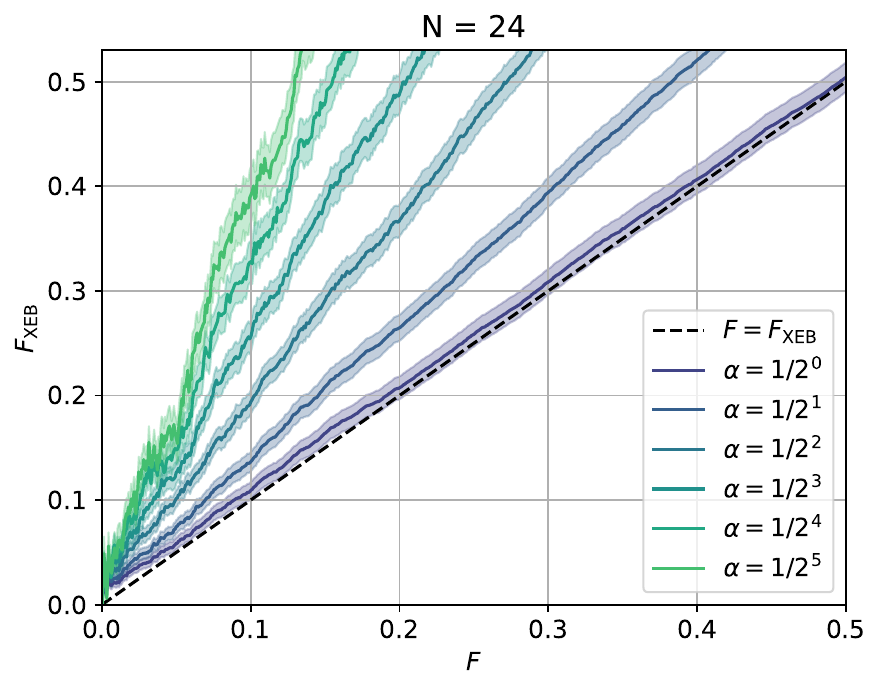}
\caption{Achievable $F_{\rm XEB}$ by sampling from bitstrings whose amplitudes have been calculated using a closed MPS simulation at fidelity $F$. The various curves represent the achievable improvements by post-processing the sampled bitstrings, showing that fidelity improvements by post-selecting on large probabilities require exponential post-selection overhead (i.e.\ the ratio of $F_{\rm XEB}/F$ increases roughly linearly as the fraction of retained data, $\alpha$, is decreased exponentially.)
\label{fig:spoofing}}
\end{figure}

Discrepant accounts of the relationship between the achievable MPS fidelity and the associated $F_{\rm XEB}$ have been reported in the literature (see Refs. \cite{PRXQuantum.4.020304} and \cite{morvan2023phase}), though we note that the fidelities here are sufficiently large that the discrepancies between these two references are relatively inconsequential. We also performed numerical simulations for our circuits in the range of fidelities relevant to the experiment, and find that $F=f^2$ serves as an excellent proxy for the achievable values of $F_{\rm XEB}$ from sampling based on closed-simulation amplitude calculations.  For these calculations, we consider RG circuits on $24$ qubits at depth $10$ (at which point the output distribution of the exact circuits is well converged to the Porter-Thomas distribution).  We sample a random bitstring $\bm{x}$, and then evolve $\ket{0}$ (forward) and $\ket{\bm{x}}$ (backwards) to the midpoint of the circuit.  Both states are then optimally approximated by a bond dimension $\chi$ MPS via direct singular-value decomposition (SVD), and we record both the estimated probability for bitstring $\bm{x}$ (squared overlap of the two MPSs) and the fidelity $F=f_0 f_{\bm{x}}$, where $f_{0({\bm x})}$ is computed from the discarded weight of the SVD for the time-evolved $\ket{0(\bm{x})}$ state.  We compute the amplitudes of 1000 randomly chosen bitstrings for 100 random circuits (for a total of 100,000 amplitudes), and then for each circuit we estimate the achievable $F_{\rm XEB}$ by resampling from the associated list of 1000 estimated probabilities and computing $F_{\rm XEB}$ from those samples. In \fref{fig:spoofing}, the lowest (bluest, $\alpha=1$) solid curve is a parametric plot of $F(\chi)$ versus $F_{\rm XEB}(\chi)$ averaged over the 100 sampled circuits (with the shaded region indicating the standard error on the mean over those circuits), showing excellent correlation.
In general, various post-processing strategies might be employed to increase the achievable $F_{\rm XEB}$ given the list of estimated probabilities. The other curves ($\alpha<1$) in \fref{fig:spoofing} are obtained by post-processing the MPS estimates of the 1000 bitstring probabilities computed for each circuit by retaining just the largest $\alpha\times1000$ largest samples (similar to the top-$k$ post-processing method described in Ref.~\cite{gao2021limitations}).  While such post-processing can produce $F_{\rm XEB}$ values larger than the achieved simulation fidelity $F$, the linear growth of $F_{\rm XEB}$ apparent in \fref{fig:spoofing} as $\alpha$ is decreased exponentially suggests that the retained fraction of bitstrings must shrink exponentially in the desired multiplicative enhancement of the fidelity.  This behavior is expected for exact sampling given the exponential decay of the exact Porter-Thomas distribution, which makes sampling bitstrings with probabilities $R$ times larger than $1/2^N$ exponentially unlikely in $R$. We attribute the persistence of this behavior in approximate sampling to the distribution of approximated probabilities also having an exponential tail.

\begin{figure}[!t]
\includegraphics[width=\columnwidth]{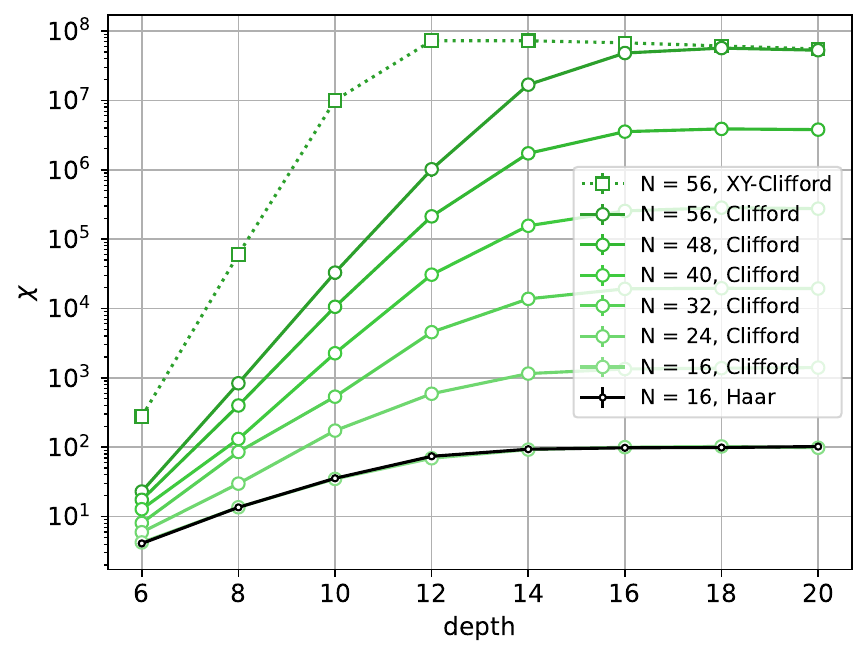}
\caption{Clifford simulations of the average bipartite purity (converted into a minimum bond dimension $\chi$ using \eref{eq:purity_bnd_bound}) for our circuits at various system sizes.  All curves are averaged over 100 random geometries and 10 random assignments of 1Q gates per geometry, with the optimal blocking determined for each geometry independently. The black curve shows simulations with Haar-random 1Q gates for $N=16$, confirming the equivalence with averages over Clifford circuits.
\label{fig:clifford_purity}}
\end{figure}

Given the exponentially large cost of enhancing $F_{\rm XEB}$ via post-processing, we see no obvious better strategy to achieve (on average across circuits) a given XEB score $F_{\rm XEB}$ than simply choosing a bond dimension large enough that the MPS fidelity $F=f_0f_{\bm{x}}$ is comparable to $F_{\rm XEB}$. Returning to the analysis of Ref.~\cite{morvan2023phase} and the bound $
F\leq \chi\,{\rm Tr}(\rho^2)$ derived therein, one should therefore use a bond dimension bounded below as
\begin{align}
\label{eq:purity_bnd_bound}
\chi\geq F/\overline{{\rm Tr}(\rho^2)},
\end{align}
where the overline denotes an average over circuits. As a result of the 2-design property of the Clifford group, $\overline{{\rm Tr}(\rho^2)}$ can be computed efficiently for our circuits by replacing the Haar random 1Q gates with i.i.d.\ random 1Q Clifford gates (see for example Ref.~\cite{PhysRevX.7.031016}), and then performing the average numerically using efficient Clifford simulations. Lower bounds on the bond dimension from such simulations are shown for $n\in\{16,24,32,40,48,56\}$ in \fref{fig:clifford_purity}. For $56$ qubits, this bound is also shown in \fref{fig:chi_extrap} at depths $d=10,20$ (blue and magenta stars, respectively). In practice, a DMRG simulation may under perform relative to this bound for several reasons, including: (1) The bound is only tight for states with a flat Schmidt spectrum, which will not be encountered for the non-Clifford circuits considered in this manuscript, and (2) DMRG is a greedy algorithm that may not produce the globally optimal MPS.  Indeed, we see that for a variety of blocking strategies, extrapolation of the DMRG results to determine what bond dimension is required to achieve the experimentally measured fidelities indicates that the bound is \emph{not} saturated in practice (at least in our implementation of the DMRG algorithm), requiring bond dimensions in the neighborhood of $10^5$ even for depth-$10$ circuits.  While it is not entirely clear how far the bond dimension in DMRG simulations could be pushed in practice, we note that the largest MPS-based simulation performed to date (by a large margin) used a bond dimension of $2^{16} =65,536$ \cite{ganahl2023density}. Utilizing approximately $1000$ tensor-processing-unit cores, this calculation took on the order of minutes for a single MPS optimization, of which many are required to simulate even a depth-$10$ circuit. While this time could surely be brought down to some extent, we note that the deeper circuits here require \emph{much} larger bond dimensions (e.g. the magenta lines and star in \fref{fig:chi_extrap}).  Moreover, slight modifications of the 1Q gate set can lower the depth at which the purity saturates to its Haar-random ($\sim 1/2^{N/2-1}$) value.  For example, fixing the 1Q gates to be $\pi/2$ rotations about axes in the $xy$ plane of the Bloch sphere (as in Ref.~\cite{morvan2023phase}) avoids situations~---~encountered early in circuits with the Haar-random SU(2) gates used here~---~in which the 1Q gates leave the $\ket{0}$ state close to the south or north pole of the Bloch sphere (which then reduces the amount of entanglement the $U_{ZZ}(\pi/2)$ gates can generate). Simulations using a Clifford 1Q gate set restricted to such rotations yield the dotted line in \fref{fig:clifford_purity}, showing that even depth-$12$ circuits (the same depth at which the cost of memory-unconstrained contraction saturates) can require near-maximal bond dimensions.

\bibliography{RCS}

\section*{Disclaimer}
This paper was prepared for informational purposes with contributions from the Global Technology Applied Research center of JPMorgan Chase \& Co. This paper is not a product of the Research Department of JPMorgan Chase \& Co. or its affiliates. Neither JPMorgan Chase \& Co. nor any of its affiliates makes any explicit or implied representation or warranty and none of them accept any liability in connection with this position paper, including, without limitation, with respect to the completeness, accuracy, or reliability of the information contained herein and the potential legal, compliance, tax, or accounting effects thereof. This document is not intended as investment research or investment advice, or as a recommendation, offer, or solicitation for the purchase or sale of any security, financial instrument, financial product or service, or to be used in any way for evaluating the merits of participating in any transaction.

\end{document}